\pdfoutput=1

\documentclass[%
 reprint,
 amsmath,amssymb,
 aps,
 longbibliography,
]{revtex4-1}

\usepackage[T1]{fontenc}
\usepackage[latin9]{inputenc}
\usepackage{textcomp}
\usepackage{mathtools}
\usepackage{amsmath}
\usepackage{amssymb}
\usepackage{stackrel}
\usepackage{graphicx}
\usepackage{url}

\pdfpageheight\paperheight
\pdfpagewidth\paperwidth

\usepackage{babel}

\begin{document}
\title{Error correction of parity-encoding-based annealing through post-readout decoding}
\author{Yoshihiro Nambu}
\affiliation{NEC-AIST Quantum Technology Cooperative Research Laboratory~\\
 National Institute of Advanced Industrial Science and Technology }
\begin{abstract}
Lechner, Hauke, and Zoller proposed a parity-encoded spin-embedding
scheme for quantum annealing (QA) with all-to-all connectivity
to avoid the issue of limited connectivity in near-term QA hardware
and to enable the implementation thereof using only geometrically 
local interactions between spins fabricated on the planar substrate. 
Nevertheless, the redundant
encoding of logical information, i.e., using a large number of spins
to embed the logical information, 
increases the computational cost and reduces the efficiency.
In this study, we show through Monte Carlo simulation that this redundant
encoding may be exploited to solve the problems of  the inefficiency and 
computational cost of the parity-encoded scheme by incorporating appropriate
decoding, namely classical post-processing, of the spins to retrieve the logical
information. Our findings open up the possibility of parity-encoded schemes 
for realizing the QA with near-term quantum technologies.
\end{abstract}
\maketitle

\section{Introduction\label{sec:1}}

In recent years, there has been a growing demand to solve industry-relevant
combinatorial optimization problems. It is well known that such problems
can be recast as a sampling of the ground state of the Ising spin Hamiltonian,
where the problem is encoded in the coupling coefficients of the spin
pairs \cite{Kirkpatrick83}. Many studies have been devoted to nature-inspired
approaches to sample the ground state efficiently, such as quantum
annealing (QA) \cite{Kadowaki98}. However, coding an arbitrary optimization
problem into Ising models requires the dataset of the coupling coefficients
between arbitrary pairs of spins in the Ising spin Hamiltonian, and
solving them using QA computers is highly demanding for near-term QA
devices. This is because it requires a physical implementation of
the long-range interaction between any pair of spins. To avoid
this problem, several schemes, such as minor embedding (ME) \cite{Choi2008,Choi2011}
and parity-encoded (PE) schemes, have been proposed \cite{Lechner15},
which logically embed the spins into a larger physical spin system to avoid
long-range interactions. 

In the ME scheme, each fully connected spin (logical spin)
is encoded on an associated chain of physical spins, and physical
spins belonging to the same chain are connected by strong ferromagnetic
interactions to align them energetically. Long-range interactions
between logical spins are replaced by short-range interactions between
chains of physical spins. As a result, a general Ising problem that is defined
on a fully connected graph of many nodes can be described by an equivalent
problem that is defined on a sparse graph with a larger number of nodes \cite{Vinci2015}. 
In contrast, in the PE scheme, which was first proposed by Sourlas \cite{Sourlas2005}
and later by Lechner, Hauke, and Zoller (LHZ) \cite{Lechner15}, the parity
of each pair of logical spins is encoded on an associated physical
spin that is arranged on a two-dimensional lattice. Four nearest-neighbor
physical spins are coupled by four-body interactions and constitute
a PE spin network. Although this approach can avoid long-range interactions, it
requires four-body interactions, which appears physically unnatural
and unfeasible. However, Puri et al. later suggested that the PE spin
network is realizable by mapping it to a network of two-photon-driven
Kerr parametric oscillators (KPOs) with only local interactions, where
a four-body interaction is mapped to a four-wave mixing of four-mode oscillator
fields in the Josephson junction coupler \cite{Puri2017}. Such a
network of KPOs can be implemented using a circuit QED platform
employing superconducting devices, which has been studied extensively 
to realize quantum computing with the current noisy intermediate-scale quantum devices \cite{Albash16,Lechner15}. 
The KPOs have been experimentally realized by using trapped ions \cite{Ding2017} and Josephson parametric oscillators \cite{Grimm2020,Frattini2022,Wang2019,Yamaji2022}.
A network of two KPOs has been recently realized using a static capacitive coupling \cite{Yamaji2023}.
Chancellor, Zohren, and Warburton independently designed a four-body coupling of a
superconducting flux quantum bit (qubit), which is a more naive model of the quantum spin,
using additional ancillary qubits \cite{Chancellor2017}. In the
PE scheme, the optimization problem is encoded in the dataset of
the magnitudes as well as orientations of the local fields that act independently
on each physical spin, whereas problem-independent four-body couplings
among the neighboring physical spins enforce the parity constraints.

In this study, we investigated another aspect of the PE scheme, that is,
its error-correcting capability, by comparing it with that of the quantum annealing
correction (QAC) scheme \cite{Jordan2006,Young2013,Pudenz2014,Pudenz2015,Vinci2015,Bookatz2015,Matsuura2016,Matsuura2017,Pearson2019}.
QAC is a QA system that is merged with the classical repetition
code to mitigate the errors arising in the QA algorithm. Its
error-correcting capability has been discussed in terms of that
of the classical repetition code. The error-correcting capability
of the PE scheme is more subtle. LHZ suggested that the PE scheme includes
intrinsic fault tolerance because the logical spins are encoded
redundantly and nonlocally in the physical spins. Pastawski and Preskill
suggested that the PE scheme can be viewed as a classical low-density
parity-check (LDPC) code that makes the scheme highly robust against weakly
correlated spin-flip noise \cite{Pastawski2016}. However, a numerical
study using simulated QA and parallel tempering showed
no clear evidence that intrinsic fault tolerance
can be used to improve the sampling efficiency of the ground state of logical spins
\cite{Albash16}. To elucidate this inconsistency, we conducted discrete-time
Markov chain Monte Carlo (DT-MCMC) simulations to sample the sequence of
state of physical spins under various constant simulation parameters.
As our MCMC is reversible, the sequence converges with the stationary
distribution related to the Boltzmann distribution. Through the MCMC
simulation, we stored the states of the physical spins sampled
during the MCMC iteration in the buffer memory. After
completing the simulation, we decoded the physical
state using one-step majority vote decoding (one-step MVD) \cite{Massey1963,Massey1968,Laferriere1977,Clerk1981,Shu2004}
to infer the state of the logical spins and calculated their energies
for all the elements in the stored sequence of the physical states.
Their energy spectra indicated that a substantial number of non-code
states contributed to the logical ground state after the decoding. In
particular, we found that the optimal simulation parameters for sampling
the physical ground state using no post-readout decoding and the logical ground states using one-step
MVD are very different.

We investigated our one-step MVD in detail and unfortunately found a drawback in that our method is not robust, that is, its performance is dependent on the problem instance. However, this drawback can be addressed by modifying the algorithm to consider high-weight parity-check constraints. We explain the necessary modification to solve the problem and demonstrate that the error-correcting capability of the algorithm is improved. We speculate that an unfair choice of a member of spin variables for majority voting or inadequate selection of simulation parameters may account for the inconsistent conclusions of previous studies and our study.

The remainder of thie paper is organized as follows. In Sec. \ref{sec:2}, we introduce
the formulation known as soft annealing \cite{Sourlas2005}, which
explains how to incorporate error-correcting codes with logical spin
systems and obtain larger physical spin systems. In this formalism,
prior knowledge of the parity constraints that characterize the
codes are naturally incorporated into the Hamiltonian of the physical
spin systems. In Sec. \ref{sec:3}, we discuss the post-readout decoding.
We consider one-step MVD as an example and explain how the prior knowledge about parity constraints can be used to infer
the correct result. In Sec. \ref{sec:4}, we demonstrate the results of the simulation
of one-step MVD when it is applied to the PE scheme. We also propose
a novel method for improving the efficiency and robustness of one-step
MVD for PE schemes. Section \ref{sec:5} concludes the paper. In addition,
we describe our simulation method, i.e., rejection-free
Markov chains, in detail in the Appendix.

\section{Classical error-correcting code and soft annealing\label{sec:2}}

\begin{figure*}
\begin{centering}
\includegraphics[viewport=80bp 120bp 880bp 410bp,clip,scale=0.55]{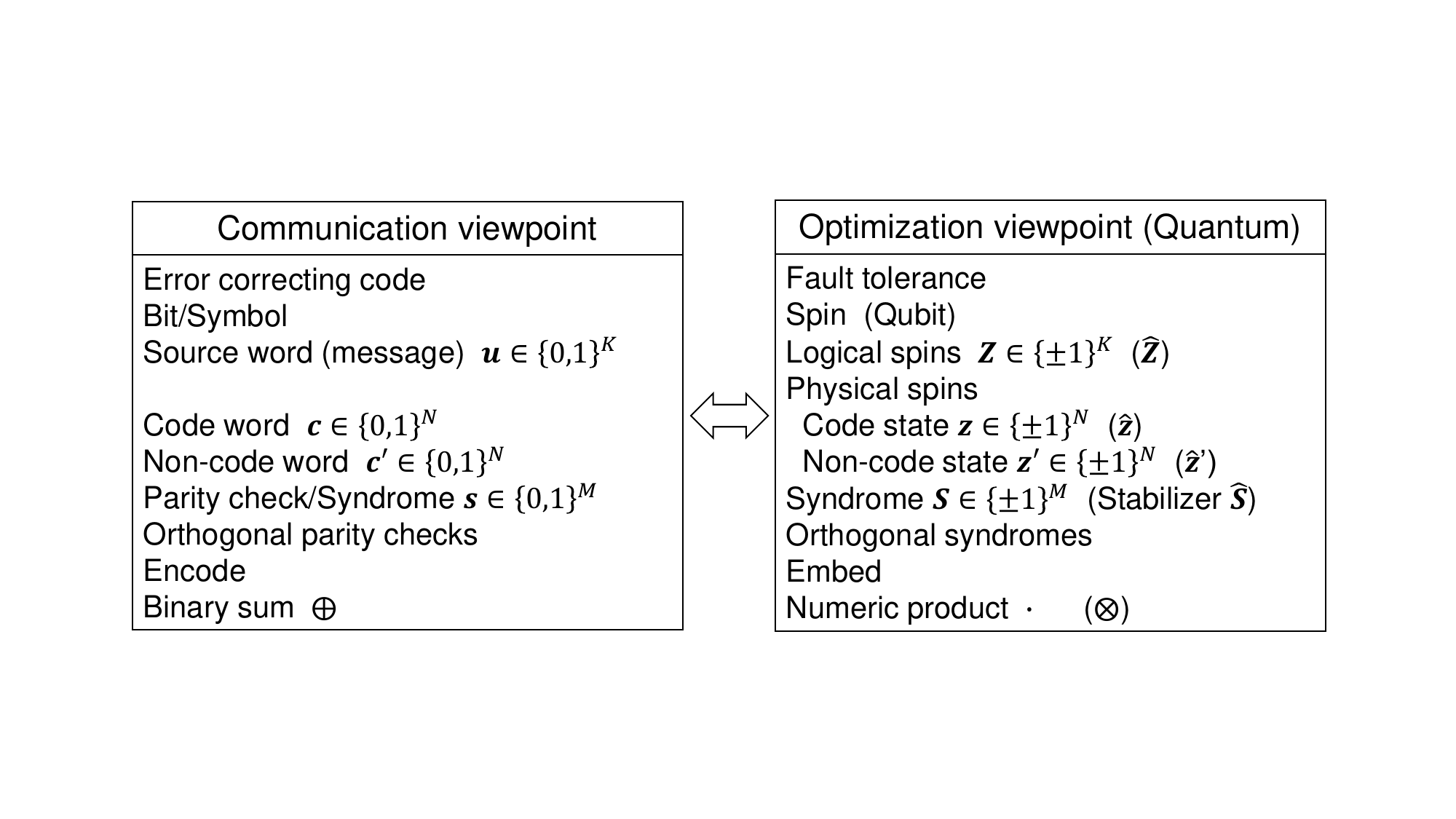}
\par\end{centering}
\caption{One-to-one correspondence between the concepts used in the two viewpoints
discussed in this paper. Left hand side: concepts used in the
communication application viewpoint. Right hand side: concepts used in the
optimization application viewpoint. The symbolic notations used in
this paper are also shown for the reader's reference. The symbols
in parentheses are used in the quantum mechanical version.\label{fig:1}}
\end{figure*}

Before starting our discussion, we must address two major viewpoints
regarding the error-correcting codes considered in this study. One is the communication
application viewpoint, where the mathematical model is discussed in terms of a collection of bits or symbols. The other is the optimization
application viewpoint, where the mathematical model is discussed in terms of a system of spins or qubits. The purpose of
this study is to examine how the findings obtained by the former can be
extended to the latter. The
concepts used in these two viewpoints have one-to-one correspondence. Figure \ref{fig:1} provides
an overview of the correspondence between the concepts used in the two
viewpoints for the reader's reference. For example, ``source word''
and ``code word,'' which are concepts used in the communication
application viewpoint, are associated with ``logical spins'' and
``physical spins'' used in the optimization application
viewpoint, respectively. We begin our discussion by introducing
the concept of soft annealing proposed by Sourlas \cite{Sourlas2001,Sourlas2002,Sourlas2005}. In this
proposal, the formulation of the error-correcting codes given by a
collection of bits or symbols is mapped to the formulation given
by a system of spins or qubits. Based on this mapping, we discuss
the constructive contribution of error-correcting codes 
to the optimization application. 

\subsection{Soft annealing}

Sourlas established a mathematical equivalence between the classical
error-correcting codes and spin-glass models for an arbitrary code \cite{Sourlas2001,Sourlas2002,Sourlas2005}. In this study,
we address two error-correcting codes relevant to QA: the classical repetition code and the PE scheme.
We briefly introduce the formulation proposed by Sourlas. Let us focus on linear
codes, which comprise the most important family of error-correcting codes \cite{Clerk1981,Shu2004}.
Suppose that each information message consists of a sequence of $K$
bits and is represented by the binary vector $\boldsymbol{u}=\left(u_{1},\ldots,u_{K}\right)\in\left\{ 0,1\right\} ^{K}$,
which is called the source word. We assume that we have no prior information
on the choice of the source word, so that a prior probability $P\left(\boldsymbol{u}\right)$
is uniform over $\boldsymbol{u}$ (ignorance prior distribution).
The source word is encoded into a redundant sequence of $N$$\left(>K\right)$
bits represented by the binary vector $\boldsymbol{c}=\left(c_{1},\ldots,c_{N}\right)=\boldsymbol{u}\boldsymbol{G}\,\left(\mathrm{mod}\,2\right)\in\left\{ 0,1\right\} ^{N}$.
The vector $\boldsymbol{c}$ is called code word and a $K\times N$
Boolean matrix $\boldsymbol{G}$ is the generating matrix of the code.
The matrix equation $\boldsymbol{c}=\boldsymbol{u}\boldsymbol{G}\,\left(\mathrm{mod}\,2\right)$
consists of $N$ independent scalar equations 
\begin{equation}
c_{i}=\stackrel[j=1]{K}{\bigoplus}u_{j}G_{ji}\in\left\{ 0,1\right\} \;\left(i=1,\ldots,N\right),
\end{equation}
where $\oplus$ denotes the Boolean (modulo-2) sum. We consider the linear
block code $C$, where the code word $\boldsymbol{c}$ consists
of a source word $\boldsymbol{u}$ and $N-K$ redundant parity
check bits.
We assume that $c_{i}=u_{i}\left(i=1,\ldots,K\right)$,
that is, $\boldsymbol{c}$ is arranged so that its first $K$ bits
contain the source word $\boldsymbol{u}$ and the remaining $N-K$ bits contain the
redundant parity check bits. Thus, $\boldsymbol{G}$ has the following
form: $\boldsymbol{G}=\left[I_{K},P\right]$, where $I_{K}$ is a $K\times K$
identity matrix and $P$ is a $K\times\left(N-K\right)$ matrix \cite{Clerk1981,Shu2004}.
The ratio $R=K/N<1$ characterizes the redundancy of the code and
is called the rate of the code. The $N$ components $c_{i}$ in the code
word $\boldsymbol{c}$ are not independent. Among all $2^{N}$ code 
words, only $2^{K}$ words have one-to-one correspondence
with the source words. Code words must satisfy the constraint $\boldsymbol{cH}^{T}=\left(0,\ldots,0\right)=\boldsymbol{0}\left(\mathrm{mod}\,2\right)$,
where $\boldsymbol{H}$ is an $\left(N-K\right)\times N$ Boolean matrix
and is called a parity check matrix. The constraints are called 
parity check equations and comprise $N-K$ independent scalar
equations.
\begin{equation}
\stackrel[j=1]{N}{\bigoplus}H_{ij}c_{j}=0\;\left(i=1,\ldots,N-K\right).
\end{equation}
Owing to these constraints, the prior probability $P\left(\boldsymbol{c}\right)$
may depend on the parity check bits in the code word $\boldsymbol{c}$,
and its distribution reflects the information regarding the choice of code words
$\boldsymbol{c}$. 

Suppose that $\boldsymbol{c}$ is transmitted through a noisy communication
channel and received as $\boldsymbol{r}=\left(r_{1},\ldots,r_{N}\right)=\boldsymbol{c}\oplus\boldsymbol{e}\in\left\{ 0,1\right\} ^{N}$,
where $\boldsymbol{e}=\left(e_{1},\ldots,e_{N}\right)\in\left\{ 0,1\right\} ^{N}$
denotes the error pattern. We call $\boldsymbol{s}=\left(s_{1},\ldots,s_{N-K}\right)=\boldsymbol{rH}^{T}\in\left\{ 0,1\right\} ^{N}$
the syndrome pattern. It should be noted that the syndrome pattern $\boldsymbol{s}$
depends only on the error pattern $\boldsymbol{e}$ as $\boldsymbol{s}=\boldsymbol{eH}^{T}$, 
and any element in $\boldsymbol{s}$ is $0$ if and only if  $\boldsymbol{r}$ is a
code word \cite{Massey1963,Clerk1981,Shu2004}.

The connection between the error-correcting codes and statistical physics 
of a spin network is established based on isomorphism between the additive Boolean
group $\left(\left\{ 0,1\right\} ,\oplus\right)$ and multiplicative
bipolar (Ising) group $\left(\left\{ \pm1\right\} ,\cdot\right)$. The source
word $\boldsymbol{u}$ is mapped to the bipolar vector $\boldsymbol{Z}=\left(Z_{1},\ldots,Z_{K}\right)\in\left\{ \pm1\right\} ^{K}$
by $Z_{i}=\left(-1\right)^{u_{i}}\in\left\{ \pm1\right\} $, and $Z_{i}Z_{j}=(-1)^{u_{i}\oplus u_{j}}$.
Similarly, the code word $\boldsymbol{c}$ is mapped to the bipolar vector
$\boldsymbol{z}=\left(z_{1},\ldots,z_{N}\right)\in\left\{ \pm1\right\} ^{N}$
by $z_{i}=\left(-1\right)^{c_{i}}$, and $z_{i}z_{j}=(-1)^{c_{i}\oplus c_{j}}$.
It follows that 
\begin{eqnarray}
z_{i}&=&\left(-1\right)^{\stackrel[j=1]{K}{\bigoplus}u_{j}G_{ji}}=C_{k_{1}\cdots k_{p}}^{i}\stackrel[l=1]{p}{\prod}Z_{k_{l}}\in\left\{ \pm1\right\}\nonumber\\
&&\;\left(i=1,\ldots,N;1\leqq p\leqq K\right).
\label{eq:3}
\end{eqnarray}
Note that $z_{i}=\left(-1\right)^{u_{i}}$ for $i=1,\ldots,K$ according to
our convention. As noted previously, we regard $Z_{i}$ and
$\boldsymbol{Z}$ as a logical spin variable and state of $K$ logical
spins, respectively. Similarly, we regard $z_{i}$ and $\boldsymbol{z}$
as a physical spin variable and state of $N$ physical spins, respectively.
Equation (\ref{eq:3}) defines the physical spin state $\boldsymbol{z}$
in terms of the Boolean matrix $\boldsymbol{C}$ and logical spin
state $\boldsymbol{Z}$; it describes the embedding of the $K$ logical spin
system into a larger $N\left(>K\right)$ physical spin system. The matrix
defined by the elements $C_{k_{1}\cdots k_{p}}^{i}$ is called the ``connectivity
matrix.'' It defines the codes and describes how the $i$th physical
spin state $z_{i}$ depends on the logical spin state $\boldsymbol{Z}$.
The element $C_{k_{1}\cdots k_{p}}^{i}$ is 1 if $c_{i}$ is the Boolean sum of $u_{k_{l}}\left(l=1,\ldots,p\right)$. Thus, a parity-check
bit in the linear code $C$ that is formed by the Boolean sum of $p$
bits of the form $\stackrel[l=1]{p}{\bigoplus}u_{k_{l}}$ is mapped
onto a $p$-spin coupling $\stackrel[l=1]{p}{\prod}Z_{k_{l}}$. (Each
logical spin variable $Z_{k_{l}}$ is understood to appear only once in the
$i$th physical spin variable $z_{i}$.)

Similarly, the parity check equations can be rewritten in terms of
the physical state $\boldsymbol{z}$: 
\begin{eqnarray}
S_{i}\left(\boldsymbol{z}\right)&=&\left(-1\right)^{\stackrel[j=1]{N}{\bigoplus}H_{ij}c_{j}}=M_{k_{1}\cdots k_{q}}^{i}\stackrel[l=1]{q}{\prod}z_{k_{l}}=+1\nonumber\\
&&\;\left(i=1,\ldots,N-K;1\leqq q\leqq N\right).
\label{eq:4}
\end{eqnarray}
This defines the Boolean matrix $\boldsymbol{M}$ called the ``parity
constraint matrix'' that is associated with the parity check matrix $\boldsymbol{H}$
of the code. $M_{k_{1}\cdots k_{q}}^{l}$ is $1$ if each parity
equation consists of the Boolean sum of $c_{k_{l}}\left(l=1,\ldots,p\right)$. 
The physical state $\boldsymbol{z}$ that satisfies Eq. (\ref{eq:4}) is called a code state. 
For the general physical state $\boldsymbol{z}'$, we refer to $S_{i}\left(\boldsymbol{z}'\right)$ 
syndrome, which is the classical counterpart of the stabilizer
operator defined in terms of the quantum correcting codes \cite{Gottesman1996,Jordan2006}.
In some studies, it is also called the parity check in the context of the communication
application \cite{Massey1963,Massey1968}.

Sourlas discussed decoding of the classical error-correcting codes
based on statistical inference according to the Bayes theorem. Based on the above connection, the physical spin state $\boldsymbol{z}$
before being exposed to a noisy environment can be inferred from the
physical spin state that is corrupted by the environment by analyzing the statistical
physics of the associated spin network. Suppose that the $i$th physical spin
is prepared in the state $z_{i}\in\left\{ \pm1\right\} $
and exposed to an environment. Then, the spin may be corrupted and
measured to be, in general, a real number $J_{i}$. This process is
stochastic and statistical by nature. The purpose of the error-correcting
codes is to infer the logical spin state $\boldsymbol{Z}$ from the
readout $\boldsymbol{J}=\left(J_{1},\ldots,J_{N}\right)$ of the physical
spin state as faithfully and efficiently as possible by exploiting
the information on the statistical property of the environment and redundancy
in the physical spin state $\boldsymbol{z}$. To this end, we consider
the conditional probability $P\left(\boldsymbol{z}|\boldsymbol{J}\right)d\boldsymbol{J}$
that the prepared state is $\boldsymbol{z}$ when the observed readout
is between $\boldsymbol{J}$ and $\boldsymbol{J}+d\boldsymbol{J}$.
According to the Bayesian inference, the Bayes-optimal estimate is obtained when we maximize $P\left(\boldsymbol{z}|\boldsymbol{J}\right)$,
called the posterior probability, or its sidelight that is discussed
later. The Bayes theorem states that 
\begin{eqnarray}
P\left(\boldsymbol{z}|\boldsymbol{J}\right)&=&\frac{P\left(\boldsymbol{J}|\boldsymbol{z}\right)P\left(\boldsymbol{z}\right)P\left(\boldsymbol{Z}\right)}{\sum_{\boldsymbol{z}}P\left(\boldsymbol{J}|\boldsymbol{z}\right)P\left(\boldsymbol{z}\right)P\left(\boldsymbol{Z}\right)}\nonumber\\
&=&\kappa P\left(\boldsymbol{J}|\boldsymbol{z}\right)P\left(\boldsymbol{z}\right)P\left(\boldsymbol{Z}\right),
\end{eqnarray}
where $\kappa$ is a constant independent of $\boldsymbol{z}$, which
is to be determined by the normalization condition $\mathop{\sum_{\boldsymbol{z}}P\left(\boldsymbol{z}|\boldsymbol{J}\right)=1}$.
$P\left(\boldsymbol{z}\right)$ and $P\left(\boldsymbol{Z}\right)$
are the prior probabilities for the physical spin state $\boldsymbol{z}$
and logical spin state $\boldsymbol{Z}$, respectively \cite{Vicente2003}.
In the present scenario, we assume an informative prior for the parity 
constraints of the physical spin state $\boldsymbol{z}$ but ignorance
prior for the logical spin state $\boldsymbol{Z}$, that is, 
\begin{equation}
P\left(\boldsymbol{z}\right)=\mu\stackrel[l=1]{N-K}{\prod}\delta\left(S_{i}\left(\boldsymbol{z}\right),+1\right),
\end{equation}
and $P\left(\boldsymbol{Z}\right)=\mathrm{const.},$ where $\mu$ is
a normalization constant and $\delta\left(i,j\right)$ is Kronecker's
delta. $P\left(\boldsymbol{J}|\boldsymbol{z}\right)$ is called the likelihood,
which describes the statistical property of the environmental noise and
is supposed to be known. In addition, the environmental noise is supposed
to be identical and independent of $z_{i}$ (a ``memoryless or Markovian
noise model''), i.e.,
\begin{equation}
P\left(\boldsymbol{J}|\boldsymbol{z}\right)=\mathop{\underset{i=1}{\stackrel{N}{\prod}}P\left(J_{i}|z_{i}\right)},\label{eq:7}
\end{equation}
where $P\left(J_{i}|z_{i}\right)$ is called the extrinsic likelihood. 

Maximizing the posterior probability $P\left(\boldsymbol{z}|\boldsymbol{J}\right)$
is equivalent to maximizing log posterior probability expressed by 
\begin{eqnarray}
L\left(\boldsymbol{z},\boldsymbol{J}\right)&=&\log P\left(\boldsymbol{z}|\boldsymbol{J}\right)\nonumber\\
&=&\stackrel[i=1]{N}{\mathop{\sum}}L_{c}\left(J_{i},z_{i}\right)+\stackrel[i=1]{N-K}{\mathop{\sum}}L_{p}\left(z_{i}\right)+\mathrm{const.},
\end{eqnarray}
 where $L_{c}\left(J_{i},z_{i}\right)$ and $L_{p}\left(z_{i}\right)$
are the extrinsic log-likelihood and log-posterior probability,
respectively, and are defined as 
\begin{eqnarray}
L_{c}\left(J_{i},z_{i}\right) & = & \log P\left(J_{i}|z_{i}\right)\nonumber \\
 & = & \frac{1}{2}\log P\left(J_{i}|z_{i}=+1\right)P\left(J_{i}|z_{i}=-1\right)\nonumber \\
 && + \frac{z_{i}}{2}\log\frac{P\left(J_{i}|z_{i}=+1\right)}{P\left(J_{i}|z_{i}=-1\right)}\nonumber \\
 & = & K_{i}z_{i}+c2,
\end{eqnarray}
\begin{equation}
L_{p}\left(z_{i}\right)=\log P\left(z_{i}\right)=\log\delta\left(S_{i}\left(\boldsymbol{z}\right),+1\right),
\end{equation}
\begin{equation}
K_{i}=K_{i}\left(J_{i}\right)=\frac{1}{2}\log\frac{P\left(J_{i}|z_{i}=+1\right)}{P\left(J_{i}|z_{i}=-1\right)},
\end{equation}
where c2 is a constant independent of $z_{i}$. $K_{i}$ is called
the extrinsic log-likelihood ratio (LLR) in the coding theory. It
should be noted that maximizing $P\left(\boldsymbol{z}|\boldsymbol{J}\right)$
is equivalent to maximizing $L\left(\boldsymbol{z},\boldsymbol{J}\right)$
because $\log x$ is a monotonically increasing function of $x$. $L_{p}\left(z_{i}\right)$
describes the parity constraints for $z_{i}\textrm{ }\left(i=1,\ldots,N-K\right)$
and is independent of $J_{i}$ that may be corrupted by environmental
noise, whereas $L_{c}\left(J_{i},z_{i}\right)$ describes the noise
property and depends on the relationship between the two types of information, $z_{i}$
and $J_{i}$, which were prepared and measured, respectively.

To present the concrete expression for $K_{i}$, we need to specify
a model for the environmental noise. We address two examples of a symmetric
channel having a common property 
\begin{equation}
P\left(J_{i}|z_{i}\right)=P\left(-J_{i}|-z_{i}\right).\label{eq:12}
\end{equation}
The first is the binary symmetric channel 
\begin{equation}
\left\{ \begin{array}{c}
P\left(J_{i}|z_{i}\right)=\left(1-p_{i}\right)\delta\left(J_{i},z_{i}\right)+p_{i}\delta\left(J_{i},-z_{i}\right)\\
K_{i}=\frac{J_{i}}{2}\ln\frac{1-p_{i}}{p_{i}}=\beta_{i}J_{i}
\end{array}\right.,
\end{equation}
where $0\leqq p_{i}\leqq1$. The other example is the additive white
Gaussian noise channel 
\begin{equation}
\left\{ \begin{array}{c}
P\left(J_{i}|z_{i}\right)=\frac{1}{\sqrt{2\pi}w_{i}}\exp\left[-\frac{\left(J_{i}-v_{i}z_{i}\right)^{2}}{2w_{i}^{2}}\right]\\
K_{i}=\frac{v_{i}}{w_{i}^{2}}J_{i}=\beta_{i}J_{i}
\end{array}\right.,
\end{equation}
where $v_{i}$ and $w_{i}^{2}$ are the amplitude of the prepared
signal $v_{i}z_{i}$ and variance of the Gaussian noise for the
$i$th physical spin, respectively. The parameter $\beta_{i}$ is called the inverse
temperature.

Sourlas eventually deduced two different types of spin Hamiltonians
associated with the error-correcting codes. The first is written in
terms of only the variables associated with the logical spins $\boldsymbol{Z}=\left(Z_{1},\ldots,Z_{K}\right)=\left(z_{1},\ldots,z_{K}\right)$, as follows:
\begin{eqnarray}
H^{logi}\left(\boldsymbol{Z}\right)&=&-L\left(\boldsymbol{z},\boldsymbol{J}\right)\nonumber \\
&=&-\stackrel[i=1]{N}{\mathop{\sum}}L_{c}\left(C_{k_{1}\cdots k_{p}}^{i}\stackrel[l=1]{p}{\prod}z_{k_{l}},J_{i}\right)\nonumber \\
&=&-\stackrel[i=1]{N}{\sum}\beta_{i}J_{i}C_{k_{1}\cdots k_{p}}^{i}\stackrel[l=1]{p}{\prod}Z_{k_{l}},
\label{eq:15}
\end{eqnarray}
which is obtained by expressing the physical spin state $\boldsymbol{z}$
in terms of the logical spin state $\boldsymbol{Z}$. In Eq. (\ref{eq:15}),
we have omitted an unimportant constant term for the present discussion.
$H^{logi}\left(\boldsymbol{Z}\right)$ obviously has the form of a
spin Hamiltonian with $p$-body interactions. The distribution of
the coupling coefficients is determined by the extrinsic LLR $K_{i}$.
In contrast, the second type of spin Hamiltonian is expressed in terms of the variables associated
with the physical spins $\boldsymbol{z}$: 
\begin{eqnarray}
H^{phys}\left(\boldsymbol{z}\right) & = & -L\left(\boldsymbol{z},\boldsymbol{J}\right)\nonumber \\
 & = & -\stackrel[i=1]{N}{\mathop{\sum}}L_{c}\left(z_{i},J_{i}\right)-\stackrel[i=1]{N-K}{\mathop{\sum}}L_{p}\left(z_{i}\right)\nonumber \\
 & = & -\stackrel[i=1]{N}{\sum}\beta_{i}J_{i}z_{i}\nonumber \\
 &&+\underset{\gamma\rightarrow\infty}{\lim}\gamma\stackrel[i=1]{N-K}{\sum}\frac{1-S_{i}\left(\boldsymbol{z}\right)}{2},\label{eq:16}
\end{eqnarray}
where we have used the identity 
\begin{equation}
\delta\left(x,+1\right)=\underset{\gamma\rightarrow\infty}{\lim}\exp\left[-\gamma\frac{1-x}{2}\right]
\end{equation}
to replace the $\delta$ in $L_{p}\left(z_{i}\right)$ with an associated
soft constraint given by the second term of Eq. (\ref{eq:16}). We have also omitted an unimportant constant term in 
Eq. (\ref{eq:16}).
$H^{phys}\left(\boldsymbol{z}\right)$ describes a spin system with
infinite ferromagnetic couplings and couplings with local magnetic
fields. It should be noted that the space of the logical spin state
$\boldsymbol{Z}$ and that of the physical spin state $\boldsymbol{z}$ differ, e.g.,
the dimension of $\boldsymbol{Z}$ is $K$, whereas that of $\boldsymbol{z}$
is $N\left(>K\right)$. Therefore, the state space associated with
$H^{phys}\left(\boldsymbol{z}\right)$ is larger than that of $H^{logi}\left(\boldsymbol{Z}\right)$.
The code space is the space of the degenerated ground states of the
Hamiltonian $H^{pen}\left(\boldsymbol{z}\right)=\stackrel[i=1]{N-K}{\sum}\frac{1-S_{i}\left(\boldsymbol{z}\right)}{2}$
in Eq. (\ref{eq:16}), which satisfies $N-K$ soft constraints $S_{i}\left(\boldsymbol{z}\right)=+1$
$\left(i=1,\ldots,N-K\right)$. It has $2^{K}$-degeneracy, which
agrees with the state space of $H^{logi}\left(\boldsymbol{Z}\right)$.
This penalty Hamiltonian $H^{pen}\left(\boldsymbol{z}\right)$
originates from the informative prior for parity constraints
and suppresses the excitation outside the code space.

We note that a set of the syndromes has one-to-one correspondence
with a set of mutually commuting stabilizer generators $\hat{\boldsymbol{S}}\left(\hat{\boldsymbol{z}}\right)=\left\{ \hat{S}_{1}\left(\hat{\boldsymbol{z}}\right),\ldots,\hat{S}_{N-K}\left(\hat{\boldsymbol{z}}\right)\right\} $
of the stabilizer group in a quantum-mechanical context, if we identify
the classical bipolar variables $Z_{i}$ and $z_{i}$ with the quantum mechanical
Pauli-$z$ operators $\hat{Z}_{i}$ and $\hat{z}_{i}$. (We designate 
the quantum mechanical non-commuting operator as $\hat{\;}$.) Consider
the logical basis state $\left\{ \left|0_{L}\right\rangle ,\left|1_{L}\right\rangle \right\} ^{K}$, which
is the eigenspace of $\hat{S}_{i}\left(\hat{\boldsymbol{z}}\right)$ associated with the eigenvalue $+1$
for $i=1,\ldots,N-K$. The code space is spanned by the basis state
$\left\{ \left|0_{L}\right\rangle ,\left|1_{L}\right\rangle \right\} ^{K}$
and is a $2^{K}$-dimensional subspace of the space of the physical
spin state as well as the $2^{K}$-degenerated ground states of the
quantum mechanical Hamiltonian $\hat{H}^{pen}\left(\hat{\boldsymbol{z}}\right)=\stackrel[i=1]{N-K}{\sum}\frac{1-\hat{S}_{i}(\hat{\boldsymbol{z}})}{2}$.
The code space projector can be written in terms of the stabilizer generators: 
$\hat{P}_{c}\left(\hat{\boldsymbol{z}}\right)=\stackrel[i=1]{K}{\prod}\left(\left|0_{L}\right\rangle \left\langle 0_{L}\right|+\left|1_{L}\right\rangle \left\langle 1_{L}\right|\right)_{i}=\stackrel[i=1]{N-K}{\prod}\frac{1+\hat{S}_{i}(\hat{\boldsymbol{z}})}{2}$.
Its orthogonal complement $\hat{P}_{nc}\left(\hat{\boldsymbol{z}}\right)=\hat{I}-\hat{P}_{c}\left(\hat{\boldsymbol{z}}\right)$ defines
the complementary non-code space. The family of stabilizers $\hat{\boldsymbol{S}}\left(\hat{\boldsymbol{z}}\right)$
defines the stabilizer codes, which can correct only the spin-flip
error, as the spin-flip error operator (Pauli-$x$ operator) $\hat{x}_{k}$
for any physical spin $k$ anti-commutes with at least one stabilizer
generator $\hat{S}_{i}\left(\hat{\boldsymbol{z}}\right)$. A collection of the syndrome subspaces is defined
according to the set of eigenvalues of $\hat{\boldsymbol{S}}\left(\hat{\boldsymbol{z}}\right)$. 
Note that because the stabilizer operator $\hat{S}_{i}\left(\hat{\boldsymbol{z}}\right)$ multiplied
by another operator $\hat{S}_{j}\left(\hat{\boldsymbol{z}}\right)$ is also a stabilizer operator (any
binary sum of parity checks is also a parity check from the classical communication
perspective), it follows that the parity check matrix $\boldsymbol{H}$
and associated parity constraint matrix $\boldsymbol{M}$ are not
unique. This gives us freedom of choice of the stabilizer operator $\hat{S}_{i}\left(\hat{\boldsymbol{z}}\right)$
as well as its classical counterpart, i.e., syndrome $S_{i}\left(\boldsymbol{z}\right)$.
Later, we shall use this fact to discuss
the method for encoding and decoding the physical spin state.

\subsection{Classical repetition code}

As an illustrative example of soft annealing, let us consider the classical repetition code that is a simple but important
example of linear block codes. The classical repetition code encodes
one bit of information into $n$ bits. From the optimization
application viewpoint, one bit of information is embedded into $n$
physical spins. The minimum code distance can be increased by repeating
the logical information several times, which helps to mitigate errors
in the physical spins. An $\left(n,k,d\right)$ linear block
code (encoding $k$ logical spins (source word) into $n$ physical
spins (code word) and with a code distance $d$) can be defined by
a set of syndromes $\boldsymbol{S}$. We consider the case of the $\left(n,k\right)=\left(N,1\right)$
repetition code, where a state $Z$ of a single logical spin 
is encoded to form $N$ physical spin states $\boldsymbol{z}=\left(z_{1},\ldots,z_{N}\right)$
with $z_{1}=Z$. This code can be defined by the set of syndromes
$S_{i}\left(\boldsymbol{z}\right)=z_{1}z_{i+1}$$\left(i=1,\ldots,N-1\right)$. Note that we
can consider another set of syndromes $S'_{i}\left(\boldsymbol{z}\right)=z_{i}z_{i+1}$$\left(i=1,\ldots,N-1\right)$
under freedom of choice as $S'_{i}\left(\boldsymbol{z}\right)=S_{i-1}\left(\boldsymbol{z}\right)S_{i}\left(\boldsymbol{z}\right)$. This
code can correct up to $t=\left\lfloor \frac{d}{2}\right\rfloor $
errors and can be reliably decoded using one-step MVD if the number
of errors does not exceed $t$, which is discussed later. In
the quantum-mechanical context, $z_{i}$ is identified with the Pauli-$z$
operator $\hat{z}_{i}$ of the $i$th physical spin and $\hat{S}_{i}\left(\hat{\boldsymbol{z}}\right)=\hat{z}_{1}\hat{z}_{i+1}$
$\left(i=1,\ldots,N-1\right)$. Such a quantum repetition bit-flip
code has the logical basis state 
\begin{equation}
\left|0_{L}\right\rangle =\stackrel[i=1]{N}{\prod}\left|0\right\rangle _{i};\left|1_{L}\right\rangle =\stackrel[i=1]{N}{\prod}\left|1\right\rangle _{i},
\end{equation}
where $\left|0\right\rangle _{i}$ and $\left|1\right\rangle _{i}$ 
denote the eigenstates of $\hat{z}_{i}$ associated with the eigenvalues $-1$ and $+1$, respectively.
The basis
$\left\{ \left|0_{L}\right\rangle ,\left|1_{L}\right\rangle \right\} $
spans the space of the code state for the repetition spin-flip code, which is the
eigenspace of $\hat{S}_{i}\left(\hat{\boldsymbol{z}}\right)$ associated with the eigenvalue $+1$
for $i=1,\ldots,N-1$, as well as the degenerate ground states of
$\hat{H}^{pen}\left(\hat{\boldsymbol{z}}\right)=\stackrel[i=1]{N-1}{\sum}\frac{1-\hat{S}_{i}\left(\hat{\boldsymbol{z}}\right)}{2}$.
The logical Pauli operators are defined as operators that preserve
the space of the code state and have the same effect on the logical basis states
as their corresponding Pauli operators have on the corresponding basis
states. They are expressed as  
\begin{equation}
\hat{Z}_{L}=\hat{z}_{1}\textrm{ and }\hat{X}_{L}=\stackrel[i=1]{N}{\prod}\hat{x}_{i},\label{eq:19}
\end{equation}
which commute with all stabilizers $\hat{S}_{i}$ as well as 
the projector for the space of the code states $\hat{P}_{c}\left(\hat{\boldsymbol{z}}\right)=\left|0_{L}\right\rangle \left\langle 0_{L}\right|+\left|1_{L}\right\rangle \left\langle 1_{L}\right|=\stackrel[i=1]{N-1}{\prod}\frac{1+\hat{S}_{i}\left(\hat{\boldsymbol{z}}\right)}{2}$,
and satisfy the anti-commutation relation $\left\{ \hat{X}_{L},\hat{Z}_{L}\right\} =0$.
The logical Pauli-$x$ operator $\hat{X}_{L}$ is the product of $N$
Pauli-$x$ operators ($N$-body interaction) acting on the physical
spins, which is to be expected because $d=N$ \cite{Bookatz2015,Vinci2015}.

The merit of this code is that its physical Hamiltonian $H^{phys}\left(\boldsymbol{z}\right)$ can be realized using up to two-body 
interactions. These operators should be compared with the logical
operators for the PE scheme, as demonstrated in the following discussion.
We note that QAC \cite{Jordan2006,Young2013,Pudenz2014,Pudenz2015,Vinci2015,Bookatz2015,Matsuura2016,Matsuura2017,Pearson2019}
uses this quantum repetition spin-flip code as its building block.
We revisit this point later. 

\subsection{Parity encoding}

We consider another interesting example. Sourlas proposed an advantage
of using $H^{phys}\left(\boldsymbol{z}\right)$ instead of $H^{logi}\left(\boldsymbol{Z}\right)$
to solve optimization problems \cite{Sourlas2005}. It was suggested that
the computational difficulties of some hard optimization problems may
be alleviated by embedding the original system into a larger system
and imposing necessary soft constraints, where the weight $\gamma$ of
the constraints term in $H^{phys}\left(\boldsymbol{z}\right)$ is assumed to
be a large yet finite value, to reduce the enlarged system to the original
system. This proposal, called soft annealing, was based on the empirical
observation that it is easier to minimize $H^{phys}\left(\boldsymbol{z}\right)$
rather than $H^{logi}\left(\boldsymbol{Z}\right)$, despite the fact
that the number of variables is much larger for $H^{phys}\left(\boldsymbol{z}\right)$.
Sourlas provided intuitive explanations; for example, in a large-dimensional space, 
it is possible to circumvent barriers to find the minimum, which makes the
optimization easier. It was pointed out that the PE schemes 
\begin{equation}
H^{phys}\left(\boldsymbol{z}\right)=\beta H^{loc}\left(\boldsymbol{z}\right)+\gamma H^{pen}\left(\boldsymbol{z}\right),\label{eq:20}
\end{equation}
where $\beta,\gamma>0$, and
\begin{equation}
H^{loc}\left(\boldsymbol{z}\right)=\underset{\left\langle i,j\right\rangle }{\sum}J_{ij}z_{ij},\label{eq:21}
\end{equation}
\begin{equation}
H^{pen}\left(\boldsymbol{z}\right)=\underset{\mathop{plaquettes}}{\sum}\frac{1-S_{ij}\left(\boldsymbol{z}\right)}{2},\label{eq:22}
\end{equation}
with $S_{ij}\left(\boldsymbol{z}\right)=S_{ij}^{4w}\left(\boldsymbol{z}\right)=z_{i\,j}z_{i\,j+1}z_{i+1\,j}z_{i+1\,j+1}$,
where $1\leq i<j\leq K-1$  \cite{Nambu22}, are nontrivial examples of the soft annealing
of a spin-glass model with all-to-all two-body interactions, i.e., 
\begin{equation}
H^{logi}\left(\boldsymbol{Z}\right)=\beta\underset{\left\langle i,j\right\rangle }{\sum}J_{ij}Z_{i}Z_{j},\label{eq:23}
\end{equation}
where $\underset{\mathop{plaquettes}}{\sum}$ runs over all plaquettes,
i.e., the elementary squares, of the lattice of physical spins, whereas
$\underset{\left\langle i,j\right\rangle }{\sum}$ runs over all 
spin pairs $\left(i,j\right)$ taken from all logical spins. 

\begin{figure}[h]
\centering{}
\includegraphics[viewport=275bp 170bp 670bp 420bp,clip,scale=0.62]{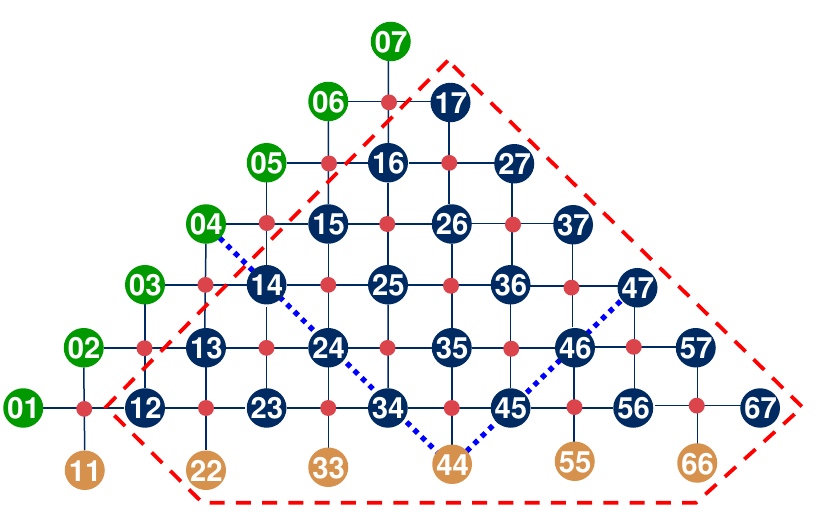}
\caption{Original and extended LHZ layouts. The blue and brown circles indicate
physical and ancillary spins, respectively. The green circles correspond
to logical spins. The label in the circle corresponds to the label specifying 
a physical spin. The small red solid circles indicate
plaquettes associated with $S_{ij}^{4w}$. The red broken line indicates
the original LHZ layout. The blue dotted line indicates the logical line $Q_{4}$.\label{fig:2}}
\end{figure}

Here, we consider an extended layout shown in Fig. \ref{fig:2} with
an additional row of physical spins, which have a direct correspondence
to logical spins, that is, $z_{0i}=Z_{i}$ $\left(i=1,\ldots,K\right)$
for the case of $K=7$ \cite{Rocchetto2016,Ender2022,Fellner2022}.
Without this row, the layout of the physical spins is reduced to the
original LHZ layout that is fully consistent with the Hamiltonian
$H^{phys}\left(\boldsymbol{z}\right)$ \cite{Lechner15}. With the addition
of this row, we can incorporate the local interaction terms originating
from the local field $h_{i}$ for $1\leq i\leq K$. In this figure, the
green circles indicate the spins associated with the $K$ logical spins $z_{0i}$ $\left(1\leq i\leq K\right)$,
whereas the blue circles indicate the physical spins $z_{ij}$ $\left(1\leq i<j\leq K\right)$.
Note that the brown circles are ancillary spins whose states are fixed to
the spin-up ($z_{ii}=+1$) states. If we omit the ancillary spins, the system
has a total of $N=\frac{K\left(K+1\right)}{2}$ spins with $K$ logical
spins and $N-K=\frac{K\left(K-1\right)}{2}$ physical spins, so that
$\left(n,k\right)=\left(N,K\right)$. The system should have $N-K$
parity constraints for the logical and physical spins. For example,
we can consider $N-K$ weight-three parity-check equations $S_{ij}^{3w}\left(\boldsymbol{z}\right)=z_{0i}z_{0j}z_{ij}=+1\textrm{ }\left(1\leq i<j\leq K\right)$.
Although $S_{ij}^{3w}$ differs from $S_{ij}^{4w}$, we can observe that
the parity constraints $S_{ij}^{3w}=+1$ and $S_{ij}^{4w}=+1$ are
consistent because $S_{ij}^{4w}=S_{ij}^{3w}S_{i\,j+1}^{3w}S_{i+1\,j}^{3w}S_{i+1\,j+1}^{3w}$
for any $1\leqq i<j\leq K-1$. This is a consequence of the freedom of
choice of syndromes. The physical spins of the four corners around each
small red solid circle in Fig. \ref{fig:2} define all elements
$S_{ij}^{4w}$, which are called plaquettes. In addition, it should be remembered that if $h_{i}=0$ for $1\leq i\leq K$, only a random local
field $J_{ij}$ acts on the physical spins, 
which justifies the use of the ignorance prior for the readout of the logical
spin state $\boldsymbol{Z}$ for error correction. Therefore, provided that 
the local fields are absent in Eq. (\ref{eq:23}), it is sufficient
to consider only the physical spins bounded by the red broken line in Fig. \ref{fig:2}, which is simply the original LHZ layout.
In this case, we have the $\left(n,k\right)=\left(N-K,K-1\right)=\left(K\left(K-1\right)/2,K-1\right)$ linear block code, which means that only $K-1$ physical variables
are logically independent, i.e., we can determine $\boldsymbol{Z}$
only up to a global spin flip. 

Let us examine how the PE spin network is described in the quantum-mechanical
context. We introduce the concept of logical lines $Q_{i}$, or constraint-preserving
driver lines \cite{Ender2022}, which define the subsets of the physical
operators that preserve the space of the code state and have one-to-one correspondence
with the Pauli operators on the logical spins \cite{Rocchetto2016,Fellner2022,Ender2022,Fuchs2022,Weidinger2023,Messinger2023}.
Consider the Pauli-$z$ operators $\hat{z}_{ij}$ that are associated
with the physical spins in the PE spin network.
We consider the following $\left(K+1\right)\times\left(K+1\right)$
symmetric matrix with Pauli-$z$ operators as the entries for the relevant
spins arranged consistently with the PE spin network shown in
Fig. \ref{fig:2},
\begin{equation}
\hat{\boldsymbol{z}}=
{\footnotesize
\left[\begin{array}{ccccccc}
\hat{1} & \hat{z}_{01} & \hat{z}_{02} & \hat{z}_{03} & \cdots & \hat{z}_{0\,K-1} & \hat{z}_{0\,K}\\
\hat{z}_{10} & \hat{1} & \hat{z}_{12} & \hat{z}_{13} & \cdots & \hat{z}_{1\,K-1} & \hat{z}_{1K}\\
\hat{z}_{20} & \hat{z}_{21} & \hat{1} & \hat{z}_{23} & \cdots & \hat{z}_{2\,K-1} & \hat{z}_{2K}\\
\hat{z}_{30} & \hat{z}_{31} & \hat{z}_{32} & \hat{1} & \cdots & \hat{z}_{3\,K-1} & \hat{z}_{3K}\\
\vdots & \vdots & \vdots & \vdots & \ddots & \cdots & \vdots\\
\hat{z}_{K-1\,0} & \hat{z}_{K-1\,1} & \hat{z}_{K-1\,2} & \hat{z}_{K-1\,3} & \cdots & \hat{1} & \hat{z}_{K-1\,K}\\
\hat{z}_{K\,0} & \hat{z}_{K\,1} & \hat{z}_{K\,2} & \hat{z}_{K\,3} & \cdots & \hat{z}_{K-1\,K} & \hat{1}
\end{array}\right],
}
\label{eq:24}
\end{equation}
where $\hat{1}$ is an identity operator acting on a physical spin and $\hat{z}_{ij}=\hat{z}_{ji}$
by construction. $\hat{\boldsymbol{z}}$ is a matrix with entries
given by $N=\frac{K\left(K+1\right)}{2}$ projectors associated
with the measurement of the configuration of the PE spin network on
the computational basis. Each entry $\hat{z}_{ij}$ is observable
for the physical spins having two possible eigenstates with eigenvalues
$\pm1$. Let us consider the Pauli-$x$ operators $\hat{X}_{i}$ and
$\hat{x}_{ij}$ acting on the logical and physical spins, respectively.
These Pauli operators satisfy the following anti-commutation and commutation
relations:
\begin{equation}
\left\{ \hat{X}_{i},\hat{Z}_{i}\right\} =\left[\hat{X}_{i},\hat{Z}_{j}\right]=0,
\end{equation}
for $j\neq i$ and 
\begin{equation}
\left\{ \hat{x}_{ij},\hat{z}_{ij}\right\} =\left[\hat{x}_{ij},\hat{z}_{kl}\right]=0,
\end{equation}
\begin{equation}
\left[\hat{X}_{i},\hat{x}_{jk}\right]=\left[\hat{Z}_{i},\hat{z}_{jk}\right]=\left[\hat{X}_{i},\hat{z}_{jk}\right]=\left[\hat{Z}_{i},\hat{x}_{jk}\right]=0
\end{equation}
for $k\neq i$ or $\ensuremath{l\neq j}$. We also define $\hat{\boldsymbol{Z}}=\left(\hat{Z}_{1},\ldots,\hat{Z}_{K}\right)$.
Then, if we replace $Z_{i}\rightarrow\hat{Z}_{i}$ and $z_{ij}\rightarrow\hat{z}_{ij}$
for the classical Hamiltonians, the resultant Hamiltonians $\hat{H}^{logi}=\hat{H}^{logi}\left(\hat{\boldsymbol{\boldsymbol{Z}}}\right)$
and $\hat{H}^{phys}=\hat{H}^{phys}\left(\hat{\boldsymbol{z}}\right)$ are
identified with the Hamiltonians in the quantum-mechanical context.
Then, we can consider the logical Pauli operators similar to the
repetition spin-flip code. It is easy to show that 
\begin{equation}
\hat{Z_{L}}_{i}=\hat{Z}_{i}=\hat{z}_{0i}=\hat{z}_{i0}\textrm{ and }\hat{X_{L}}_{i}=\stackrel[j=0]{K}{\prod}\hat{x}_{ij}=\stackrel[j=0]{K}{\prod}\hat{x}_{ji}
\end{equation}
commute with the stabilizer operators 
\begin{equation}
\hat{S}_{ij}\left(\hat{\boldsymbol{z}}\right)=\hat{z}_{i\,j}\hat{z}_{i\,j+1}\hat{z}_{i+1\,j}\hat{z}_{i+1\,j+1}\textrm{ }\left(0\leq i<j\leq K-1\right)\label{eq:29}
\end{equation}
as well as the penalty Hamiltonian 
\begin{equation}
\hat{H}^{pen}\left(\hat{\boldsymbol{z}}\right)=\underset{\mathop{plaquettes}}{\sum}\frac{1-\hat{S}_{ij}(\hat{\boldsymbol{z}})}{2},
\end{equation}
and the projector to the space of the code states 
\begin{eqnarray}
\hat{P}_{c}\left(\hat{\boldsymbol{z}}\right)&=&\underset{\mathop{plaquettes}}{\prod}\frac{1+\hat{S}_{ij}(\hat{\boldsymbol{z}})}{2}\nonumber \\
&=&\stackrel[i=1]{K}{\prod}\left(\left|0_{L}\right\rangle \left\langle 0_{L}\right|+\left|1_{L}\right\rangle \left\langle 1_{L}\right|\right)_{i}\nonumber \\
&=&\stackrel[i=1]{K}{\prod}\left(\left|0\right\rangle \left\langle 0\right|+\left|1\right\rangle \left\langle 1\right|\right)_{i},
\end{eqnarray}
where $\left|0\right\rangle _{i}$ and $\left|1\right\rangle _{i}$ denote the eigenstates of $\hat{Z}_{i}=\hat{z}_{0i}=\hat{z}_{i0}$ associated with the eigenvalues 
$-1$ and $+1$, respectively. The logical Pauli operators satisfy the (anti-) commutation relations 
\begin{equation}
\left\{ \hat{X_{L}}_{i},\hat{Z_{L}}_{i}\right\} =\left[\hat{X_{L}}_{i},\hat{Z_{L}}_{j}\right]=0
\end{equation}
for $i=1,\ldots,K$ with $j\neq i$. We can also confirm that 
\begin{equation}
\left\{ \hat{X_{L}}_{i},\hat{z}_{ij}\right\} =\left\{ \hat{X_{L}}_{j},\hat{z}_{ij}\right\} =0.
\end{equation}
The logical Pauli-$x$ operator $\hat{X_{L}}_{i}$ is determined by the product
of $K$ Pauli-$x$ operators associated with physical spins with a label
 involving $i$. Such a chain of spins in the PE spin network is called the logical line (or constraint-preserving
driver line \cite{Ender2022}). The stabilizer operator $\hat{S}_{ij}\left(\hat{\boldsymbol{z}}\right)$
commutes with $\hat{H}^{phys}\left(\hat{\boldsymbol{z}}\right)$ and anti-commutes with the spin-flip operators $\hat{x}_{ij}$ 
for the physical spins involved in the plaquette given in Eq. (\ref{eq:29}).

Now, let us consider the action of $\hat{X_{L}}_{i}$ 
on every element of $\hat{\boldsymbol{z}}$, i.e., $\hat{\boldsymbol{z}}\rightarrow\hat{\boldsymbol{z}'}=\hat{X_{L}}_{i}\hat{\boldsymbol{z}}\hat{X_{L}}_{i}$.
The operator $\hat{X_{L}}_{i}$ inverts the sign of the matrix elements
of $\hat{\boldsymbol{z}}$ along the $\left(i+1\right)$th row and
$\left(i+1\right)$th column except for the diagonal elements collectively.
Note that the diagonal element should be preserved as it belongs
to both the $\left(i+1\right)$th row and column and its sign is inverted
twice. In the Heisenberg picture, this implies that the logical Pauli-$x$ 
operator $\hat{X_{L}}_{i}$ collectively flips the orientation of
the set of $2K$ physical spins in the logical lines. The logical
Pauli-$x$ operators $\hat{X_{L}}_{i}$ are counterparts of the logical
Pauli-$x$ operator $\hat{X}_{L}$ of the repetition spin-flip code
given in Eq. (\ref{eq:19}). Such high-weight operators flip all the
physical spins associated with a logical spin $\hat{Z}_{i}$ collectively.
The resultant collective operation maps a code state to another code
state for both the repetition spin-flip code and PE scheme, as it commutes
with the projector to the space of the code states. This implies that the code distance
$d$ of $K$ logical spins is expressed by $d=K$ for this extended
LHZ layout. Similarly, we obtain $d=K-1$ for the original LHZ layout.
The repetition spin-flip code and parity encoding guarantees correction
of up to $t=\left\lfloor \frac{d-1}{2}\right\rfloor $ spin-flip errors.
We can describe the QA of the PE scheme by combining the standard
transverse-field driver Hamiltonian 

\begin{equation}
\hat{H}_{1}^{tr}\left(\hat{\boldsymbol{x}}\right)=\kappa\underset{\left\langle i,j\right\rangle }{\sum}\hat{x}_{ij}
\end{equation}
with $\hat{H}^{phys}\left(\hat{\boldsymbol{z}}\right)$ and introducing a temporal dependence for $\beta$,
$\gamma$, and $\kappa$, where $\hat{x}_{ji}=\hat{x}_{ij}$ is always
assumed for any $1\leq i<j\leq K$. We may select another transverse-field
driver Hamiltonian based on $K$ logical Pauli-$x$ operators \cite{Konz2019,Ender2022,Fuchs2022,Weidinger2023,Messinger2023}
\begin{equation}
\hat{H}_{2}^{tr}\left(\hat{\boldsymbol{x}}\right)=\kappa\stackrel[i=1]{K}{\sum}\hat{X_{L}}_{i}\neq\hat{H}_{1}^{tr}\left(\hat{\boldsymbol{x}}\right),
\end{equation}
which requires $K$-body interactions and is difficult
to implement practically using current technologies.

\subsection{QAC}

\begin{figure*}
\begin{centering}
\includegraphics[viewport=250bp 20bp 800bp 510bp,clip,scale=0.6]{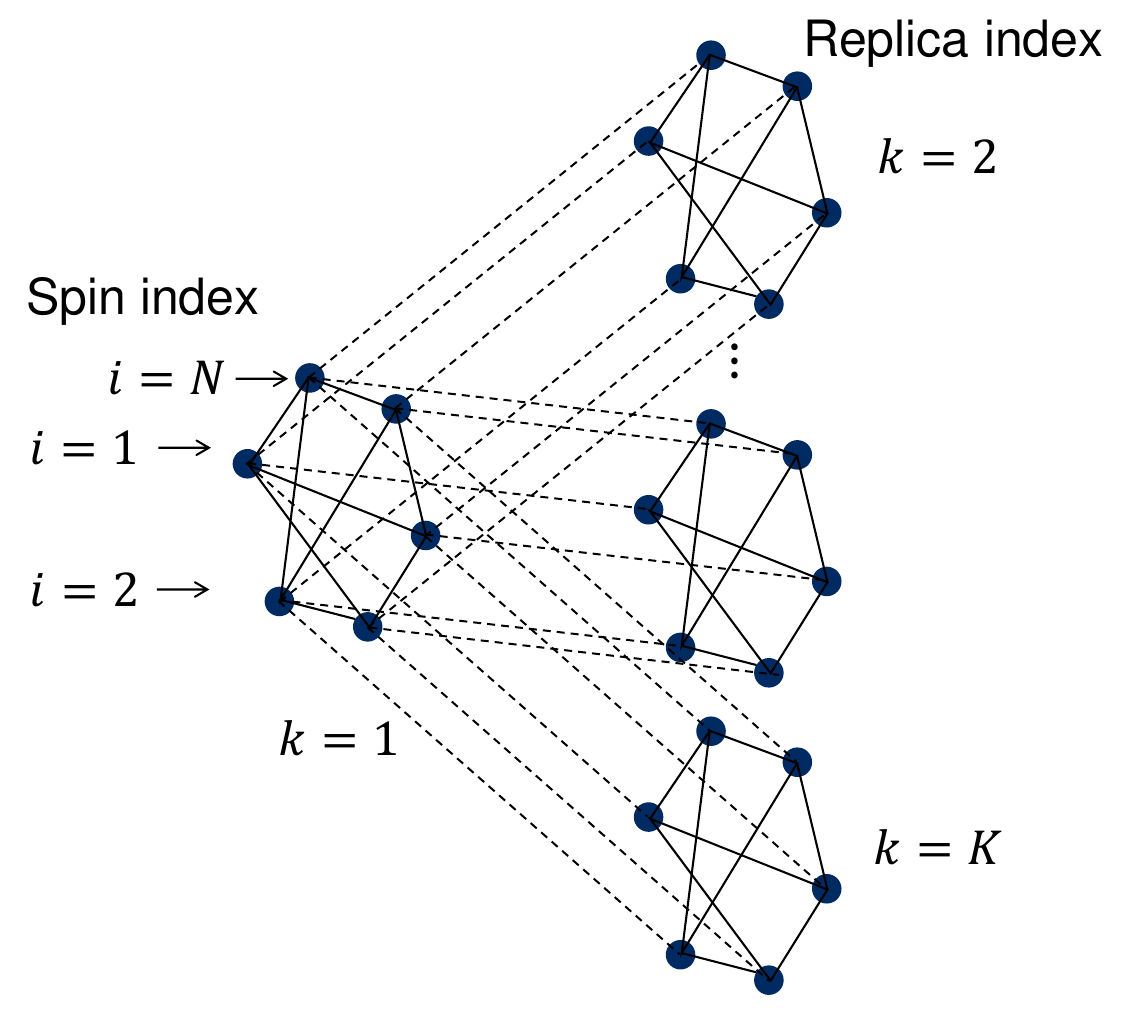}
\par\end{centering}
\caption{Layout of physical spins in QAC when $\hat{S}_{i\,k}(\hat{\boldsymbol{z}})$
is given by Eq. (\ref{eq:37}). The spins specified with the same index
$i$ are ferromagnetically coupled and constitute the logical group associated
with the logical spin $i$. The index $k$ specifies the replica to which the spin
belongs. The solid lines indicate mutual spin couplings within a replica.
The broken lines indicate ferromagnetic couplings.\label{fig:3}}
\end{figure*}

\begin{figure*}
\begin{centering}
\includegraphics[viewport=140bp 180bp 820bp 420bp,clip,scale=0.6]{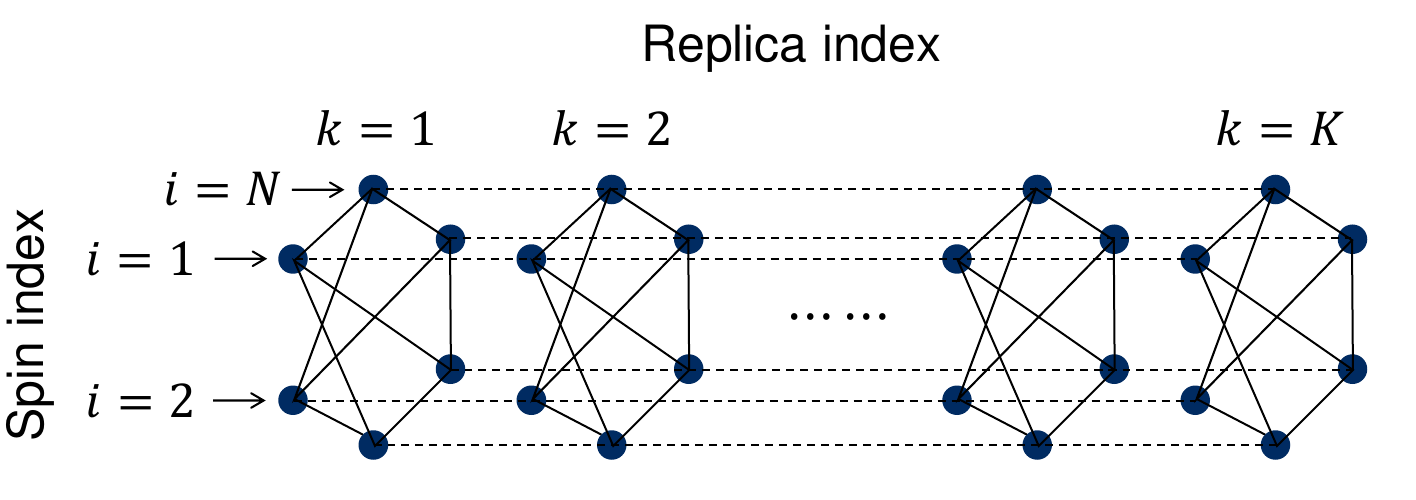}
\par\end{centering}
\caption{Layout of physical spins in QAC with $\hat{S}_{i\,k}(\hat{\boldsymbol{z}})$
given by Eq. (\ref{eq:38}). In this case, the ferromagnetic
couplings in each logical group differ from those in Fig. \ref{fig:3}.\label{fig:4}}
\end{figure*}

QAC is a heuristic method that was designed to protect the logical
spin systems from spin-flip errors by introducing a classical repetition 
code independently for each spin \cite{Jordan2006,Young2013,Pudenz2014,Pudenz2015,Vinci2015,Bookatz2015,Matsuura2016,Matsuura2017,Pearson2019}.
We investigated QAC to compare it with the PE scheme, as the PE scheme somewhat resembles QAC.
The typical layouts of the
physical spins in QAC are illustrated in Figs. \ref{fig:3}
and \ref{fig:4}. We consider $K$ copies of the spin systems, each
comprising $N$ spins described by the logical spin Hamiltonian, which we
call replicas. Hence, every spin in the logical spin system is accompanied
by its $K-1$ replicas. To form a $K$ repetition spin-flip code,
the $K$ physical spins are ferromagnetically coupled by two-body
interaction to form a logical group associated with the logical spin. 

The Hamiltonian for the physical spin system is considered as follows. Suppose
we designate a Pauli-$z$ operator for the spin $i$ $\left(=1,\ldots,,N,\right)$
in the logical Hamiltonian as $\hat{Z}_{i}$. The encoding introduces
a new index, $k$ $\left(=1,\ldots,K\right)$, specifying the replica to which
the spin $i$ belongs, so that $\hat{Z}_{i}\mapsto\hat{z}_{i\,k}$. 
Let $\hat{\boldsymbol{z}}=\left\{ \hat{z}_{i\,k}\right\} _{i=1,\ldots,,N,k=1,\ldots,K}$ be the set of $N\times K$ physical spin operators. 
Then, the subsets $\left\{ \hat{z}_{i\,k}\right\} _{k=1,\ldots,K}\subset\hat{\boldsymbol{z}}$ and $\left\{ \hat{z}_{i\,k}\right\} _{i=1,\ldots,N}\subset\hat{\boldsymbol{z}}$ 
correspond to the physical spin operators in the logical group $i$ and in the replica $k$, respectively.
If there is no coupling
between the logical spins with different indices $i$, the logical
groups specified by the different $i$ are mutually independent.
Then, the physical Hamiltonian of the repetition spin-flip code is
expressed as 
\begin{eqnarray}
\hat{H}^{phys}\left(\hat{\boldsymbol{z}}\right) & = & \stackrel[i=1]{N}{\sum}\left(\beta h_{i}\stackrel[k=1]{K}{\sum}\hat{z}_{i,k}+\gamma\stackrel[k=2]{K}{\sum}\frac{1-\hat{S}_{i\,k}(\hat{\boldsymbol{z}})}{2}\right)\nonumber \\
 & = & \beta\stackrel[k=1]{K}{\sum}\stackrel[i=1]{N}{\sum}h_{i}\hat{z}_{i,k}\nonumber \\
 &&+\gamma\stackrel[i=1]{N}{\sum}\stackrel[k=2]{K}{\sum}\frac{1-\hat{S}_{i\,k}(\hat{\boldsymbol{z}})}{2},
\end{eqnarray}
where $\beta,\gamma>0$, and
\begin{equation}
\hat{S}_{i\,k}(\hat{\boldsymbol{z}})=\hat{z}_{i\,1}\hat{z}_{i\,k+1}\;\left(k=1,...,K-1\right),\label{eq:37}
\end{equation}
or
\begin{equation}
\hat{S}_{i\,k}(\hat{\boldsymbol{z}})=\hat{z}_{i\,k}\hat{z}_{i\,k+1}\;\left(k=1,...,K-1\right)\label{eq:38}
\end{equation}
denotes the $K-1$ stabilizer operator, which is the origin of the ferromagnetic
coupling between the spins in each logical group. The stabilizer operator
$\hat{S}_{i\,k}\left(\hat{\boldsymbol{z}}\right)$ commutes with $\hat{H}^{phys}\left(\hat{\boldsymbol{z}}\right)$ and anti-commutes
with the spin-flip errors for the physical spins involved in Eq. (\ref{eq:37})
or (\ref{eq:38}).

Note that, although this Hamiltonian is applicable to the communication application where the physical spins within a replica are assumed to be mutually independent, it is not necessarily applicable to QA because coupling among the spins within a replica is crucial in QA. To extend this Hamiltonian to describe QA,
the two-body coupling has been heuristically introduced between the spins within each replica in some studies, that is,  
\begin{eqnarray}
\hat{H}^{two}\left(\hat{\boldsymbol{z}}\right)&=&\stackrel[k=1]{K}{\sum}\left(\beta\stackrel[\left\langle i,j\right\rangle ]{N}{\sum}J_{ij}\hat{z}_{i,k}\hat{z}_{j,k}\right)\nonumber \\
&=&\beta\stackrel[k=1]{K}{\sum}\stackrel[\left\langle i,j\right\rangle ]{N}{\sum}J_{ij}\hat{z}_{i,k}\hat{z}_{j,k},
\end{eqnarray}
into the physical Hamiltonian, and the use of 
\begin{eqnarray}
\hat{H}^{QAC}\left(\hat{\boldsymbol{z}}\right)&=&\hat{H}^{phys}\left(\hat{\boldsymbol{z}}\right)+\hat{H}^{two}\left(\hat{\boldsymbol{z}}\right)\nonumber \\
&=&\beta\hat{H}^{enc}\left(\hat{\boldsymbol{z}}\right)+\gamma\hat{H}^{pen}\left(\hat{\boldsymbol{z}}\right)\label{eq:40}
\end{eqnarray}
was proposed for the Hamiltonian of the QAC, where $\left\langle i,j\right\rangle $
indicates that the sum runs over the coupled spins within the replica, and
$\beta$ and $\gamma$ are positive weight parameters \cite{Young2013,Pudenz2014,Pudenz2015,Vinci2015,Pearson2019}.
We refer to $\hat{H}^{QAC}\left(\hat{\boldsymbol{z}}\right)$ as the QAC Hamiltonian. The last
equation indicates that $\hat{H}^{QAC}\left(\hat{\boldsymbol{z}}\right)$ comprises two components.
The first is an encoded Hamiltonian 
\begin{equation}
\hat{H}^{enc}\left(\hat{\boldsymbol{z}}\right)=\stackrel[k=1]{K}{\sum}\left.\hat{H}^{logi}\left(\hat{\boldsymbol{Z}}\right)\right|_{\hat{Z}_{i}\mapsto\hat{z}_{i,k}},
\label{eq:41}
\end{equation}
where 
\begin{equation}
\hat{H}^{logi}\left(\hat{\boldsymbol{Z}}\right)=\stackrel[\left\langle i,j\right\rangle ]{N}{\sum}J_{ij}\hat{Z}_{i}\hat{Z}_{j}+\stackrel[i=1]{N}{\sum}h_{i}\hat{Z}_{i},
\label{eq:42}
\end{equation}
and $\hat{\boldsymbol{Z}}=\left\{ \hat{Z}_{i}\right\} _{i=1,\ldots,N}$ is the set of $N$ logical spin operators. $\hat{H}^{logi}\left(\hat{\boldsymbol{Z}}\right)$ is the embedded Hamiltonian for $N$ logical spins, which is our main interest. The second is the penalty Hamiltonian
\begin{equation}
\hat{H}^{pen}\left(\hat{\boldsymbol{z}}\right)=\stackrel[i=1]{N}{\sum}\stackrel[k=1]{K-1}{\sum}\frac{1-\hat{S}_{i\,k}(\hat{\boldsymbol{z}})}{2},
\end{equation}
which commutes with $\hat{H}^{enc}\left(\hat{\boldsymbol{z}}\right)$ and is introduced to take
advantage of the prior knowledge of $K-1$ sets of the
parity constraints on the $K$ physical spin variables $\hat{z}_{i\,k}$
for $k=1,\ldots,K$ in the Bayes inference. $\hat{H}^{pen}\left(\hat{\boldsymbol{z}}\right)$ enforces
the alignment of the $K$ physical spins in each logical group. The
Hamiltonian $\hat{H}^{two}\left(\hat{\boldsymbol{z}}\right)$ introduces correlations
between the physical spins in different logical groups,
which are absent in the Hamiltonian $\hat{H}^{phys}\left(\hat{\boldsymbol{z}}\right)$
in the communication scenario. In the limit $\gamma\rightarrow0$,
$\hat{H}^{QAC}\left(\hat{\boldsymbol{z}}\right)$ approaches the Hamiltonian
for $K$ independent copies of the $N$ spin system, each of which is described by the
logical Hamiltonian $\left.\hat{H}^{logi}\left(\hat{\boldsymbol{Z}}\right)\right|_{\hat{Z}_{i}\mapsto\hat{z}_{i,k}}$. Hence, we refer to a set
of $N$ physical spins specified by $k$ as a replica. In contrast,
in the limit $\beta\rightarrow0$, $\hat{H}^{QAC}\left(\hat{\boldsymbol{z}}\right)$
approaches the Hamiltonian $\hat{H}^{pen}\left(\hat{\boldsymbol{z}}\right)$ that energetically penalizes
the non-code state of the $\left[N,1\right]$ repetition spin-flip
code. In this case, it should be noted that if we select the weight-two stabilizer
in Eq. (\ref{eq:38}), $\hat{H}^{QAC}\left(\hat{\boldsymbol{z}}\right)$
is formally identical to that in simulated QA (SQA)
using the path-integral Monte Carlo method \cite{Marto?ak2002,Heim2015,Waidyasooriya2020,Hu2021}.
This implies that SQA may be re-investigated from the viewpoint
of the classical error-correcting code. We can describe the quantum dynamics
of QAC by combining the standard transverse-field driver Hamiltonian
based on local Pauli-$x$ operators
\begin{equation}
\hat{H}_{1}^{tr}\left(\hat{\boldsymbol{x}}\right)=\kappa\stackrel[i=1]{N}{\sum}\stackrel[k=1]{K}{\sum}\hat{x}_{i,k}
\end{equation}
with $\hat{H}^{QAC}\left(\hat{\boldsymbol{z}}\right)$ and introducing
the temporal dependence for $\beta$, $\gamma$, and $\kappa$. Note that
the logical Pauli operators for QAC are expressed as  
\begin{equation}
\hat{Z_{L}}_{i}=\hat{z}_{i,1}\textrm{ and }\hat{X_{L}}_{i}=\stackrel[k=1]{K}{\prod}\hat{x}_{i,k}.\label{eq:45}
\end{equation}
We can select another transverse-field driver Hamiltonian based on
the $K$-body logical Pauli-$x$ operator \cite{Vinci2015,Vinci2016,Konz2019,Dodds2019,Ender2022,Fuchs2022,Weidinger2023}
\begin{equation}
\hat{H}_{2}^{tr}\left(\hat{\boldsymbol{x}}\right)=\kappa\stackrel[i=1]{N}{\sum}\hat{X_{L}}_{i}\neq\hat{H}_{1}^{tr}\left(\hat{\boldsymbol{x}}\right),
\end{equation}
which requires $K$-body interactions and is difficult
to implement practically using current technologies.

\section{Decoding of post-readout spin state\label{sec:3}}

The purpose of decoding is to infer the correct logical and physical target states $\boldsymbol{Z}$
and $\boldsymbol{z}$, respectively, given the coupling $\boldsymbol{J}$ (and local
field $\boldsymbol{h}$). From the communication application viewpoint,
we identify $\boldsymbol{J}$ (and $\boldsymbol{h}$) as the readout
associated with the physical state $\boldsymbol{z}$. If $\boldsymbol{J}$
is corrupted by environmental noise, the target states are expected to
be the state maximizing the posterior probability $P\left(\boldsymbol{Z}|\boldsymbol{J}\right)$
or $P\left(\boldsymbol{z}|\boldsymbol{J}\right)$ if the environmental
noises in the spins are mutually weakly correlated \cite{Nishimori1993,Pastawski2016}.
Then, the optimal estimators for $\boldsymbol{Z}$ and $\boldsymbol{\boldsymbol{z}}$
are expressed as 
\begin{widetext}
\begin{equation}
\boldsymbol{Z}^{*}=\underset{\boldsymbol{Z}\in\left\{ \pm1\right\} ^{K}}{\arg\max}P\left(\boldsymbol{Z}|\boldsymbol{J}\right)=\underset{\boldsymbol{Z}\in\left\{ \pm1\right\} ^{K}}{\arg\max}L\left(\boldsymbol{Z},\boldsymbol{J}\right)=\underset{\boldsymbol{Z}\in\left\{ \pm1\right\} ^{K}}{\arg\min}H^{logi}\left(\boldsymbol{Z}\right)=\boldsymbol{Z}_{g}\label{eq:47}
\end{equation}
\end{widetext}
in terms of the logical spins, or equivalently, 
\begin{widetext}
\begin{equation}
\boldsymbol{z}^{*}=\underset{\boldsymbol{z}\in\left\{ \pm1\right\} ^{N}}{\arg\max}P\left(\boldsymbol{z}|\boldsymbol{J}\right)=\underset{\boldsymbol{z}\in\left\{ \pm1\right\} ^{N}}{\arg\max}L\left(\boldsymbol{z},\boldsymbol{J}\right)=\underset{\boldsymbol{z}\in\left\{ \pm1\right\} ^{N}}{\arg\min}H^{phys}\left(\boldsymbol{z}\right)=\boldsymbol{z}_{g}\label{eq:48}
\end{equation}
\end{widetext}
in terms of the physical spins, where $\boldsymbol{Z}_{g}$ and $\boldsymbol{z}_{g}$
are the ground states of the perturbed Hamiltonians $H^{logi}\left(\boldsymbol{Z}\right)$
and $H^{phys}\left(\boldsymbol{z}\right)$, respectively. 
 It should be noted that the two formulations are equivalent if and only if the one-to-one correspondence between $\boldsymbol{Z}_{g}$ and $\boldsymbol{z}_{g}$ is preserved, that is, $\boldsymbol{z}_{g}$ needs to be the code state. Therefore, Eq. (\ref{eq:48}) should be complemented with the parity constraints $S_{i}\left(\boldsymbol{z}_{g}\right)=+1 \left(i=1,\ldots,N-K\right)$. To this end, one needs to select an appropriate set of weight parameters $\left\{ \beta,\lambda\right\}$ in the physical Hamiltonian $H^{phys}\left(\boldsymbol{z}\right)$, as discussed later. Equations (\ref{eq:47}) and (\ref{eq:48}) are called the maximum a posteriori (MAP) estimators.
The target state $\boldsymbol{z}_{g}$ should be a code state that is nearest
to $\boldsymbol{J}$ and minimizes the block-wise error probability $p_{B}$
in the communication application. Similarly, in the optimization application, the target states
are the logical ground state $\boldsymbol{Z}_{g}$ and the associated physical
code states $\boldsymbol{z}_{g}=\boldsymbol{Z}_{g}\otimes\boldsymbol{Z}_{g}$ given $\boldsymbol{J}$ (and $\boldsymbol{h}$),
which encode the combinatorial optimization problem to be solved.
Therefore, they are common target states to be inferred from both viewpoints.

Nevertheless, it should be noted that if the readout $\boldsymbol{J}$ is significantly perturbed, 
the target states may be one of many excited states that
are neighboring states of the ground states of the perturbed Hamiltonian \cite{Rujan1993,Nishimori1993}.
In such a case, we can alternatively select the marginal posterior maximizer
(MPM) estimators, i.e., 
\begin{eqnarray}
\boldsymbol{Z}_{i}^{*}&=&\underset{Z_{i}\in\left\{ \pm1\right\} }{\arg\max}P\left(Z_{i}|\boldsymbol{J}\right)=\mathrm{sgn}\left[\underset{Z_{i}=\pm1}{\sum}Z_{i}P\left(Z_{i}|\boldsymbol{J}\right)\right]\nonumber \\
&=&\mathrm{sgn}\left[\left\langle Z_{i}\right\rangle _{P\left(\boldsymbol{Z}|\boldsymbol{J}\right)}\right]
\label{eq:49}
\end{eqnarray}
in terms of the logical spins and 
\begin{eqnarray}
z_{i}^{*}&=&\underset{z_{i}\in\left\{ \pm1\right\} }{\arg\max}P\left(z_{i}|\boldsymbol{J}\right)=\mathrm{sgn}\left[\underset{z_{i}=\pm1}{\sum}Z_{i}P\left(z_{i}|\boldsymbol{J}\right)\right]\nonumber \\
&=&\mathrm{sgn}\left[\left\langle z_{i}\right\rangle _{P\left(\boldsymbol{z}|\boldsymbol{J}\right)}\right]\label{eq:50}
\end{eqnarray}
iin terms of the physical spins, where 
\begin{equation}
P\left(Z_{i}|\boldsymbol{J}\right)=\underset{\left\{ Z_{k}:k\neq i\right\} }{\sum}P\left(\boldsymbol{Z}|\boldsymbol{J}\right)
\end{equation}
and 
\begin{equation}
P\left(z_{i}|\boldsymbol{J}\right)=\underset{\left\{ z_{k}:k\neq i\right\} }{\sum}P\left(\boldsymbol{z}|\boldsymbol{J}\right)
\end{equation}
are the marginal posterior distributions. The angular brackets in
Eqs. (\ref{eq:49}) and (\ref{eq:50}) denote the average over the thermal
distributions $P\left(\boldsymbol{Z}|\boldsymbol{J}\right)=\exp\left(-H^{logi}\left(\boldsymbol{Z}\right)\right)$
and $P\left(\boldsymbol{z}|\boldsymbol{J}\right)=\exp\left(-H^{phys}\left(\boldsymbol{z}\right)\right)$,
respectively, where the inverse temperature $\beta_{i}$ has been absorbed in
the perturbed Hamiltonians $H^{logi}\left(\boldsymbol{Z}\right)$
and $H^{phys}\left(\boldsymbol{z}\right)$. They may offer a better
estimate of the target states given the prior information regarding 
the environmental noise through the parameter $\beta_{i}$. It is
known that MPM decoding is better than MAP decoding if we
calculate the thermal average of the spin variables in Eqs. (\ref{eq:49})
and (\ref{eq:50}) using the perturbed Hamiltonian at the Nishimori temperature \cite{Rujan1993,Nishimori1993,Nishimori1999}.
We can infer the optimal estimator $\boldsymbol{Z}^{*}$ by reconstructing
the sets of optimal estimators $Z_{i}^{*}$ for $i=1,\ldots,K$ as
well as $\boldsymbol{z}^{*}$ by reconstructing $z_{i}^{*}$ for $i=1,\ldots,N$.
This is called maximal entropy decoding \cite{Rujan1993,Chancellor2016}.

\begin{figure*}
\centering{}
\includegraphics[viewport=200bp 100bp 720bp 460bp,clip,scale=0.7]{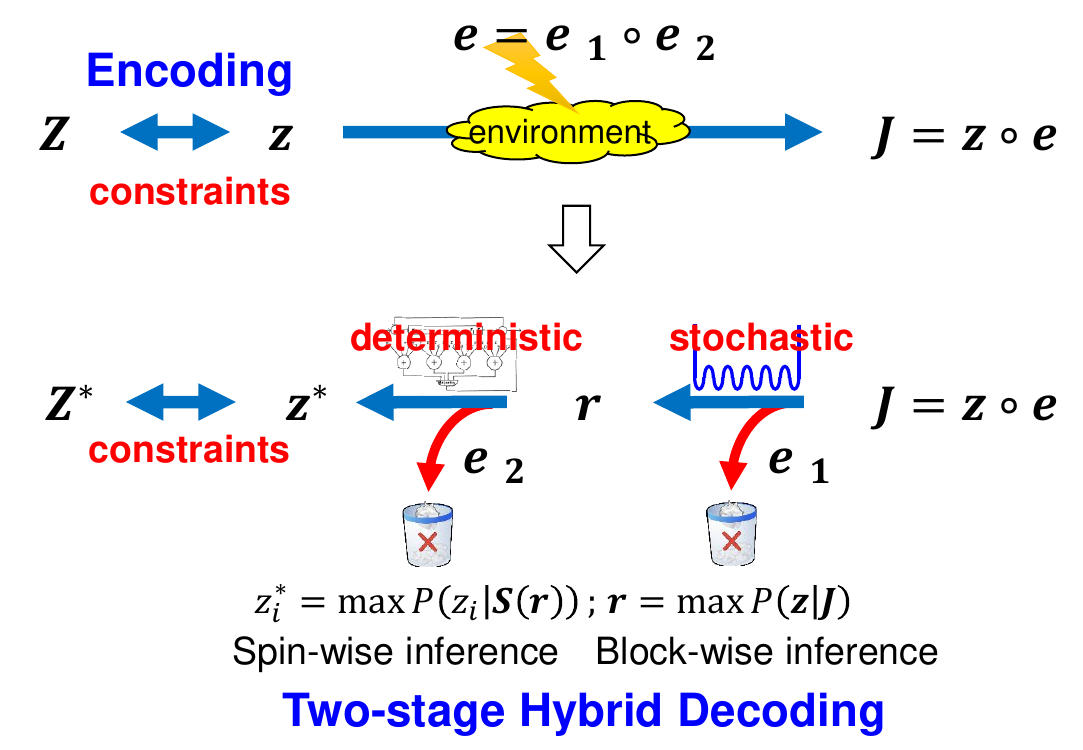}
\caption{Two-stage hybrid decoding scheme considered in this study. In the first stage, we decode the readout $\boldsymbol{J}$ using a stochastic
decoding method and obtain the interim readout $\boldsymbol{r}$.
In the second stage, we decode the readout $\boldsymbol{r}$
using a deterministic decoding method and obtain the optimal estimator $\boldsymbol{z}^{*}$.
The deterministic decoding is the one-step MVD in this study. The
errors in $\boldsymbol{J}$ are entirely corrected in the sequence
of the two decoding stages. \label{fig:5}}
\end{figure*}

In the following, we focus on the case where the target states are
the logical ground states $\boldsymbol{Z}_{g}$ and associated physical code state $\boldsymbol{z}_{g}=\boldsymbol{Z}_{g}\otimes\boldsymbol{Z}_{g}$. This is because our main objective is the development of an
efficient solver for combinatorial optimization problems by taking advantage
of nature-inspired approaches. It is assumed that we have a method
that can statistically sample the neighboring states of the physical
target state $\boldsymbol{r}\sim\boldsymbol{z}_{g}$, such as simulated
annealing based on Monte Carlo simulation and QA. The
readout state $\boldsymbol{r}$ may be a non-code state. We address
the decoding of the interim readout $\boldsymbol{r}$ to infer the target
state $\boldsymbol{z}^{*}$ by taking advantage of the informative
prior for parity constraints incorporated by the error-correcting codes. Let
us consider the following scenario. Given $\boldsymbol{J}$ and $\boldsymbol{h}$,
we statistically sample the interim readout $\boldsymbol{r}$ in the
first stage. Successively, we decode $\boldsymbol{r}$ with some deterministic
algorithms in the second stage and attempt to improve it by taking advantage
of the information involved in the syndrome that is derived from $\boldsymbol{r}$.
If the inferred state $\boldsymbol{z}^{*}$ is better than $\boldsymbol{r}$,
we select $\boldsymbol{z}^{*}$ as the estimate of $\boldsymbol{z}$.
We can iterate this post-readout decoding until saturation,
that is, no further improvement occurs. Therefore, our algorithm can be considered as a hybrid computation of earlier stochastic and later
deterministic algorithms, as shown schematically in Fig. \ref{fig:5}.
If the stochastic algorithm is executed by a quantum computer including
a quantum annealer, our algorithm is merely a hybrid quantum and classical 
computation. We studied how such decoding can be implemented from
the classical viewpoint. We analyzed the similarities and differences between decoding
for the PE scheme and QAC. For deterministic decoding in
the second stage, we focus on one-step MVD,
which is the simplest conventional decoding method for classical repetition
codes. We compare the one-step MVD for QAC and that for the PE
scheme. The one-step MVD relies on a priori probability (APP) decoding,
which resembles MPM decoding. Its theoretical foundation was first
described by Massey \cite{Massey1963,Massey1968}, who considered
a special set of syndromes called orthogonal parity checks and their correlation
with errors. Massey's theory suggests that we should be careful
to compare the correct set of readouts of the physical states
in order to perform fair majority voting and derive the correct inference.
Let us briefly review Massey's theory, called threshold decoding, using 
spin representation.

Let us consider the $\left(n,k\right)$ linear block code $C$. Let the vectors
$\boldsymbol{Z}=\left(Z_{1},\ldots,Z_{k}\right)\in\left\{ \pm1\right\} ^{k}$
and $\boldsymbol{z}=\left(z_{1},\ldots,z_{n}\right)\in\left\{ \pm1\right\} ^{n}$
denote the classical logical and physical spin states, respectively.
The logical state $\boldsymbol{Z}$ is assumed to be encoded by the
code $C$ in the physical state $\boldsymbol{z}$ according to Eq.
(\ref{eq:3}). The physical state $\boldsymbol{z}$ is assumed to
be constrained by the $n-k$ parity check equations, as shown in Eq. (\ref{eq:4}). 
Suppose that we sample $\boldsymbol{r}=\left(r_{1},\ldots,r_{n}\right)\in\left\{ \pm1\right\} ^{n}$
in the first stage, each element of which is hard-decided to bipolar values. The readout $\boldsymbol{r}$ may be corrupted by
noise and expressed using a set of random noise $\boldsymbol{e}=\left(e_{1},\ldots,e_{n}\right)\in\left\{ \pm1\right\} ^{n}$
as an element-wise (Hadamard) product $\boldsymbol{r}=\boldsymbol{z}_{g}\circ\boldsymbol{e}$,
that is, $e_{i}=\left(z_{g}\right)_{i}r{}_{i}$ $\left(i=1,\ldots,n\right)$,
where $e_{i}$ is the error in the readout $r_{i}$. We consider
the syndrome pattern $\boldsymbol{S}\left(\boldsymbol{r}\right)=\left(S_{1}\left(\boldsymbol{r}\right),\ldots,S_{n-k}\left(\boldsymbol{r}\right)\right)\in\left\{ \pm1\right\} ^{n-k}$
associated with the parity check equations as $S_{i}\left(\boldsymbol{r}\right)=M_{k_{1}\cdots k_{q}}^{i}\stackrel[l=1]{q}{\prod}r_{k_{l}}\in\left\{ \pm1\right\} $.
The key point of one-step MVD is forming a special subset of $J$
composite syndromes $\boldsymbol{A}_{m}=\left(A_{m1},A_{m2},\ldots,A_{mJ}\right)\in\left\{ \pm1\right\} ^{J}$
with $J\leqq n-k$ from the product of the appropriately selected $S_{i}$ values
to check the error $e_{m}$ of the $m$th physical spin $\left(m=1,\ldots,n\right)$.
This is defined as 
\begin{equation}
A_{mi}=r_{m}\underset{j\in C_{i}\left(m\right)}{\mathop{\prod}}r_{j}=e_{m}\underset{j\in C_{i}\left(m\right)}{\prod}e_{j},\label{eq:53}
\end{equation}
where $C_{i}$$\left(m\right)$ is an $n_{mi}=\left|C_{i}\left(m\right)\right|$
tuple of the indices of $\boldsymbol{r}$ comprising the syndrome $A_{mi}$.
Note that the second equality in Eq. (\ref{eq:53}) holds because $A_{mi}$
is a product of $S_{i}$ and is itself a syndrome. In addition, because any code
word $\boldsymbol{c}$ satisfies $\boldsymbol{cH}^{T}=\boldsymbol{0}$, 
\begin{equation}
z_{m}\left(\underset{j\in C_{i}\left(m\right)}{\mathop{\mathop{\prod}}}z_{j}\right)=+1
\end{equation}
holds for any code state $\boldsymbol{z}$. Thus, $A_{mi}$ is a function
of only $\boldsymbol{e}$. $\boldsymbol{A}_{m}$ has another notable
property: every $A_{mi}$ involves the noise component $e_{m}$, but no
other noise component $e_{i}$ $\left(i\neq m\right)$ is involved
in more than one syndrome in the set, i.e., every $e_{i}$
$\left(i\neq m\right)$ appears in only one of the sets $\left\{ C_{1}\left(m\right),\ldots,C_{J}\left(m\right)\right\} $.
Such a subset $\boldsymbol{A}_{m}$ of composite syndromes is said
to be orthogonal on $e_{m}$. Massey showed that a set $\boldsymbol{A}_{m}$
of syndromes orthogonal on $e_{m}$ is a good APP estimator for $e_{m}$ \cite{Massey1963,Massey1968}. 

Consider the posterior probability distribution $P\left(\boldsymbol{e}|\boldsymbol{A}_{m}\right)$,
which describes the conditional probability that $\boldsymbol{e}$
is the noise pattern when the syndrome $\boldsymbol{A}_{m}$ orthogonal
on $e_{m}$ has been observed, and its marginals $P\left(e_{m}|\boldsymbol{A}_{m}\right)=\underset{\boldsymbol{e}\neq e_{m}}{\sum}P\left(\boldsymbol{e}|\boldsymbol{A}_{m}\right)$.
We focus on $e_{m}$ instead of $\sigma_{m}$ because 
we can expect a nontrivial statistical correlation between $e_{m}$
and $\boldsymbol{A}_{m}$, as $A_{mi}$ is a function of only $\boldsymbol{e}$,
as shown in Eq. (\ref{eq:53}). Once the Bayes optimal estimator $e_{m}^{*}$
for $e_{m}$ is obtained for $m=1,\ldots,n$, we can determine the optimal
estimator $z_{m}^{*}$ for $z_{m}$ by $z_{m}^{*}=r_{m}e_{m}^{*}$.
In the non-algebraic decoding scheme, Bayesian inference
can be used to obtain the most probable error $e_{m}$ of the $m$th physical
state $z_{m}$, similar to MPM decoding. Then, the optimal estimator
$e_{m}^{*}$ is expressed as  
\begin{eqnarray}
e_{m}^{*}&=&\underset{e_{m}=\pm1}{\arg\max}\left[P\left(e_{m}|\boldsymbol{A}_{m}\right)\right]=\mathrm{sgn}\left[\underset{e_{m}=\pm1}{\sum}e_{m}P\left(e_{m}|\boldsymbol{A}_{m}\right)\right]\nonumber \\
&=&\mathrm{sgn}\left[\left\langle e_{m}\right\rangle _{P\left(e_{m}|\boldsymbol{A}_{m}\right)}\right],
\label{eq:55}
\end{eqnarray}
where the angular bracket in Eq. (\ref{eq:55}) indicates the average
over the distribution $P\left(e_{m}|\boldsymbol{A}_{m}\right)$. Equation (\ref{eq:55})
can be interpreted as the decoding strategy, where $e_{m}^{*}$ is
determined by the majority vote of $e_{m}$ associated with the statistically
independent ensemble with the posterior distribution $P\left(e_{m}|\boldsymbol{A}_{m}\right)$.
In general, the evaluation of such an average requires the computation of the summations
of $2^{n-1}$ terms for determining the two-valued marginal posterior,
which may require computational resources that scale exponentially
with $n$. Otherwise, we need to use approximate methods, e.g., Monte
Carlo sampling or a computationally more efficient probability propagation
algorithm \cite{Pastawski2016}. In contrast, if we can construct
a syndrome $\boldsymbol{A}_{m}$ orthogonal on $e_{m}$ successfully
from the syndrome $\boldsymbol{S}$ and use a simplifying assumption,
we can evaluate $\left\langle e_{m}\right\rangle _{P\left(e_{m}|\boldsymbol{A}_{m}\right)}$
cheaply using the knowledge of $P\left(e_{m}|\boldsymbol{A}_{m}\right)$.

Based on the orthogonality on $e_{m}$ of $\boldsymbol{A}_{m}$ and the assumption
of spin-by-spin independence of the noise, any set of distinct states
in the syndromes $\boldsymbol{A}_{m}$ is disjoint. Then, because of
the Bayes theorem and monotonicity of logarithmic functions, it follows
that 
\begin{equation}
e_{m}^{*}=\mathrm{sgn}\left[w_{m0}+\stackrel[i=1]{J}{\sum}w_{mi}A_{mi}\right],\label{eq:56}
\end{equation}
where 
\begin{eqnarray}
P\left(A_{mi}=-1|e_{m}=+1\right)&=&P\left(A_{mi}=+1|e_{m}=-1\right)\nonumber \\
&=&p_{i},
\label{eq:57}
\end{eqnarray}
\begin{eqnarray}
P\left(A_{mi}=-1|e_{m}=-1\right)&=&P\left(A_{mi}=+1|e_{m}=+1\right)\nonumber \\
&=&1-p_{i},\label{eq:58}
\end{eqnarray}
$w_{m0}=\log\frac{1-\xi_{m}}{\xi_{m}}$, $w_{mi}=\log\frac{1-p_{i}}{p_{i}}$,
and $\xi_{m}=P\left(e_{m}=-1\right)=1-P\left(e_{m}=+1\right)$ for
$i=1,\ldots,J$. Here, $e_{i}$ is assumed to be a statistically random
variable independent of $i$. In Eqs. (\ref{eq:57}) and (\ref{eq:58}),
$p_{i}$ is the probability of an odd number of \textquotedbl$-1$\textquotedbl{}
among all $e_{j}$ exclusive of $e_{m}$ in the products that constitute
$A_{mi}$, the $i$th syndrome orthogonal on $e_{m}$. It follows
that 
\begin{equation}
p_{i}=\frac{1}{2}\left[1-\underset{j\in B_{i}\left(m\right)}{\prod}(1-2\xi_{j})\right].\label{eq:59}
\end{equation}
It should be noted that if $\xi_{i}<\frac{1}{2}$ for any $i=1,\ldots,n$,
it follows that $0<w_{mi}\leq w_{m0}$ for $i=1,\ldots,J$. The positive
value of $w_{mi}$ implies that there should be a positive statistical
correlation between $A_{mi}$ and $e_{m}$, i.e., $1-p_{i}>p_{i}$.
Equation (\ref{eq:56}) indicates that the Bayes optimal estimator $e_{m}^{*}$
for $e_{m}$ can be evaluated from the syndrome $\boldsymbol{A}_{m}$
orthogonal on $e_{m}$ if we can calculate the weight $w_{mi}$
from that of $\xi_{i}$ for $i=1,\ldots,J$. 

The key points for deriving Eqs. (\ref{eq:57})--(\ref{eq:59}) are
as follows. The syndrome $\boldsymbol{A}_{m}=\left(A_{m1},A_{m2},\ldots,A_{mJ}\right)$
is selected in such a manner that all errors $e_{i}$ $\left(m=1,\ldots,n\right)$
are involved in at least one of the elements of $\boldsymbol{A}_{m}$
but the errors $e_{m}$ and $e_{i\neq m}$ are asymmetrical; $e_{m}$ is
involved in all elements of $\boldsymbol{A}_{m}$, whereas the other
errors $e_{i\neq m}$ are involved in only one of the elements of $\boldsymbol{A}_{m}$.
Comprehensive examples are presented later. Of course, $\boldsymbol{A}_{m}=+\boldsymbol{1}$
if and only if $\boldsymbol{e}=+\boldsymbol{1}$, where $+\boldsymbol{1}$
is a constant vector with the element as $+1$. The right-hand side of
Eq. (\ref{eq:56}) can be considered to be a function of the syndromes
$\boldsymbol{S}$, as $\boldsymbol{A}_{m}$ is a function of $\boldsymbol{S}$.
That is, we can infer the error pattern for all spins by
analyzing the syndrome pattern, which is very reasonable. It should
be noted that a choice in Eq. (\ref{eq:53}) might be fair if the
set $C_{i}$$\left(m\right)$ treats all readouts $r_{i}$ of
the spin variables symmetrically and the majority voting defined by Eq. (\ref{eq:55}) uses all syndromes \cite{Pastawski2016}. The
fair choice of the set $C_{i}$$\left(m\right)$ might be crucial
for realizing efficient one-step MVD. 

Equation (\ref{eq:56}) can be further simplified by modifying $\boldsymbol{A}_{m}$
in the following manner \cite{Massey1963,Tanaka1980,Clerk1981}. First,
we introduce $\boldsymbol{B}_{m}=\left(B_{m0},B_{m1},\ldots,B_{mJ}\right)\in\left\{ \pm1\right\} ^{J+1}$,
where $B_{m0}=r_{m}$ and 
\begin{eqnarray}
B_{mi}&=&r_{m}A_{mi}=\underset{j\in C_{i}\left(m\right)}{\mathop{\prod}}r_{j}\nonumber \\
&=&z_{m}\underset{j\in C_{i}\left(m\right)}{\prod}e_{j}\;\left(m=1,\ldots,k;i=1,\ldots,J\right).
\end{eqnarray}
We state that the $\boldsymbol{B}_{m}$ estimators are orthogonal on $z_{m}$. 
Evidently, $B_{mi}$ satisfies $z_{m}B_{mi}=z_{m}r_{m}A_{mi}=e_{m}A_{mi}$
for $m=1,\ldots,k$ and $i=1,\ldots,J$. Similar to Eq. (\ref{eq:56}),
the Bayes optimal estimator $z_{m}^{*}$ for $z_{m}$ is expressed as 
\begin{eqnarray}
z_{m}^{*}&=&\underset{z_{m}=\pm1}{\arg\max}\left[P\left(z_{m}|\boldsymbol{B}_{m}\right)\right]=\mathrm{sgn}\left[\underset{z_{m}=\pm1}{\sum}z_{m}P\left(z_{m}|\boldsymbol{B}_{m}\right)\right]\nonumber \\
&=&\mathrm{sgn}\left[\left\langle z_{m}\right\rangle _{P\left(z_{m}|\boldsymbol{B}_{m}\right)}\right].
\end{eqnarray}
If we assume spin-by-spin independence of the environmental noise,
it follows that 
\begin{equation}
z_{m}^{*}=\mathrm{sgn}\left[\stackrel[i=0]{J}{\sum}w_{mi}B_{mi}\right],\label{eq:62}
\end{equation}
where we have assumed an ignorance prior for the physical spin state
$\boldsymbol{z}$, i.e., $P\left(z_{m}=+1\right)=P\left(z_{m}=-1\right)$.
The equality $z_{m}B_{mi}=e_{m}A_{mi}$ implies $P\left(z_{m}B_{mi}=\pm1|z_{m}\right)=P\left(e_{m}A_{mi}=\pm1|e_{m}\right)$
for $m=1,\ldots,k$; that is, $P\left(B_{mi}=\pm V|z_{m}=V\right)=P\left(A_{mi}=\pm V|e_{m}=V\right)$,
where $V\in\left\{ \pm1\right\} $. It is easy to confirm that $w_{mi}$
in Eqs. (\ref{eq:56}) and (\ref{eq:62}) are the same, because for $m=1,\ldots,k$,
\begin{eqnarray}
P\left(B_{mi}=-1|z_{m}=+1\right)&=&P\left(B_{mi}=+1|z_{m}=-1\right)\nonumber \\
&=&\left\{ \begin{array}{cl}
\xi_{m} & \mathrm{for}\;i=0\\
p_{i} & \mathrm{for}\;i=1,\ldots,J
\end{array}\right.
\end{eqnarray}
and 
\begin{eqnarray}
P\left(B_{mi}=-1|z_{m}=-1\right)&=&P\left(B_{mi}=+1|z_{m}=+1\right)\nonumber \\
&=&\left\{ \begin{array}{cl}
1-\xi_{m} & \mathrm{for}\;i=0\\
1-p_{i} & \mathrm{for}\;i=1,\ldots,J
\end{array}\right.\nonumber \\
\end{eqnarray}
hold. Equation (\ref{eq:62}) indicates that $z_{m}^{*}$ is determined
by the one-step majority vote of the $J+1$ elements $B_{mi}$$\;\left(i=0,\ldots,J\right)$
in the $\boldsymbol{B}_{m}$ weighted by $w_{mi}$, that is, 
\begin{equation}
z_{m}^{*}=\mathrm{maj}\left(w_{m0}B_{m0},w_{m1}B_{m1},\ldots,w_{mJ}B_{mJ}\right),\label{eq:65}
\end{equation}
where we have used the following definition for the majority function of bipolar
variables $y_{0},y_{1},\ldots,y_{J}\in\left\{ \pm1\right\} ^{J+1}$ \cite{Rudolph1972}:
\begin{equation}
\mathrm{maj}\left(y_{0},y_{1},\ldots,y_{J}\right)=\begin{cases}
+1, & \mathrm{if}\;\stackrel[i=0]{J}{\sum}y_{i}>0\\
-1, & \mathrm{if}\;\stackrel[i=0]{J}{\sum}y_{i}<0\\
\mathrm{undefined} & \mathrm{otherwise}
\end{cases}.
\end{equation}
The sign of the weight $w_{mi}$ describes the positive $\left(w_{mi}>0\right)$
or negative $\left(w_{mi}<0\right)$ statistical correlation that is 
expected between the orthogonal estimators $B_{mi}$ $\left(A_{mi}\right)$
and $z_{m}$ $\left(e_{m}\right)$ in Eq. (\ref{eq:62}) (Eq.(\ref{eq:56})) on which Bayes inference relies. The magnitude of
the weight $\left|w_{mi}\right|$ describes the measure of reliability
of the estimator $B_{mi}$ $\left(A_{mi}\right)$, which is determined
by the error probability $\xi_{j}$ for the readout $r_{j}$ that
constitutes the estimator $B_{mi}$ $\left(A_{mi}\right)$. When 
$\left|w_{mi}\right|$ is larger, the estimator $B_{mi}$
$\left(A_{mi}\right)$ is more reliable. The optimal estimate for the remaining spin
states $z_{i}^{*}$ $\left(i\neq m\right)$ can be performed by sifting
the index of the spin cyclically and repeating the one-step MVD in the
same manner. This is because if it is possible to form a set of $J$
syndromes orthogonal on $e_{m}$, it is possible to form a set of
$J$ syndromes orthogonal on any spin $e_{i}\;\left(i\neq m\right)$
if $C$ is a cyclic code. Correct decoding of $z_{m}$ is guaranteed
if there are $\left\lfloor \frac{J}{2}\right\rfloor $ or fewer errors
in the error of the readout
$\boldsymbol{r}$. 

Note that the calculation of the weight $w_{mi}$ becomes computationally
expensive when $n_{mi}=\left|C_{i}\left(m\right)\right|\gg1$, even
if all $\xi_{j}$ $\left(j=1,\ldots,n\right)$ are known, because
$w_{mi}$ is a complicated nonlinear function of $\xi_{j}$ for the
received state $r_{j}$, which is a member of $C_{i}$$\left(m\right)$.
Therefore, it is often necessary to introduce some approximations in the calculation
of Eq. (\ref{eq:59}), which is beyond the scope of our study and can be referenced in \cite{Tanaka1980,Shu2004}.
Here, we note that the most simplified approximation assumes that $w_{mi}=const.$
for $i=0,\ldots,J$, which amounts to neglecting the imbalance in the
reliability of the estimators owing to imbalance in the error probabilities
$\xi_{i}$ of the readout $r_{i}$. In this simplified approximation,
$z_{m}^{*}$ is determined by the one-step majority vote of the $J+1$
elements $B_{mi}$ in $\boldsymbol{B}_{m}$; that is, 
\begin{equation}
z_{m}^{*}=\mathrm{maj}\left(B_{m0},B_{m1},\ldots,B_{mJ}\right).\label{eq:67}
\end{equation}

\subsection{Classical repetition code}
We discuss two illustrative
examples, namely the classical repetition code and PE scheme, to understand the one-step MVD comprehensively. Let
us first consider the $\left(n,k\right)=\left(N,1\right)$ classical repetition
code, which encodes a logical spin variable $Z=z_{1}$ into $N$ physical
spin states $\boldsymbol{z}=\left(z_{1},\ldots,z_{N}\right)\in\left\{ \pm1\right\} ^{N}.$
The minimum code distance $d_{min}$ can be increased by repeating the
logical information $N$ times; thus, $d_{min}=N$, which enables the correction of 
errors in the $N$ physical spin states. In principle, the repetition
code should be capable of correcting up to $t_{RP}=\left\lfloor \frac{d_{min}-1}{2}\right\rfloor $
spin-flip errors. 
The orthogonal syndrome appropriate for one-step MVD is expressed as $A_{i}=r_{1}r_{i+1}=e_{1}e_{i+1}$ for $i=1,\ldots,N-1$. 
Therefore, we can use the syndrome $\boldsymbol{S}$ in Eq. (\ref{eq:37}) as the estimator for the error $e_{1}$; that is, 
$\boldsymbol{A}=\boldsymbol{S}=\left(S_{1},S_{2},\ldots,S_{N-1}\right)\in\left\{ \pm1\right\} ^{N-1}$.
Thus, the repetition code is a self-orthogonal code
\cite{Clerk1981,Shu2004}. If we apply the second type of one-step
MVD in Eq. (\ref{eq:62}), the set of $J+1=N$ estimators $\boldsymbol{B}$ orthogonal on $z_{1}$
associated with the set of $J=N-1$ syndromes $\boldsymbol{A}$ orthogonal
on $e_{1}$ is identified as $\boldsymbol{B}=\boldsymbol{r}=\left(r_{1},\ldots,r_{N}\right)\in\left\{ \pm1\right\} ^{N}$.
The Bayes-optimal estimator $Z^{*}$ for $Z$ is given
by the one-step majority vote of the set of readouts $\boldsymbol{r}$, that is, 
\begin{equation}
Z^{*}=z_{1}^{*}=\mathrm{maj}\left(r_{1},\ldots,r_{N}\right),\label{eq:68}
\end{equation}
if we neglect the imbalance for the reliability
of the estimator $r_{i}$. 
Equation (\ref{eq:68}) implies that the one-step MVD can correct up to
$\left\lfloor \frac{N-1}{2}\right\rfloor $ errors. Because a set $\boldsymbol{A}$
of $J=d_{min}-1$ syndromes orthogonal on error $e_{1}$ of the logical
spin state can be formed, one-step MVD can achieve the error-correcting
capability $t_{RP}$ of the repetition code. Therefore, the repetition
code can be completely orthogonalized in one step \cite{Massey1963,Massey1968,Laferriere1977,Clerk1981,Shu2004}. 

\subsection{QAC}
The repetition spin-flip code is used for the stabilizer subspace codes \cite{Lidar2019}
in QAC to suppress the local spin-flip errors in QA \cite{Jordan2006,Young2013,Pudenz2014,Pudenz2015,Vinci2015,Bookatz2015,Matsuura2016,Matsuura2017,Pearson2019}.
We focus on the classical model of $N$ logical spins described by
the logical Hamiltonian $\hat{H}^{logi}\left(\hat{\boldsymbol{Z}}\right)$ (Eq. (\ref{eq:42})), which are
embedded into the $N\times K$ physical spin system described by the QAC Hamiltonian
$\hat{H}^{QAC}\left(\hat{\boldsymbol{z}}\right)$ (Eq. (\ref{eq:40})). To this end, the quantum mechanical operators
$\hat{z}_{i}$ are replaced with the classical spin variables $z_{i}\in\left\{ \pm1\right\}$
and their associated readouts are considered to be $r_{i}\in\left\{ \pm1\right\}$. To describe
the state of the physical spins and its readout, we introduce the $N\times K$
matrices $\boldsymbol{z}\in\left\{ \pm1\right\} ^{N\times K}$ and $\boldsymbol{r}\in\left\{ \pm1\right\} ^{N\times K}$
($\left\{ \pm1\right\} ^{N\times K}$ denotes an $N\times K$ matrix
of bipolar variables), whose elements $z_{ij}$ and $r_{ij}$ describe
the state and its readout, respectively, of the physical spin specified by the index
$i\;\left(=1,\ldots,N\right)$ in the replica $j\;\left(=1,\ldots,K\right)$.
\begin{figure}[h]
\begin{centering}
\includegraphics[viewport=320bp 150bp 650bp 410bp,clip,scale=0.8]{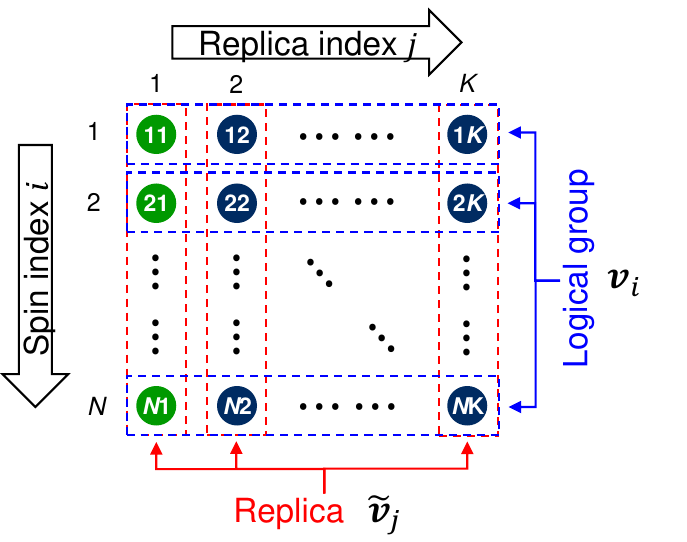}
\caption{Structure of the matrices $\boldsymbol{v}=\boldsymbol{z}\textrm{ or }\boldsymbol{r}$.
The green circles are the elements corresponding directly to the logical spins. \label{fig:6}}
\par\end{centering}
\end{figure}

To simplify the notation, we call $\boldsymbol{z}$ and $\boldsymbol{r}$
the physical state and readout, respectively. As shown in Fig. \ref{fig:6}, each 
column of $\boldsymbol{v}=\boldsymbol{z}$ $\left(\boldsymbol{v}=\boldsymbol{r}\right)$ 
describes the states (readouts) of $N$ 
physical spins in each replica, whereas each row describes the states (readouts) of $K$ physical spins 
in each logical group associated with the logical spin \cite{Vinci2015}.
To simplify the notation, let $\boldsymbol{v}_{i}=\left(v_{i\,1},\ldots,v_{i\,K}\right)$
denote the $i$th row vector of $\boldsymbol{v}$ and let $\tilde{\boldsymbol{v}}_{j}=\left(v_{1\,j},\ldots,v_{N\,j}\right)^{T}$
denote the $j$th column vector of $\boldsymbol{v}$ so that $\boldsymbol{v}=\left(\tilde{\boldsymbol{v}}_{1},\ldots,\tilde{\boldsymbol{v}}_{K}\right)=\left(\boldsymbol{v}_{1},\ldots,\boldsymbol{v}_{N}\right)^{T}$, where $\boldsymbol{v}=\boldsymbol{z}\textrm{ or }\boldsymbol{r}$.
The vectors $\boldsymbol{z}_{i}$ and $\tilde{\boldsymbol{z}}_{j}$
describe the state of the $i$th logical group associated with the $i$th
logical spin and state of the $j$th replica, respectively. The spins in every replica
share the intra-replica couplings $J_{ij}$ that are defined in the logical
Hamiltonian $H^{logi}\left(\boldsymbol{Z}\right)$. Each logical spin variable $Z_{i}$ is encoded
by the classical $\left[K,1\right]$ repetition code and constitutes the logical
group $\boldsymbol{z}_{i}$ $\left(i=1,\ldots,N\right)$, each of which consists
of $K$ physical spins. We impose the parity constraints $z_{i,1}z_{i,j+1}=+1$
$\left(j=1,\ldots,K-1\right)$ for the physical spin variables in each
logical group $i$ independently. Because $N$ replicas share the same intra-replica
couplings, we may expect that every weight $w_{ij}$ of the estimator $r_{ij}$
$\left(j=1,\ldots,K\right)$ for the one-step MVD is independent of
$j$ based on symmetry consideration. Then, we can identify 
\begin{equation}
A_{ij}=r_{i1}r_{i\,j+1}=e_{i1}e_{i\,j+1}\;\left(j=1,\ldots,K-1\right)
\end{equation}
for $K-1$ syndromes orthogonal on the error $e_{i1}$ for $i=1,\ldots,N$.
Evidently, $A_{ij}$ satisfies the requirements for orthogonal syndromes,
as it involves a single common element $r_{i1}$ and the other
elements $r_{i\,j+1}$ $\left(j=1,\ldots,K-1\right)$ that appear
only once in the $K-1$ element $A_{ij}$. Then, we can further identify
$B_{i1}=r_{i1}$ and 
\begin{equation}
B_{i\,j+1}=r_{i1}A_{ij}=r_{i\,j+1}=z_{i1}e_{i\,j+1}\;\left(j=1,\ldots,K-1\right)
\end{equation}
for $K$ estimators that are orthogonal on every physical spin variable $z_{i1}$
for $i=1,\ldots,N$. Thus, we obtain the Bayes-optimal estimator
for the $i$th logical spin variable $Z_{i}=z_{i1}$ from the readouts
$\boldsymbol{r}$ as follows:
\begin{equation}
Z_{i}^{*}=z_{i1}^{*}=\mathrm{maj}\left(B_{i1},\ldots,B_{iK}\right)=\mathrm{sgn}\left[\left\langle \boldsymbol{r}\right\rangle _{i}\right],\label{eq:71}
\end{equation}
\begin{equation}
\left\langle \boldsymbol{r}\right\rangle _{i}=\left\langle \boldsymbol{r}_{i}\right\rangle =\frac{1}{K}\stackrel[j=1]{K}{\sum}r_{ij},\label{eq:72}
\end{equation}
where $\boldsymbol{B}_{i}=\left(B_{i\,1},\ldots,B_{i\,K}\right)=\boldsymbol{r}_{i}\in\left\{ \pm1\right\} ^{K}$
corresponds to the orthogonal estimators for $Z_{i}$ and the angular
brackets $\left\langle \cdots\right\rangle _{i}$ denote the average
of the matrix elements in the $i$th row. Equations (\ref{eq:71}) and (\ref{eq:72})
indicate that the most probable value of the logical spin variable
$Z_{i}^{*}$ is determined by the average orientation of the readouts $\boldsymbol{r}_{i}$
for the $K$ physical spins in the associated logical group $i$.
Thus, we can infer the most probable state $\boldsymbol{Z}^{*}$ for
$N$ logical spins from the sign of the averaged readouts $\left\langle \boldsymbol{r}\right\rangle _{i}$ 
for $i=1,\ldots,N$.

Although this result appears rather trivial, it may nevertheless be interesting
to compare it with the finite temperature decoding of the Sourlas
code \cite{Rujan1993,Nishimori1993,Nishimori1999}. Consider the limit $\gamma\rightarrow0$ for the QAC Hamiltonian $H^{QAC}\left(\boldsymbol{z}\right)$.
Then, the penalty Hamiltonian $H^{pen}\left(\boldsymbol{z}\right)$, which forces the spins
in each logical group $i$ to be aligned 
for $\boldsymbol{z}$ to become the code state, is absent. Thus, the $K$
replicas run independently in parallel in $\gamma\rightarrow0$. The same results should be applicable if a single replica
runs $K$ times according to the logical Hamiltonian and if the average orientation 
of every readout of the logical spin variables is evaluated, which
is equivalent to finite temperature decoding. This implies that if the
energetic penalty is sufficiently weak, thermal excitation might be beneficial
for QAC when it is combined with the post-readout classical decoding.
This aspect is discussed later in the paper.

Let us compare the following three post-readout classical decoding
strategies for the $N\times K$ physical spins to obtain the ground
state $\boldsymbol{Z}_{g}$ of the logical Hamiltonian $H^{logi}\left(\boldsymbol{Z}\right)$ (hereinafter referred to as the logical ground
state) embedded in $H^{QAC}\left(\boldsymbol{z}\right)$. (a) MAP decoding attempts 
to minimize $H^{QAC}\left(\boldsymbol{z}\right)$ and simply reports its readout $\boldsymbol{r}$.
This strategy is also called the energy minimization strategy \cite{Vinci2015}.
(b) Best replica (BR) decoding after MAP decoding reports
the readout $\left(\tilde{\boldsymbol{r}}_{j^{*}}\right)^{T}$ (the transpose of column $j^{*}$ of
$\boldsymbol{r}$) with the smallest logical energy, i.e., $j^{*}=\underset{j}{\arg\min}H^{logi}\left(\tilde{\boldsymbol{r}}_{j}\right)$, 
where $\boldsymbol{r}$ is the readout of the MAP decoding. (c) One-step MVD after MAP decoding 
reports $\boldsymbol{Z}^{*}$
obtained by one-step MVD $Z_{i}^{*}=\mathrm{sgn}\left[\left\langle \boldsymbol{r}\right\rangle _{i}\right]$
for $i=1,\ldots,N$ from the readout $\boldsymbol{r}$ of the MAP decoding. Note that $\left\langle \boldsymbol{r}\right\rangle _{i}$
can be calculated easily by averaging the $i$th row of the readout matrix
$\boldsymbol{r}$. We first focus on strategies (a) and (b).
Suppose that the readout $\boldsymbol{r}_{a}$ of the (a) MAP decoding  
and (b) BR decoding on the readout $\boldsymbol{r}_{b}$ of the MAP decoding successfully
infers the target state $\boldsymbol{Z}^{*}=\boldsymbol{Z}_{g}$.
Then, all columns of $\boldsymbol{r}_{a}$, i.e., $\left(\tilde{\boldsymbol{r}_{a}}\right)_{j}$ for $j=1,\ldots,N$, 
should agree with $\left(\boldsymbol{Z}_{g}\right)^{T}$ up to a global spin flip 
in strategy (a), as MAP decoding infers
the most probable state for the physical ground state $\boldsymbol{z}_{g}$ of $H^{QAC}\left(\boldsymbol{z}\right)$,
whereas it is sufficient for a column of $\boldsymbol{r}_{b}$, i.e., $\left(\tilde{\boldsymbol{r}_{b}}\right)_{j}$ for at least one $j$, 
to agree with $\left(\boldsymbol{Z}_{g}\right)^{T}$ up to a global spin flip
in strategy (b). This suggests that strategy (b) can sample the
logical ground state $\boldsymbol{Z}_{g}$ more efficiently
than strategy (a). For example, consider the limit $\gamma\rightarrow0$
for simplicity, where the replicas are statistically independent. Then, it is reasonable to assume that every replica succeeds with the same probability 
$p$ for sampling the logical ground state $\boldsymbol{Z}_{g}$.
This implies that the probability for successfully sampling the target state $\boldsymbol{Z}_{g}$ is $P_{a}=p^{K}$ for strategy (a), 
whereas the probability that the best of the $K$ replicas
will succeed is $P_{b}=1-\left(1-p\right)^{K}>p$ for $K\geq2$ for strategy (b). It follows 
that $P_{A}<p<P_{B}$ for $0<p<1$ and $P_{A}\ll p\ll P_{B}$ if $K\gg2$.
Therefore, strategy (b) will significantly improve the performance
compared with the improvement when using strategy (a) for the same problem instance, at least
in the limit $\gamma\rightarrow0$. 

Next, we compare strategy (c) one-step MVD after MAP decoding and strategy
(a) MAP decoding. Suppose that one-step MVD on the readout $\boldsymbol{r}_{c}$ of the MAP decoding 
successfully infers the logical ground state as the most probable state
$\boldsymbol{Z}^{*}=\boldsymbol{Z}_{g}$. None of
the columns of $\boldsymbol{r}_{c}$, i.e., $\left(\tilde{\boldsymbol{r}_{c}}\right)_{j}$, may agree with $\left(\boldsymbol{Z}_{g}\right)^{T}$
even if the logical ground state $\boldsymbol{Z}_{g}$ is inferred successfully in strategy
(c). This is possibly because one-step MVD can infer $\boldsymbol{Z}_{g}$
if at least half of the elements in every readout $\boldsymbol{r}_{i}$ associated with logical group $i$  
(every row of $\boldsymbol{r}$) are correct.
Furthermore, QAC has a peculiar property that $H^{QAC}\left(\boldsymbol{z}\right)$, $H^{enc}\left(\boldsymbol{z}\right)$, and $H^{pen}\left(\boldsymbol{z}\right)$
share the same physical ground state $\boldsymbol{z}_{g}$. This implies
that the ground state of $H^{QAC}\left(\boldsymbol{z}\right)=\beta H^{enc}\left(\boldsymbol{z}\right)+\gamma H^{pen}\left(\boldsymbol{z}\right)$ is
independent of $\beta$ and $\gamma$. Then, the majority of the erroneous
but correctable states that contribute to the readout $\boldsymbol{r}_{c}$
should be the excited states of $H^{QAC}\left(\boldsymbol{z}\right)$, $H^{enc}\left(\boldsymbol{z}\right)$, and $H^{pen}\left(\boldsymbol{z}\right)$
with a small Hamming distance from the physical
ground state $\boldsymbol{z}_{g}$. Thus, they should be distributed in the low-energy portion
of the energy spectrum of $H^{QAC}\left(\boldsymbol{z}\right)$, which should have a density of states that is much higher than that of the ground state $\boldsymbol{z}_{g}$. QAC can sample such excited states
efficiently through sampling based on the Hamiltonian $H^{QAC}\left(\boldsymbol{z}\right)$ at finite
temperature (by local cooling \cite{Young2013-2,Sarovar2013}) even
in the limit $\gamma\rightarrow0$, where the penalty Hamiltonian
$H^{pen}\left(\boldsymbol{z}\right)$ is absent. Thus, QAC may acquire great benefit from
the post-readout classical decoding even if $\gamma=0$. This is similar to the finite
temperature decoding of the Sourlas code, even though the target states
differ. This discussion suggests that even though the energetic
penalty is sufficiently weak, QAC may have better performance by applying one-step
MVD of the readout after sampling at a finite temperature.

Our discussion also suggests that thermal excitation might be beneficial
for QAC even if $\gamma>0$ when it is combined with the post-readout
classical decoding. This is evident from the discussion that, given two elements
$H^{enc}\left(\boldsymbol{z}\right)$ and $H^{pen}\left(\boldsymbol{z}\right)$, the optimal values for the hyperparameter
set $\left\{ \beta,\gamma\right\} $ may depend on the post-readout
decoding strategy. Recall that $H^{pen}\left(\boldsymbol{z}\right)$ reflects the prior knowledge
on the parity constraints on the physical spin variables. Suppose
$\gamma>0$ in the QAC Hamiltonian $H^{pen}\left(\boldsymbol{z}\right)$ enforces $K$ physical
spins in every logical group to align by energetically
penalizing the misalignment with their ferromagnetic interactions. The
code states that satisfy all parity constraints are the degenerate
ground states of $H^{pen}\left(\boldsymbol{z}\right)$. If we consider the fact that the
physical ground state $\boldsymbol{z}_{g}$ of $H^{QAC}\left(\boldsymbol{z}\right)$ is the common ground state
of $H^{enc}\left(\boldsymbol{z}\right)$ and $H^{pen}\left(\boldsymbol{z}\right)$, it may be reasonable to balance two
contributions for the energy gap in $H^{QAC}\left(\boldsymbol{z}\right)$, i.e., $\beta\Delta H^{enc}\sim\gamma\Delta H^{pen}=\gamma$,
to sample the common ground state efficiently, where $\Delta H^{x}$
is the energy gap between the first excited and ground states
of $H^{x}$. Note that $\Delta H^{pen}=+1$. However, this might be
reasonable only for strategy (a), where the desired
readout $\boldsymbol{r}_{a}$ is the physical ground state $\boldsymbol{z}_{g}$,
but not for strategies (b) and (c), where the desired readouts $\boldsymbol{r}_{b}$
and $\boldsymbol{r}_{c}$ are very different from $\boldsymbol{r}_{a}$.
In particular, in strategy (c), the desired readout
$\boldsymbol{r}_{c}$ is one of a majority of correctable erroneous states
in which at least half of $K$ physical spins in every logical group
$\left(\boldsymbol{r}_{c}\right){}_{i}$ are aligned correctly. Because
such states are non-code states, redistributing the energy spectrum
of $H^{QAC}\left(\boldsymbol{z}\right)$ is preferable by controlling $\gamma$ so that $\gamma<\beta\Delta H^{Enc}$
to increase the sampling efficiency for the desired $\boldsymbol{r}^{c}$.
We also note that a finite energy gap remains in $\beta\Delta H^{enc}$, 
even if $\frac{\gamma}{\beta}$ approaches zero. This residual
gap $\beta\Delta H^{enc}$ might be the reason that thermal excitation
is beneficial even if $\gamma\rightarrow0$ for QAC when it is
combined with post-readout classical decoding. This situation
is in contrast with the PE scheme, where the energy gap closes as $\frac{\gamma}{\beta}$
becomes smaller, as considered later. It is important to note that this
consideration would be justified if the population of the sampling
results is distributed with a finite width so that the population of the relevant
excited states has a finite probability. This conclusion holds true regardless of whether the final state
reaches thermal equilibrium and the underlying dynamics approach
the final state \cite{Amin2015}.

Some of the above suggestions are actually observed
in the results of the QAC experiments in the literature. For example, in \cite{Pudenz2014},
Pudenz et al. demonstrated QAC using a programmable annealing processor.
They used $K=4$ replicated systems composed of a chain of $N$ logical
spins that were anti-ferromagnetically coupled, and ran the annealing
processor to sample the logical ground state. They claimed that a
substantial improvement in the performance of the processors was observed compared with the result in the absence of error correction in the QAC. They attributed
this improvement to the population of erroneous but correctable excited
states of the QAC Hamiltonian, although they did not specify the origin
of this population. It may be attributed to thermal excitation during
the course of annealing. Because the annealing processor is a real-world
machine, it is coupled to the thermal environment. Moreover, the user-specified parameters have finite strength and are controllable with finite
speed. In fact, the D-Wave annealer is likely to sample the statistical
ensemble of the readouts close to the thermal state, even though the
spin dynamics during the annealing are not necessary classical and
its temperature does not necessarily agree with (is higher than) the
operating temperature of the device \cite{Amin2015,Benedetti2016,Chancellor2016,Marshall2019}.
The results of Ref. \cite{Pudenz2014} suggest that post-readout
classical decoding can improve the success probability for sampling
the logical ground state by decoding the erroneous but correctable excited
states of the QAC Hamiltonian. Furthermore, they presented an interesting
result in their experiment: the optimal value of the penalty
strength $\gamma$ differs between our decoding strategies (b) and (c).
For example, the sampling results for the two strategies are compared in Fig. 5 of Ref. \cite{Pudenz2014}. In this figure, the success
probability for sampling the logical ground state is plotted as
a function of the anti-ferromagnetic chain length and penalty strength
$\gamma$ (penalty scale $\beta$ in their definition) for three values
of the strength $\beta$ (problem scale $\alpha$ in their definition).
Their results suggested that the optimal $\gamma$ is significantly
smaller for strategy (c) one-step MVD than strategy (b) BR
decoding, which is consistent with our consideration. In addition, the success
probability is relatively large for strategy (c) even though $\gamma\sim0$,
whereas it is very small for strategy (b), which is expected
from the finite temperature decoding.

\subsection{Parity encoding}

Annealing is a heuristic optimization method, in which the solution
is statistically inferred from an ensemble of many samples.
Such samples can be obtained either through repeated runs using
a single copy or through parallel runs using many copies. QAC can
be interpreted as a system for statistically inferring the optimal
solution through parallel runs of many copies based on the classical 
repetition codes. The solution can be obtained, for example, by selecting
the best of the sampled readouts (strategy (b)) or by majority
voting of the sampled readouts (strategy (c)). Now, let us consider the PE scheme. Although its interpretation is more subtle
than that of QAC, the comparison between the PE scheme and QAC helps us
to understand the error-correcting capability of the PE scheme. 

Consider the logical and encoded physical Hamiltonians $H^{logi}\left(\boldsymbol{Z}\right)$
and $H^{phys}\left(\boldsymbol{z}\right)$, given by Eqs. (\ref{eq:23})
and (\ref{eq:20}), respectively. We assume that the $K$-dimensional vector $\boldsymbol{Z}=\left(Z_{1},\ldots,Z_{N}\right)^{T}\in\left\{ \pm1\right\} ^{K}$
describes the state of $K$ logical spins. Then, let us consider a symmetric
$K\times K$ bipolar matrix $\boldsymbol{z}=\boldsymbol{Z}\otimes\boldsymbol{Z}$$\in\left\{ \pm1\right\} ^{K\times K}$, that is, 
$z_{ij}=Z_{i}Z_{j}$ $\left(1\leq i,j\leq K\right)$,
which describes the physical spins in the code states and satisfies the parity-check
equations $S_{ij}\left(\boldsymbol{z}\right)=z_{i\,j}z_{i\,j+1}z_{i+1\,j}z_{i+1\,j+1}=+1$
for $1\leq i<j\leq K-1$. Note that only $K-1$ physical variables
are logically independent because we can determine $\boldsymbol{Z}$
only up to the global spin flip. The matrix $\boldsymbol{z}$ contains
$N=\frac{K\left(K-1\right)}{2}$ independent variables. Therefore,
a set of physical spin variables in $\boldsymbol{z}$ constitutes
the $\left(\frac{K\left(K-1\right)}{2},K-1\right)$ block code. The matrix
$\boldsymbol{z}$ evolves according to the Hamiltonian $H^{phys}\left(\boldsymbol{z}\right)$.
We consider the general symmetric $K\times K$ bipolar matrix $\boldsymbol{r}\in\left\{\pm1\right\}^{K\times K}$
with unit diagonal elements $r_{ii}=1$ $\left(i=1,\ldots,N\right)$
to describe the readout of the physical spin state. The readout $\boldsymbol{r}$
may be the non-code state, i.e., $S_{i,j}\left(\boldsymbol{r}\right)=r_{i\,j}r_{i\,j+1}r_{i+1\,j}r_{i+1\,j+1}$$\neq+1$
for some $1\leq i<j\leq K$. In contrast, in QAC, the matrix $\boldsymbol{z}'\in\left\{ \pm1\right\} ^{N\times K}$
describes the physical spins that evolve according to the Hamiltonian
$H^{QAC}$. The matrices $\boldsymbol{z}$ in the PE scheme and $\boldsymbol{z}'$
in QAC differ in two manners. First, $\boldsymbol{z}$ is a symmetric
square matrix with unit diagonal elements, whereas $\boldsymbol{z}'$
is not. Second, the row and column of $\boldsymbol{z}'$ have
direct physical interpretations, that is, they correspond to the physical spin variables
in the same logical group and those in the same replica, respectively; however, the row
and column of $\boldsymbol{z}$ have no such direct physical interpretation. 

At present, the error-correcting capability of the PE scheme has not been
clarified specifically in terms of optimization application, 
although it was analyzed from the perspective of communication application
based on the analogy of the LDPC code \cite{Pastawski2016}.
We note that these two viewpoints need to be distinguished,
as explained previously. We consider the PE scheme from the viewpoint
of optimization application. We show that physical noise is
potentially useful in inferring the logical ground state $\boldsymbol{Z}_{g}$, 
not only for the QAC but also for the PE scheme below. 

Now, let us consider the one-step MVD for the PE scheme. Here, we consider
$K-2$ weight-three syndromes orthogonal on every error $e_{ij}$, 
\begin{eqnarray}
A_{ijk}\left(\boldsymbol{r}\right)&=&r_{ij}r_{jk}r_{ki}=e_{ij}e_{jk}e_{ki}\in\left\{ \pm1\right\}\nonumber \\
&&\;\left(\left\{ i,j,k\right\} _{c}=1,\ldots,K\right),
\end{eqnarray}
as suggested by Pastawski and Preskill \cite{Pastawski2016},
where $\left\{ i,j,k\right\} _{c}$ denotes one of 
$\dbinom{K}{3}$
combinations. Evidently, $A_{ijk}$ satisfies the requirements for
the orthogonal syndrome, as it involves the common element $e_{ij}$ and
the other element $e_{jk}e_{ki}$ that appears only once in the $K-2$
elements of $A_{ijk}$ for every $i<j$. We can easily confirm that
$A_{ijk}$ can be represented as a product of weight-four syndromes
$S_{ij}\left(\boldsymbol{r}\right)=r_{i\,j}r_{i\,j+1}r_{i+1\,j}r_{i+1\,j+1}$,
that is, a product of the readouts of the four physical spins associated
with the plaquettes in the PE spin network \cite{Rocchetto2016},
where $r_{jj}$ is fixed at $+1$ by construction. For example, one
can observed that $A_{247}=S_{24}S_{25}S_{26}S_{34}S_{35}S_{36}$ (see Fig.
\ref{fig:2}).

We can also identify $K-1$ weight-two estimators orthogonal
on $z_{ij}$: 
\begin{eqnarray}
B_{ijk}\left(\boldsymbol{r}\right)&=&r_{ij}A_{ijk}\left(\boldsymbol{r}\right)=r_{jk}r_{ki}=z_{ij}e_{jk}e_{ki}\in\left\{ \pm1\right\}\nonumber \\
&&\left(\left\{ i,j,k\right\} _{c}=1,\ldots,K\right).
\end{eqnarray}
It should be noted that $B_{iji}=B_{ijj}=r_{ij}$, which should be
a member of the $K-1$ elements of these syndromes. Then, if we simply
apply the one-step MVD, it follows that 
\begin{equation}
z_{ij}^{*}=\mathrm{maj}\left(B_{ij1},\ldots,B_{ijK}\right)=\mathrm{sgn}\left[\left\langle \boldsymbol{B}_{ij}\right\rangle \right],\label{eq:75}
\end{equation}
where $\boldsymbol{B}_{ij}=\left(B_{ij1},\ldots,B_{ijK}\right)\in\left\{ \pm1\right\} ^{K-1}$
corresponds to the estimators orthogonal on $z_{ij}$, and 
\begin{equation}
\left\langle \boldsymbol{B}_{ij}\right\rangle =\stackrel[k=1]{K}{\sum}r_{jk}r_{ki}-r_{ij}=\left(\boldsymbol{r}\left(\boldsymbol{r}-\boldsymbol{I}\right)\right)_{ij},\label{eq:76}
\end{equation}
where $\boldsymbol{I}$ is a $K$-dimensional identity matrix. Furthermore,
if $K\gg1$, Eq. (\ref{eq:76}) can be approximated to 
\begin{equation}
\left\langle \boldsymbol{B}_{ij}\right\rangle \sim\stackrel[k=1]{K}{\sum}r_{jk}r_{ki}=\left(\boldsymbol{r}^{2}\right)_{ij}.\label{eq:77}
\end{equation}
Comparisons of Eqs. (\ref{eq:71}), (\ref{eq:72}), and (\ref{eq:75})--(\ref{eq:77})
reveal a notable difference. Equation (\ref{eq:71}) indicates
that although the Bayes optimal estimate of the logical spin variable
$Z_{i}=z_{i1}$ is determined by the average of the readouts $\boldsymbol{r}_{i}$ for the $K$ physical spins in the associated logical group $i$ 
in QAC, that of the
physical spin variable $z_{ij}$ is determined by the nonlocal correlations of the readouts for two physical spin variables in the PE scheme.

These correlations are according to prior knowledge on
the parity constraints. However, to use this prior knowledge for majority
voting, we need to be careful regarding its fair use. It should
be remembered that our formula for one-step MVD in Eq. (\ref{eq:62})
has a theoretical background based on threshold decoding, where
the fair use of the prior knowledge on the parity constraints is
considered. In contrast, the previously reported one-step
MVD has no theoretical background and uses a majority vote over
different subsets of the products of physical spin variables that describe 
the spanning trees defined on the logical spins \cite{Albash16,Weidinger2023}, 
but not the orthogonal syndromes, for decoding. This spanning tree
method does not consider the fair use of the prior knowledge and
cannot be justified definitely; thus, its inference would be
unreliable as well as inefficient. It should also be noted that
Eq. (\ref{eq:75}) is justified only if each weight
for the estimator $B_{ijk}$ is independent of $k$, which cannot
be guaranteed in general. In the following section, we demonstrate
the one-step MVD for the PE scheme. 
We prove that one-step MVD is effective for the PE scheme through numerical simulation. We also propose
a novel method to improve the performance for sampling the physical
ground state $\boldsymbol{z}_{g}$. 

\section{Experiment and results of PE scheme\label{sec:4}}

\begin{figure}[h]
\begin{centering}
\includegraphics[viewport=300bp 140bp 660bp 460bp,clip,scale=0.68]{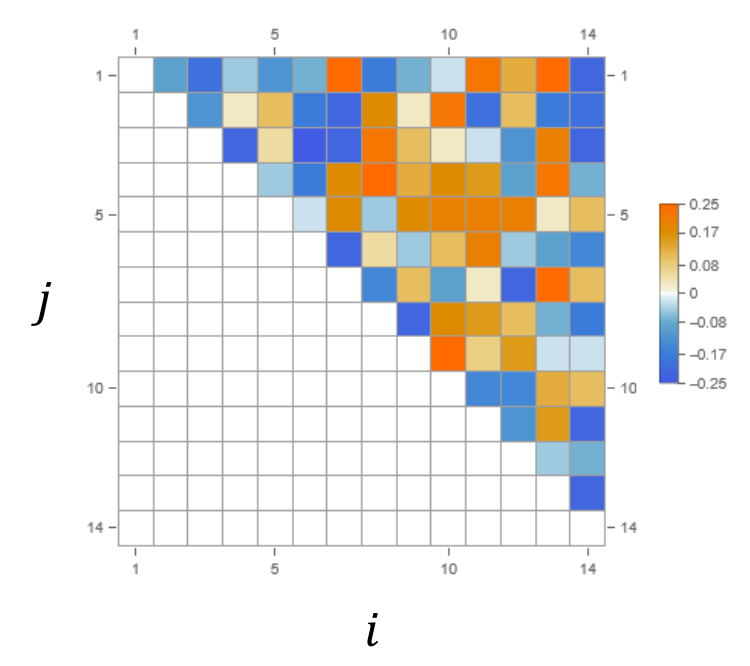}
\par\end{centering}
\caption{Spatial distribution of the elements of the assumed coupling matrix $\boldsymbol{J}$.\label{fig:7}}
\end{figure}

We performed a classical discrete-time Monte Carlo (DTMC) simulation  to verify the validity of one-step MVD for the PE scheme. 
We used the rejection-free (RF) version of the MCMC simulation \cite{Watanabe2006,Nambu22},
which is known as a jump or embedded-chain Monte Carlo \cite{Mazumdar2012,Parekh2020,Rosenthal2021}.
In this method, all diagonal elements in tht transition kernel
are eliminated. This results in the elimination of self-loop transitions, that
is, a state transitioning to itself, which is time consuming,
specifically in low-temperature simulation. Although the modified kernel
changes the stationary distribution from a Maxwell-Boltzmann (MB)
distribution to a non-MB distribution, the mapping between these
distributions is bijective so that we can always recover the MB distribution
from the stationary distribution obtained by the RF MCMC simulation.
The details of our RF MCMC algorithm are provided in the Appendix. We selected  
a spin-glass problem with $K=14$ in which the coupling matrix $\boldsymbol{J}$
is selected from random variables that are uniformly distributed in
the range $[-1/4,1/4]$ as an example. Figure \ref{fig:7}
shows the spatial distribution of the assumed coupling matrix $\boldsymbol{J}$, wherein the element $J_{ij}$ is the coupling coefficient between the spins $i$
and $j$. The simulation was performed using the Mathematica$^{\circledR}$ Ver. 13 
platform on Windows 10/11 operating systems. Please refer to Ref. \cite{Nambu22} 
for the details of our simulation. 

Because the PE scheme has currently been developed for QA
machines that are intended for solving combinatorial optimization problems,
we consider that the target state to be sampled is the ground state $\boldsymbol{Z}_{g}$
of the embedded logical spin systems described by the Hamiltonian
$H^{logi}\left(\boldsymbol{Z}\right)$ or, equivalently, the associated physical code state 
$\boldsymbol{z}_{g}=\boldsymbol{Z}_{g}\otimes\boldsymbol{Z}_{g}$.
We compared the performance for sampling these target states using
the decoding strategies (a) MAP decoding and (c) one-step MVD after MAP decoding.

For the MAP decoding, the RF MCMC attempts to sample the ground state $\boldsymbol{z}_{g}^{\prime}$ 
of the physical Hamiltonian $H^{phys}\left(\boldsymbol{z}\right)=-\beta H^{loc}\left(\boldsymbol{z}\right)-\gamma H^{pen}\left(\boldsymbol{z}\right)$
in Eqs. (\ref{eq:20})--(\ref{eq:22}) with $S_{ij}\left(\boldsymbol{z}\right)=S_{ij}^{4w}\left(\boldsymbol{z}\right)$.
Note that $\boldsymbol{z}_{g}^{\prime}$ depends on the set of weight parameters
$\left\{ \beta,\gamma\right\}$ and does not necessarily agree with $\boldsymbol{z}_{g}$; it can even be a non-code state. 
This is in contrast with QAC, where the ground
state of the Hamiltonians $H^{QAC}\left(\boldsymbol{z}\right)$ (see Eq. (\ref{eq:40}))) is necessarily a code state because it is the common ground state of $H^{enc}\left(\boldsymbol{z}\right)$ and $H^{pen}\left(\boldsymbol{z}\right)$. Therefore, the condition
$H^{pen}\left(\boldsymbol{z}\right)=0$ is crucial for sampling the
correct target state $\boldsymbol{z}_{g}$ for the PE scheme. To sample the
target state that is the lowest energy state of the local Hamiltonians
$H^{loc}\left(\boldsymbol{z}\right)$ in Eq. (\ref{eq:20}) under
the condition $H^{pen}\left(\boldsymbol{z}\right)=0$ using the Hamiltonian
$H^{phys}\left(\boldsymbol{z}\right)$, we implemented a bookkeeping
operation after one-step MVD. We calculated the logical energy $H^{logi}\left(\boldsymbol{Z}^{*}\right)$
for the logical state $\boldsymbol{Z}^{*}$ decoded from the readout $\boldsymbol{r}$ sweep by sweep and
retained only the readout with the lowest energy ever sampled in the
bookkeeping. Using this online bookkeeping operation, which can
be considered as an extra sampling or filtering process, we can
achieve successful sampling at the end of the simulation once the target
state is sampled during the annealing simulation. 

\begin{figure*}
\begin{centering}
\includegraphics[viewport=20bp 110bp 950bp 460bp,clip,scale=0.55]{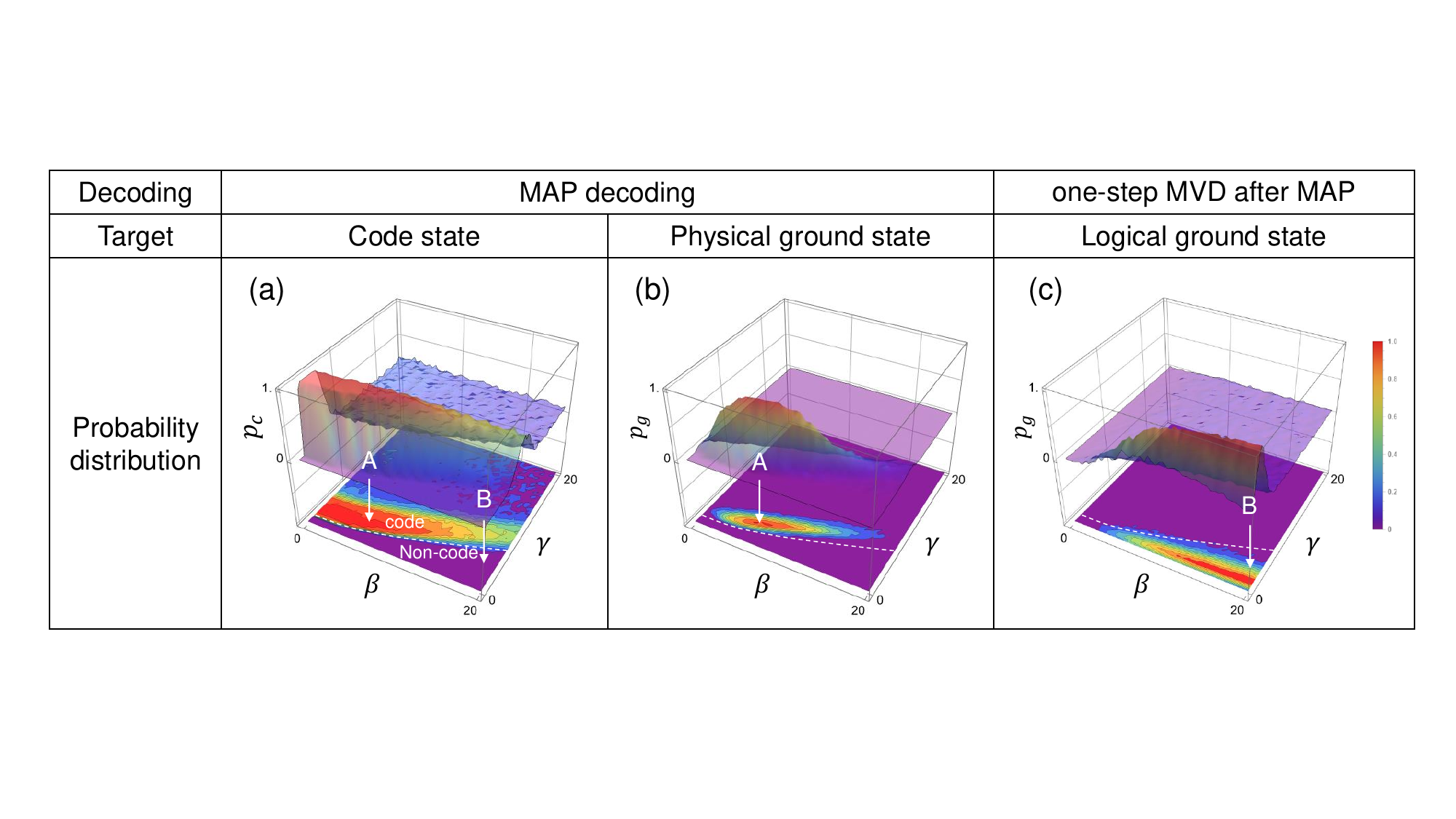}
\par\end{centering}
\caption{Landscape of the probability distribution. Subfigures (a) and (b)
are the results of the single-stage MAP decoding, whereas 
(c) is the result of one-step MVD for the readout of MAP decoding.
The target state of (a) is any code state, whereas $\boldsymbol{z}_{g}=\boldsymbol{Z}_{g}\otimes\boldsymbol{Z}_{g}$ 
and $\boldsymbol{Z}_{g}$ are the target states for (b) and
(c), respectively. The probability distributions for sampling these
target states are plotted as functions of the set of weight parameters
$\left\{ \beta,\gamma\right\} $.\label{fig:8}}
\end{figure*}

In our previous study, we investigated the performance of the PE scheme
based on only MAP decoding using the RF MCMC simulation \cite{Nambu22}.
We showed that the performance is largely dependent on the set of constant
weight parameters $\left\{ \beta,\gamma\right\} $ in the physical
Hamiltonian $H^{phys}\left(\boldsymbol{z}\right)$, which parameterizes
the coupling strength of the thermal environment and its relative weight
between two contributions $H^{loc}\left(\boldsymbol{z}\right)$ and
$H^{pen}\left(\boldsymbol{z}\right)$ in the Hamiltonian $H^{phys}\left(\boldsymbol{z}\right)$.
Figure \ref{fig:8} shows the typical landscapes of the various probability
distributions obtained by our simulation. Subfigures (a) and (b)
depict the results of the MAP decoding, whereas
(c) depicts the results of the one-step MVD inferred from 
the readout of the MAP decoding. In (a), the probability distribution
$p_{c}$ for sampling any code state that is the ground state of  $H^{pen}\left(\boldsymbol{z}\right)$, 
and in (b) and (c), the probability distribution $p_{g}$ for
sampling the target state $\boldsymbol{z}_{g}$ and $\boldsymbol{Z}_{g}$, respectively, are plotted
as functions of a set of weight parameters $\left\{ \beta,\gamma\right\}$.
These results were obtained from the statistical analysis of 300 identical
and independent simulations. We can observe from (a) that the
code state can be obtained efficiently within a confined region
on the $\left\{ \beta,\gamma\right\} $ plane. Specifically, we note
that $p_{c}$ is very small for a sufficiently large $\gamma$, which
might contradict the general belief that $p_{c}$ should be almost unity
for a sufficiently large $\gamma$. This might be a noteworthy
observation. In addition, we cannot sample the code states for the region 
where $\gamma$
is smaller than a certain limit, which is shown by the white broken line
in (a). It is reasonable that the target state $\boldsymbol{z}_{g}$
can be sampled only within the region on the $\left\{ \beta,\gamma\right\} $
plane where the code states are available in (a), as
the target state $\boldsymbol{z}_{g}$
must be one of the code states. Therefore, the probability distribution
in (b) is consistent with that in (a). However,
if we introduce the one-step MVD after the MAP decoding, the probability
distribution changes significantly. 

In this study, we re-investigated the performance of the PE scheme after introducing
one-step MVD for the readout sampled by the RF MCMC simulation, that is, the MP decoding. In
addition to the one-step MVD given by Eqs. (\ref{eq:75})--(\ref{eq:77}), 
we applied extra operations to deduce the logical state $\boldsymbol{Z^{*}}$ from the physicals state 
$\boldsymbol{z^{*}}$ inferred by one-step MVD,
as described later. Subfigure (c) in Fig. \ref{fig:8}
is a typical result showing the landscapes of the probability distribution
$p_{g}$ for inferling the target state $\boldsymbol{Z}_{g}$ successfully
using the one-step MVD as a function of a set of weight parameters $\left\{ \beta,\gamma\right\} $.
We found that the MC iterations required to achieve a comparable
success probability for infering the target state were nearly 400 times
smaller for one-step MVD after MAP decoding (subfigure (c)) than for MAP decoding
(subfigure (b)). Remarkably, such a target state could be found only
within the region $\left\{ \beta,\gamma\right\} $, where we could
not sample the code states using only MAP decoding. Note that
the difference among these simulations lies
in the post-readout one-step MVD as well as the bookkeeping operations
to keep track of the best outcome during the simulation. This result
suggests that the one-step MVD contributes sufficiently to the inference of the correct
target state and this contribution originates from the non-code state
sampled in the MAP decoding in the first stage. 

Unfortunately, we could observe only the best result 
sampled during the simulation 
owing to the bookkeeping operation, which discards the useless readouts during the RF MCMC simulation.
To understand the simulation, we performed a supplementary
simulation to clarify the function and potential of the one-step
MVD for the PE scheme. In the new simulation, we removed the bookkeeping
operations from the algorithm. Instead, we sequentially pushed all
sampled states during the RF MCMC simulation, that is, the MAP decoding, into the LIFO stack memory
and copied its contents into storage memory. We performed an offline
analysis for the stored series of readouts and investigated the function
and potential of the one-step MVD. We investigated two series of the
associated readouts when the set of weight parameters $\left\{ \beta,\gamma\right\} $
was fixed at $A$ (case A) in Fig. \ref{fig:8}, where both code and non-code
states could be sampled, and $B$ (case B) in Fig. \ref{fig:8}, where only
non-code states could be sampled by the MAP decoding. In the
following, we first show the results for case A, followed by those
for case B, and compare them. We analyzed $600K\left(K-1\right)-50$
samples with $K=14$, where the initial 50 samples in the burn-in periods
were discarded.

\begin{figure*}
\begin{centering}
\includegraphics[viewport=90bp 150bp 860bp 400bp,clip,scale=0.66]{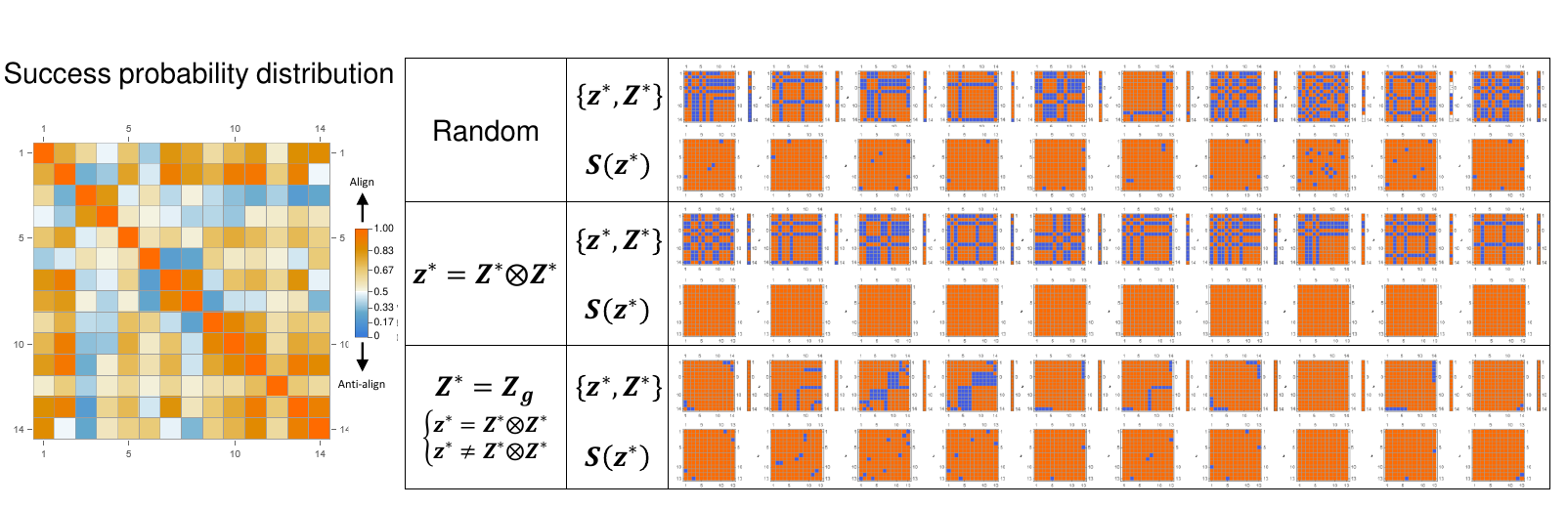}
\par\end{centering}
\caption{Result of one-step MVD for the readout  $\boldsymbol{r}$ of MAP decoding when
the weight parameters $\left\{ \beta,\gamma\right\}$ are fixed to
the value at which the code states can be frequently sampled. 
The left matrix plot shows the marginal probability distribution for every spin that demonstrated an agreement 
between the readout $\boldsymbol{r}$ and target state $\boldsymbol{z}_{g}$.
In the right panel, the error patterns $\left\{ \boldsymbol{e}(\boldsymbol{z}^{*}),\boldsymbol{E}(\boldsymbol{Z}^{*})\right\} $
for the optimal estimate $\left\{ \boldsymbol{z}^{*},\boldsymbol{Z}^{*}\right\} $
(upper results) and associated weight-four syndrome pattern $\boldsymbol{S}\left(\boldsymbol{z}^{*}\right)$
(lower results) are shown for three selected sets of 10 readouts, i.e., randomly selected 
readouts, for which the inferred physical state $\boldsymbol{z}^{*}$ agreed
with one of the code states $\boldsymbol{z}^{*}=\boldsymbol{Z}^{*}\otimes\boldsymbol{Z}^{*}$,
and for which the inferred logical state $\boldsymbol{Z}^{*}$ agreed
with the target state $\boldsymbol{Z}_{g}$. The blue element corresponds to an erroneous spin.\label{fig:9}}
\end{figure*}

Figure \ref{fig:9} shows the results for case A. The left panel shows the matrix plot
of the marginal probability distribution for every spin, demonstrating an agreement between the orientations in the readout $\boldsymbol{r}$ 
and target state $\boldsymbol{z}_{g}$, i.e., $r_{ij}=\left(z_{g}\right)_{ij}$.
The warm-colored and cold-colored elements indicate probabilities
higher and
lower than 1/2, respectively. Specifically, the orange and blue rectangles
indicate the perfect alignment and anti-alignment of the orientation of the spin
with that of the spin in the target state $\boldsymbol{z}_{g}$, respectively.
Recall that the diagonal elements are fixed to $+1$ by construction.
We can observe that the marginal probability distribution varies significantly
spin by spin, and some spins tend to anti-align their
orientations with the orientation of the spin in the target state $\boldsymbol{z}_{g}$.
This marginal probability distribution depends on the distribution
of the coupling constants as well as the values of the weight parameters
$\left\{ \beta,\gamma\right\} $. The panel on the right
shows 10 readouts in the stored series of the readouts. The upper
results are typical readouts randomly selected from the stored series,
whereas the middle and lower results are the readouts for which the
optimal estimate agrees with the code state, i.e., $\boldsymbol{z}^{*}=\boldsymbol{Z}^{*}\otimes\boldsymbol{Z}^{*}$, 
and the target state, i.e., $\boldsymbol{Z}^{*}=\boldsymbol{Z}_{g}$. In these figures, $\left\{ \boldsymbol{e}(\boldsymbol{z}^{*}),\boldsymbol{E}(\boldsymbol{Z}^{*})\right\} $
for the optimal estimates $\left\{ \boldsymbol{z}^{*},\boldsymbol{Z}^{*}\right\} $
and associated syndrome pattern $\boldsymbol{S}\left(\boldsymbol{z}^{*}\right)$
are plotted for a set of 10 selected readouts, where $\boldsymbol{e}(\boldsymbol{z})=\boldsymbol{z}\circ\boldsymbol{z}_{g}$
is an error pattern for the physical spin state $\boldsymbol{z}$, that is, $e_{ij}=z_{ij}\left(z_{g}\right)_{ij}$
$\left(i,j=1,\ldots,K\right)$, and $\boldsymbol{E}(\boldsymbol{Z})=\boldsymbol{Z}\circ\boldsymbol{Z}_{g}$
for the logical spin state $\boldsymbol{Z}$, that is, $E_{i}=Z{}_{i}\left(Z_{g}\right)_{i}$
$\left(i=1,\ldots,K\right)$. 
Note that the elements
in $\boldsymbol{e}(\boldsymbol{z})$ and $\boldsymbol{E}(\boldsymbol{Z})$ with a 
value of $-1$ imply that $z_{ij}\neq\left(z_{g}\right)_{ij}$ and $Z_{i}\neq\left(Z_{g}\right)_{i}$.
Thus, they can be regarded as errors associated with the physical state $\boldsymbol{z}$
and logical state $\boldsymbol{Z}$ relative to their target states
$\boldsymbol{z}_{g}$ and $\boldsymbol{Z}_{g}$, respectively.

\begin{figure*}
\centering{}\includegraphics[viewport=170bp 75bp 780bp 430bp,clip,scale=0.84]{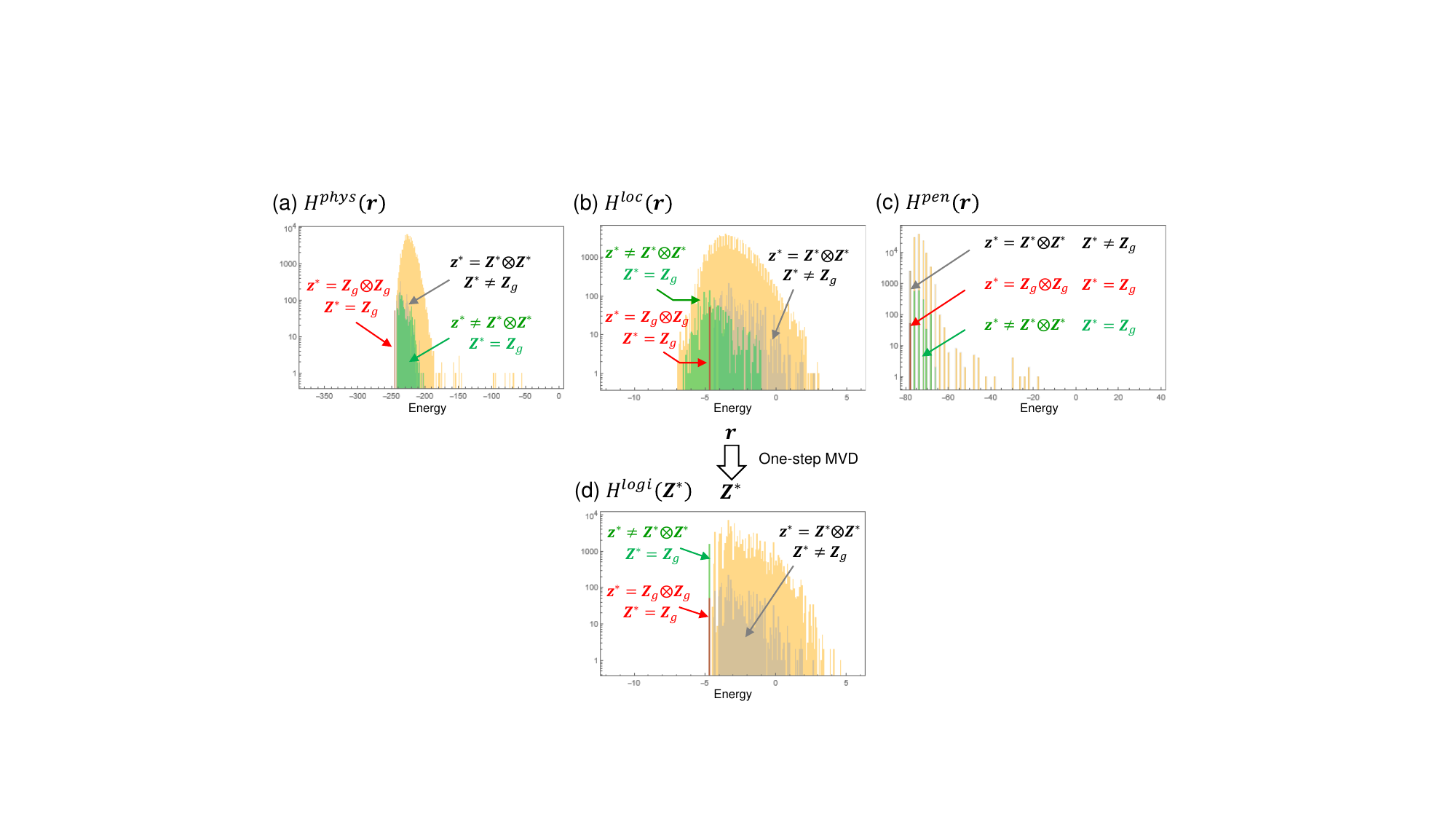}
\caption{Energy spectra of the sampled readouts $\boldsymbol{r}$ (upper histograms)
and estimate $\boldsymbol{Z}^{*}$ obtained from $\boldsymbol{r}$
(lower histogram). The upper histograms (a)-(c) indicate the energy
spectra associated with the Hamiltonian (a) $H^{phys}\left(\boldsymbol{r}\right)$,
(b) $H^{loc}\left(\boldsymbol{r}\right)$, and (c) $H^{pen}\left(\boldsymbol{r}\right)$.
The lower histogram (d) indicates the energy spectra associated with
the Hamiltonian $H^{logi}\left(\boldsymbol{Z}^{*}\right)$. The red, gray, and green bins show the contributions of the readouts for which 
$\boldsymbol{z}^{*}=\boldsymbol{z}_{g}$, $\boldsymbol{z}^{*}=\boldsymbol{Z}^{*}\otimes\boldsymbol{Z}^{*}$ but $\boldsymbol{Z}^{*}\neq\boldsymbol{Z}_{g}$, and 
$\boldsymbol{Z}^{*}=\boldsymbol{Z}_{g}$ but $\boldsymbol{z}^{*}\neq\boldsymbol{Z}^{*}\otimes\boldsymbol{Z}^{*}$, respectively.
\label{fig:10}}
\end{figure*}

To clarify the interim readout $\boldsymbol{r}$ before one-step MVD, we investigated
the energy spectra for the readouts shown in Fig. \ref{fig:9}.
Figure \ref{fig:10} depicts various energy spectra 
when $\left\{ \beta,\gamma\right\} $ is at point $A$. The upper
histograms indicate the spectra of the Hamiltonian (a) $H^{phys}\left(\boldsymbol{r}\right)$,
(b) $H^{loc}\left(\boldsymbol{r}\right)$, and (c) $H^{pen}\left(\boldsymbol{r}\right)$,
which are relative frequency distributions of the energy of the
readout $\boldsymbol{r}$ of the MAP decoding. It should be noted that $\boldsymbol{r}$ may not necessarily
be a code state. In these figures, the red bin indicates the contribution of  
the readouts for which the physical state $\boldsymbol{z}^{*}$ inferred by one-step MVD agreed with the target state, 
i.e., $\boldsymbol{z}^{*}=\boldsymbol{z}_{g}$.
In contrast, the gray bins indicate the contributions of the readouts for
which the inferred state $\boldsymbol{z}^{*}$ 
became the code state, i.e., $\boldsymbol{z}^{*}=\boldsymbol{Z}^{*}\otimes\boldsymbol{Z}^{*}$.
The green bins indicate the contributions of the readouts for which the inferred state $\boldsymbol{z}^{*}$ was not the code state, 
but the logical state $\boldsymbol{Z}^{*}$ deduced from $\boldsymbol{z}^{*}$ agreed with the target state $\boldsymbol{Z}_{g}$.
The lower histogram (d) indicates the spectrum of the Hamiltonian $H^{logi}\left(\boldsymbol{Z}^{*}\right)$ 
after the readouts $\boldsymbol{r}$ are decoded as the estimate $\boldsymbol{Z}^{*}$ by the one-step MVD.
It should be noted that if $\boldsymbol{z}$ is a code state, $\boldsymbol{z}=\boldsymbol{Z}\otimes\boldsymbol{Z}$, and $H^{loc}\left(\boldsymbol{z}\right)=H^{logi}\left(\boldsymbol{Z}\right)$ holds. 
The red, gray, and green bins in 
histogram (d) have the same meanings as those in the upper histograms.
We can observe that green bins in histogram (b)
are reduced to a single bin in histogram (d) after one-step MVD reflecting that 
decoding is a dimensionality-reducing operation. From histogram (d), we can observe that
a significant number of the target states $\boldsymbol{Z}_{g}$ is inferred by one-step MVD
from the non-code states even if $\left\{\beta,\gamma\right\}$
is at point $A$. From the comparison of gray bins in histograms
(b) and (d), we observe that the one-step MVD does not change the spectra of the code states.

\begin{figure*}
\begin{centering}
\includegraphics[viewport=90bp 160bp 860bp 370bp,clip,scale=0.66]{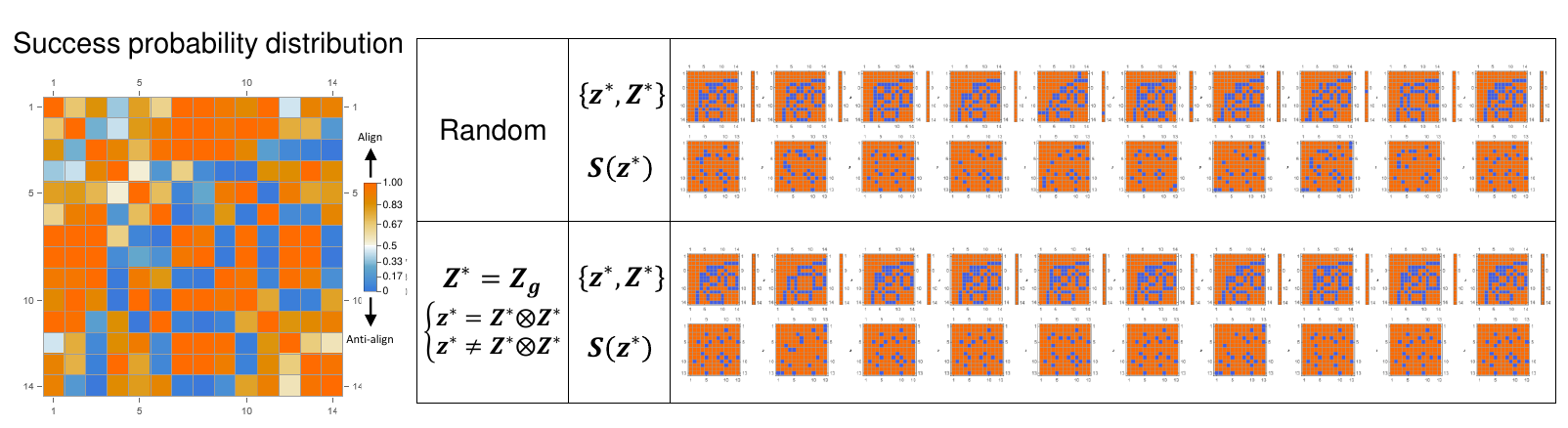}
\par\end{centering}
\caption{Results of one-step MVD for the readout of MAP decoding when
the weight parameters $\left\{ \beta,\gamma\right\} $ are fixed to
the value at which almost no code state can be sampled. 
The left matrix plot shows the marginal probability distribution for every spin exhibiting an agreement between the orientation in the readout $\boldsymbol{r}$ and the target state $z_{g}$.
In the right panel, the error patterns $\left\{ \boldsymbol{e}(\boldsymbol{z}^{*}),\boldsymbol{E}(\boldsymbol{Z}^{*})\right\} $
for the optimal estimate $\left\{ \boldsymbol{z}^{*},\boldsymbol{Z}^{*}\right\} $
(upper results) and associated weight-four syndrome pattern $\boldsymbol{S}\left(\boldsymbol{z}^{*}\right)$
(lower results) are shown for two sets of 10 readouts, i.e., for randomly
selected readouts and the readouts for which the inferred logical state $\boldsymbol{Z}^{*}$
agreed with the target state $\boldsymbol{Z}_{g}$.
\label{fig:11}}
\end{figure*}

Next, we discuss the results for case B in which $\left\{ \beta,\gamma\right\} $
is fixed at point $B$ in Fig. \ref{fig:11}. 
The matrix plot of the marginal probability distribution for every spin exhibiting an agreement between the orientations in the readout $\boldsymbol{r}$ and target state $\boldsymbol{z}_{g}$ is shown on the left.
We can observe that
some spins have strong tendencies to anti-align their orientations
with the orientation of the spin in the target state $z_{g}$, which is
similar to that when $\left\{ \beta,\gamma\right\} $ is fixed at
point $A$, although the distribution differs significantly. Specifically,
several spins have a very large probability of being anti-aligned,
with the value reaching nearly unity. The panels on the right-hand side
show 10 readouts in the stored series of the readouts. The upper
results are typical readouts that are randomly selected from the stored series,
whereas the lower results are the readouts for which the optimal estimate
agrees with the target state, i.e., $\boldsymbol{Z}^{*}=\boldsymbol{Z}_{g}$.

\begin{figure*}
\begin{centering}
\includegraphics[viewport=170bp 75bp 780bp 420bp,clip,scale=0.84]{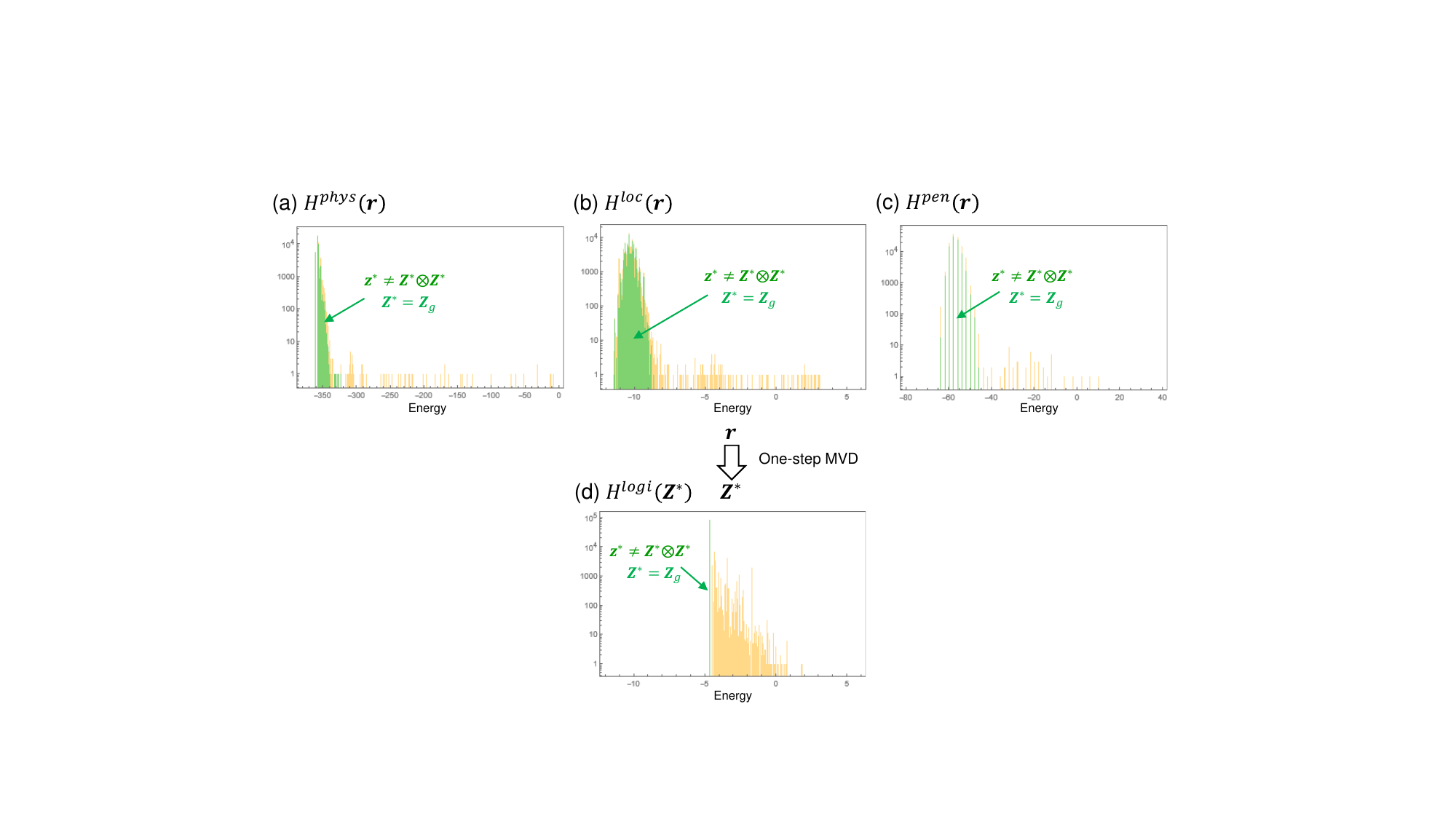}
\par\end{centering}
\caption{(a)-(c) Energy spectra of the sampled readouts $\boldsymbol{r}$ (upper histograms)
and (d) estimate $\boldsymbol{Z}^{*}$ obtained from $\boldsymbol{r}$
(lower histogram).\label{fig:12}}
\end{figure*}

Figure \ref{fig:12} presents the various energy spectra of the readouts
when $\left\{ \beta,\gamma\right\} $ is at point $B$. The upper
histograms indicate the spectra of the Hamiltonian (a) $H^{phys}\left(\boldsymbol{r}\right)$,
(b) $H^{loc}\left(\boldsymbol{r}\right)$, and (c) $H^{pen}\left(\boldsymbol{r}\right)$
for the readout $\boldsymbol{r}$, whereas the lower histogram
shows the spectrum of the Hamiltonian (d) $H^{logi}\left(\boldsymbol{Z}^{*}\right)$
for the estimate $\boldsymbol{Z}^{*}$. In these figures, the colored
bins have the same meaning as those in Fig. \ref{fig:10}. It can be observed 
that there is no red bin, which indicates that no target state $\boldsymbol{z}_{g}$
can be sampled by the MAP decoding. Nevertheless, a single and significantly
large green bin can be observed in the ground state $\boldsymbol{Z}^{*}=\boldsymbol{Z}_{g}$
of the Hamiltonian (d) $H^{logi}\left(\boldsymbol{Z}^{*}\right)$.
It is evident from the spectrum of $H^{pen}\left(\boldsymbol{r}\right)$
that the estimate $\boldsymbol{Z}^{*}$
originates from the readouts in the non-code state space. This proves
that our one-step MVD is suitable for decoding the target state $\boldsymbol{Z}_{g}$
from the readouts $\boldsymbol{r}$, that is, the non-code states sampled
by the MAP decoding in the first stage. 

\begin{figure*}
\begin{centering}
\includegraphics[viewport=160bp 120bp 730bp 440bp,clip,scale=0.72]{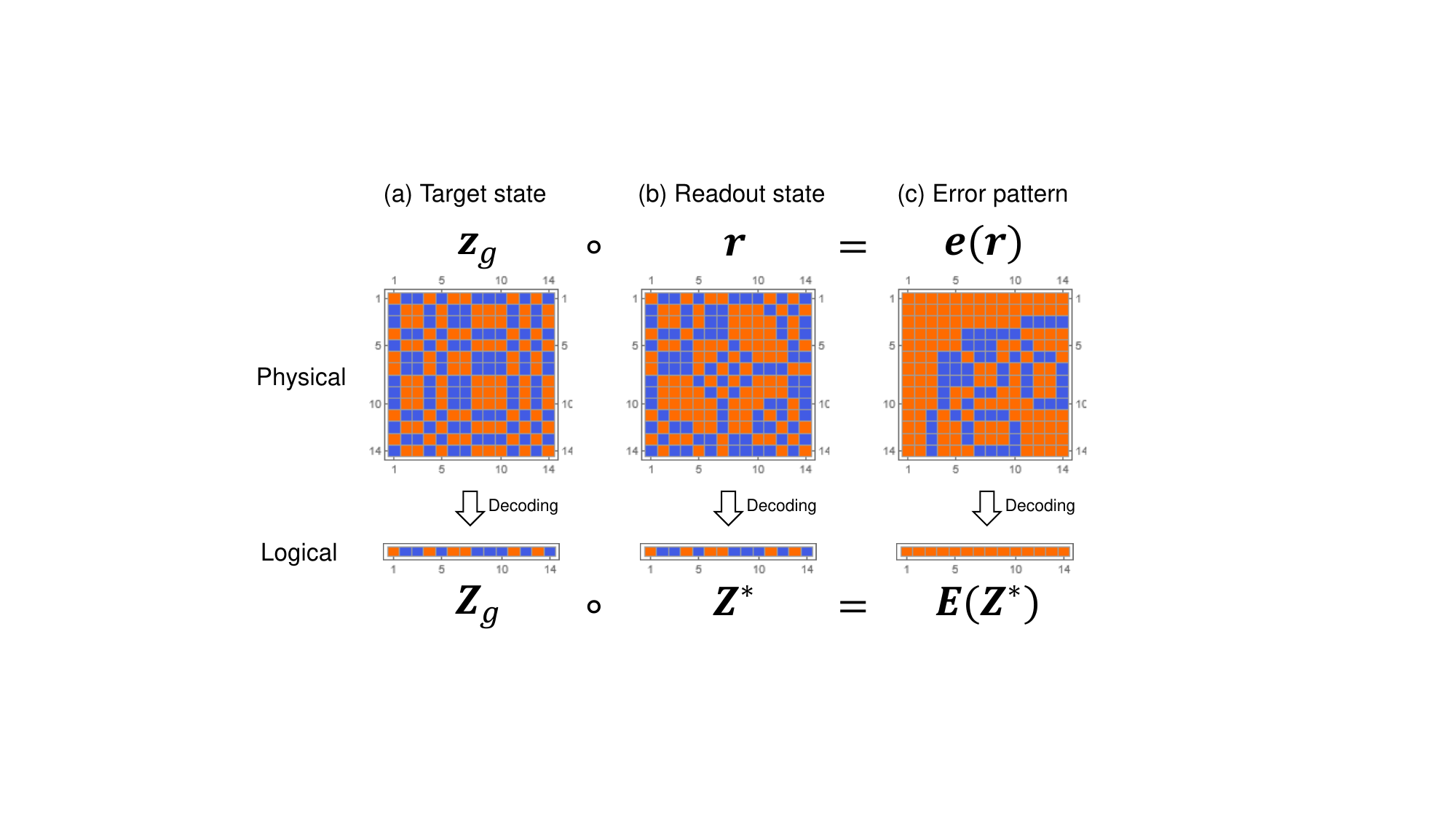}
\par\end{centering}
\caption{An example of the results of our strategy to deduce $\boldsymbol{Z}^{*}$.\label{fig:13}}
\end{figure*}

Unfortunately, we discovered later that our decoding algorithm had
a flaw, although it was subsequently solved. The issue was related to the deduction of
the estimator $\boldsymbol{Z}^{*}\in\left\{ \pm1\right\} ^{K}$ from
the optimal estimator $\boldsymbol{z}^{*}\in\left\{ \pm1\right\} ^{K\times K}$
obtained through one-step MVD. If $\boldsymbol{z}^{*}$ is the code state,
we obtain $\boldsymbol{z}^{*}=\left(\boldsymbol{z}^{*}\right)^{T}=\boldsymbol{Z}^{*}\otimes\boldsymbol{Z}^{*}$.
In this case, we can easily deduce $\boldsymbol{Z}^{*}$ up to a
global sign as 
\begin{equation}
\boldsymbol{Z}^{*}=\boldsymbol{z}_{i}^{*}=\left(z_{i\,1}^{*},\ldots,z_{i\,K}^{*}\right)\label{eq:78}
\end{equation}
for an arbitrary $i=1,\ldots,K$. However, if $\boldsymbol{z}^{*}$ is
a non-code state, $\boldsymbol{z}^{*}\neq\boldsymbol{Z}^{*}\otimes\boldsymbol{Z}^{*}$, 
and Eq. (\ref{eq:78}) does not necessarily hold. In this
case, the derivation of $\boldsymbol{Z}^{*}$ is nontrivial, although
this is resolved later. We attempted to deduce the unique $\boldsymbol{Z}^{*}$
in an ad-hoc manner, as follows. In our decoding algorithm, we first 
estimated the physical state using one-step MVD according to $\boldsymbol{z}^{*}=\mathrm{sgn}\left[\mathcal{F}\left(\boldsymbol{r}\right)\right]$,
where $\mathcal{F}\left(\boldsymbol{r}\right)=\boldsymbol{r}\left(\boldsymbol{r}-\boldsymbol{I}\right)\sim\boldsymbol{r}^{2}$.
Subsequently, we performed an extra majority vote for $\boldsymbol{z}^{*}$ by assuming that 
$\mathrm{sgn}\left[\left(\boldsymbol{z}^{*}\boldsymbol{z}^{*}\right)_{1i}\right]=\mathrm{sgn}\left[\stackrel[k=1]{K}{\sum}z_{1k}^{*}z_{ki}^{*}\right]$
gives a good estimate of $Z^{*}_{1}Z^{*}_{i}$, where $z_{1k}^{*}$ is
considered to eliminate a global sign flip of $\boldsymbol{Z}^{*}$.
Note that, if $\boldsymbol{z}^{*}=\boldsymbol{Z}^{*}\otimes\boldsymbol{Z}^{*}$ holds, we have 
$\boldsymbol{z}_{ij}^{*}=Z_{i}^{*}Z_{j}^{*}$ and $Z_{i}^{*}=Z_{1}^{*}\mathrm{sgn}\left[\left(\boldsymbol{z}^{*}\boldsymbol{z}^{*}\right)_{1i}\right]$, but this ad-hoc assumption is not justifiable theoretically.
All of the above results were obtained using this strategy.
However, we found later that this was not an appropriate approach. To comprehend the related problem, let us consider an example of the
readout $\boldsymbol{r}$ shown in Fig. \ref{fig:13} that was inferred
as the target state $\boldsymbol{Z}_{g}$ in our strategy.
The plots in (a) depict the target states: the upper and lower
traces are the target physical state $\boldsymbol{z}_{g}=\boldsymbol{Z}_{g}$$\otimes\boldsymbol{Z}_{g}$
and logical ground state $\boldsymbol{Z}_{g}$, respectively.
The plots in (b) depict the readout state $\boldsymbol{r}$
(upper) and the state $\boldsymbol{Z}^{*}$ decoded through our strategy
(lower), which is the target state $\boldsymbol{Z}_{g}$. The plots in (c) depict the error patterns $\boldsymbol{e}(\boldsymbol{r})=\boldsymbol{r}\circ\boldsymbol{z}_{g}$
for the readout $\boldsymbol{r}$ and $\boldsymbol{E}(\boldsymbol{Z}^{*})=\boldsymbol{Z}^{*}\circ\boldsymbol{Z}_{g}$
for the estimator $\boldsymbol{Z}^{*}$. The error patterns $\boldsymbol{e}(\boldsymbol{r})$
and $\boldsymbol{E}(\boldsymbol{Z}^{*})$ have elements with values of 
$-1$, i.e., erroneous spins.

\begin{figure*}
\begin{centering}
\includegraphics[viewport=140bp 150bp 860bp 410bp,clip,scale=0.65]{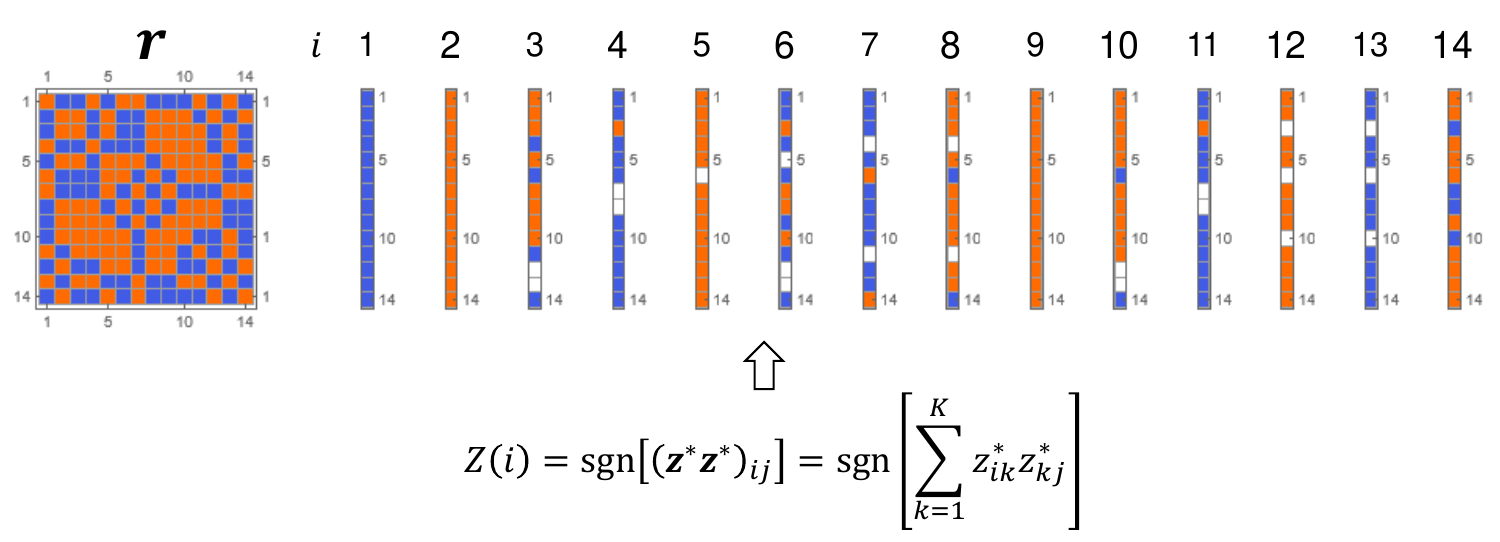}
\caption{Evidence of a flaw in our strategy. The white elements denote an undefined
state associated with the tie vote.\label{fig:14}}
\par\end{centering}
\end{figure*}

We can transform the Hamiltonian $H^{phys}\left(\boldsymbol{z}\right)$
into $H^{phys}\left(\boldsymbol{e}\right)$  through the gauge transformations
$J_{ij}\rightarrow J_{ij}\left(z_{g}\right)_{ij}$ and $z_{ij}\rightarrow e_{ij}$
if we know $\boldsymbol{z}_{g}$ a priori. The Hamiltonian $H^{phys}\left(\boldsymbol{e}\right)$
has a trivial ground state $\boldsymbol{e}_{g}$ with $\left(e_{g}\right)_{ij}=+1$
for all $i$ and $j$. We can also transform the Hamiltonian $H^{logi}\left(\boldsymbol{Z}\right)$
into $H^{logi}\left(\boldsymbol{E}\right)$ using the gauge transformations
$J_{ij}\rightarrow J_{ij}\left(Z_{g}\right)_{i}\left(Z_{g}\right)_{j}$
and $Z_{i}\rightarrow E_{i}$ if we know $\boldsymbol{Z}_{g}$ a priori. The
Hamiltonian $H^{logi}\left(\boldsymbol{E}\right)$ has a trivial ground
state $\boldsymbol{E}_{g}$ with $\left(E_{g}\right)_{i}=+1$ for
all $i$. The upper and lower results in (c) depict
the error patterns $\boldsymbol{e}(\boldsymbol{r})$ and $\boldsymbol{E}(\boldsymbol{Z}^{*})$ for the
readout state $\boldsymbol{r}$ shown in (b). We can observe 
that the error pattern $\boldsymbol{e}(\boldsymbol{r})$ reflects the marginal probability
distribution shown in Fig. \ref{fig:11}, although the error pattern
$\boldsymbol{E}(\boldsymbol{Z}^{*})$ is its ground state $\boldsymbol{E}_{g}$.
It should be noted that the error pattern $\boldsymbol{e}(\boldsymbol{r})$ has the elements
$(\boldsymbol{e}(\boldsymbol{r}))_{i1}=(\boldsymbol{e}(\boldsymbol{r}))_{1i}=+1$ and $(\boldsymbol{e}(\boldsymbol{r}))_{i2}=(\boldsymbol{e}(\boldsymbol{r}))_{2i}=+1$ for $i=1,\ldots,N$, which
implies that the readouts $r_{ij}=r_{ji}$ $\left(j=1\textrm{ or }2\right)$ are
sufficient to infer the target state $\boldsymbol{Z}_{g}$ up to its global
sign. These error-free elements with respect to the target logical
state $\boldsymbol{Z}_{g}$ may be the reason that our strategy for inferring the ground state $\boldsymbol{Z}_{g}$ succeeded.
Figure \ref{fig:14} depicts the calculated results of $Z\left(i\right)=\mathrm{sgn}\left[\left(\boldsymbol{z}^{*}\boldsymbol{z}^{*}\right)_{ij}\right]$
for $i=1,\ldots,N$ using the readout $\boldsymbol{r}$ shown
in Fig. \ref{fig:13} (b). Clearly, the target state $\boldsymbol{Z}_{g}$ can be
inferred up to its global sign only if we select $i=1$ and $2$.
This suggests that the crucial reason for the success of our strategy
is the subset of the error-free readouts $\boldsymbol{r}_{i}=\left(r_{i1},\ldots,r_{iK}\right)$
or $\tilde{\boldsymbol{r}}_{i}=\left(r_{1i},\ldots,r_{Ki}\right)^{T}$
associated with the logical lines for $i=1$ and $2$ of the readout state
$\boldsymbol{r}$. That is, the simulation results presented
thus far depended on these readouts.

\begin{figure*}
\centering{}
\includegraphics[viewport=140bp 170bp 830bp 400bp,clip,scale=0.72]{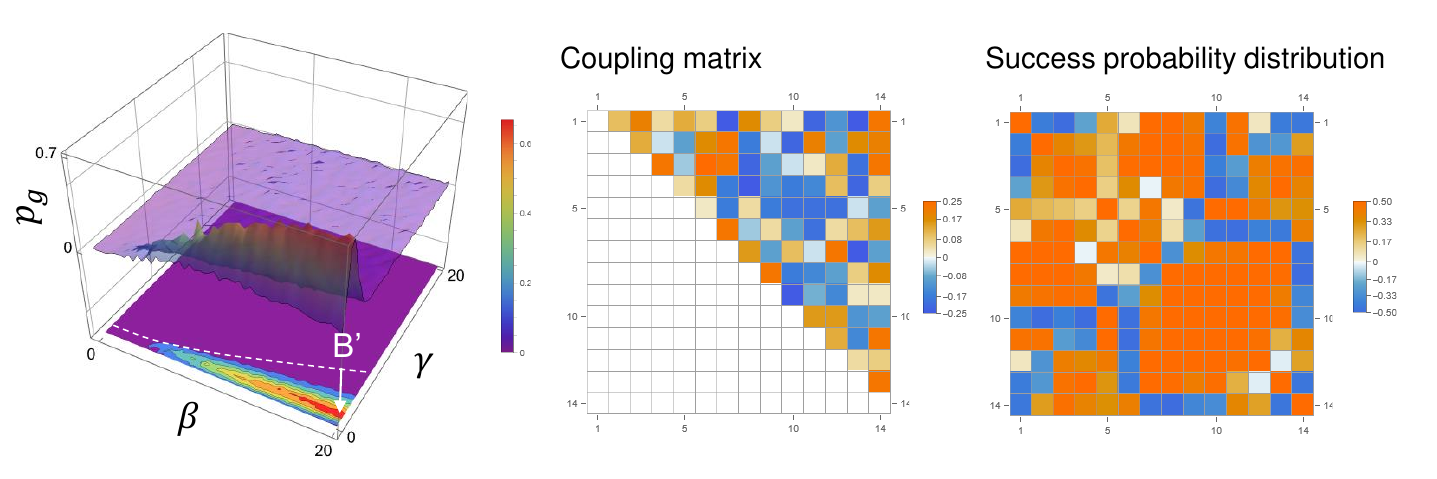}
\caption{Another problem instance demonstrating the limitation of our strategy. The success
probability for sampling the target state $\boldsymbol{Z}_{g}$ using our strategy, 
coupling matrix $\boldsymbol{J}$, and marginal probability distribution for 
every spin exhibiting an agreement between the orientation in the readout $\boldsymbol{r}$ and the target state $\boldsymbol{z}_{g}$. 
\label{fig:15}}
\end{figure*}

Our strategy is not sufficient to infer the target state for general problem instances. To demonstrate this, we
show the result for another problem
instance with the inverted coupling constants $\boldsymbol{J}\rightarrow-\boldsymbol{J}$.
The left plot in Fig. \ref{fig:15} indicates the landscape of the
probability distribution $p_{g}$ for successfully infering the target logical
state $\boldsymbol{Z}_{g}$ using our strategy, and the
right plot indicates the marginal probability distribution for 
successfully sampling the target physical state $\boldsymbol{z}_{g}=\boldsymbol{Z}_{g}$$\otimes\boldsymbol{Z}_{g}$
when the weight parameters $\left\{ \beta,\gamma\right\} $ are fixed
at point $B'$ in the left plot. In contrast to the previous result
shown in Figs. \ref{fig:7}, \ref{fig:8}, and \ref{fig:11}, we could
not find a logical line where the success probability for all involved
elements was no less than 1/2. This suggests that our strategy has 
limited validity and its success depends on the problem instances. In fact,
the success probability for sampling the target state was smaller
for this instance than in the instance shown in Fig. \ref{fig:8} (c), even though the number of readout samples was the same.

\begin{figure*}
\begin{centering}
\includegraphics[viewport=50bp 20bp 920bp 520bp,clip,scale=0.56]{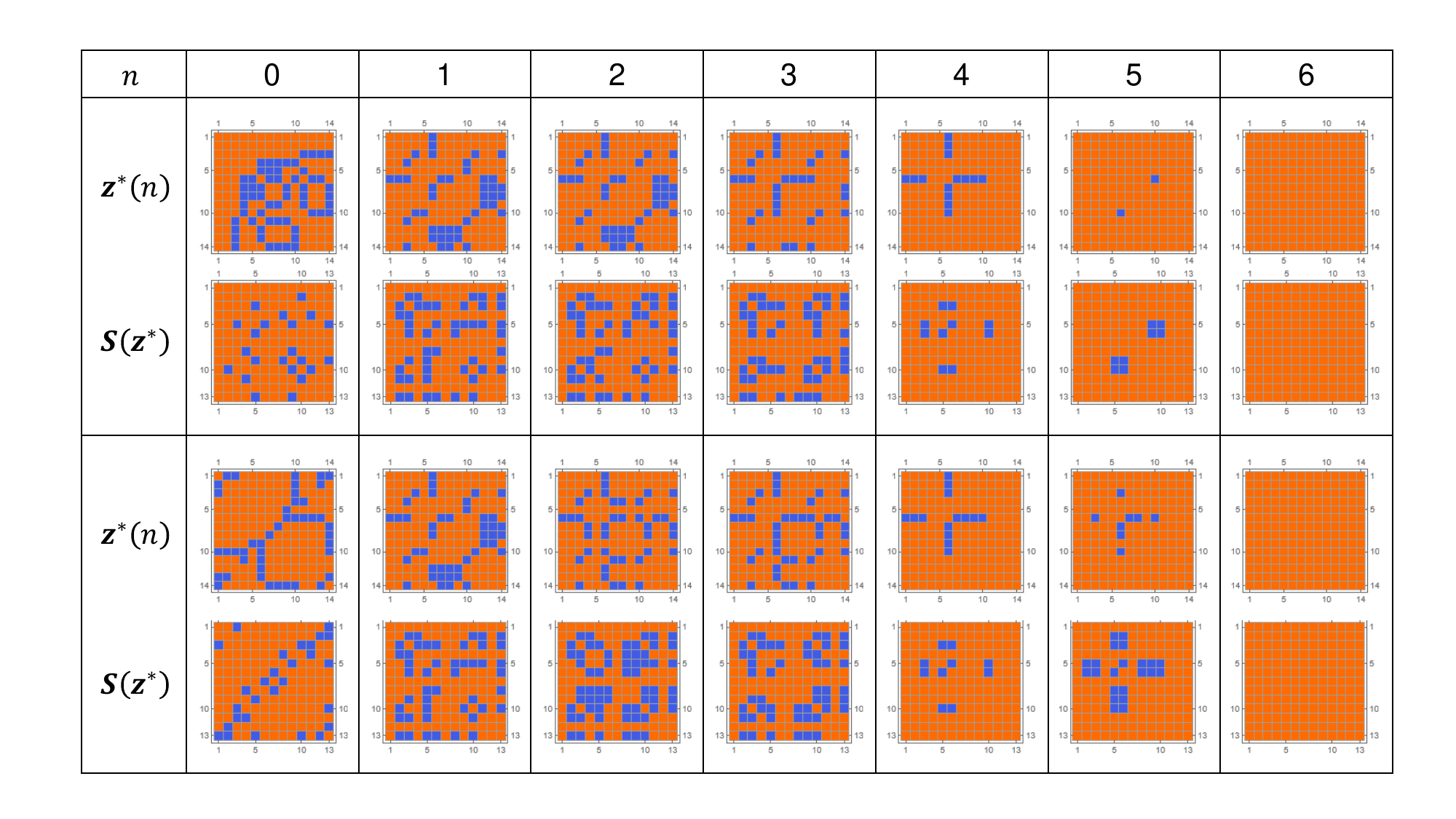}
\par\end{centering}
\caption{Matrix plots for visualizing the error patterns of $\boldsymbol{z}^{*}\left(n\right)$ 
and weight-four syndrome patterns $\boldsymbol{S}\left(\boldsymbol{z}^{*}\left(n\right)\right)$
with the element $S_{ij}^{4w}\left(\boldsymbol{z}^{*}\left(n\right)\right)$
for two readouts $\boldsymbol{r}$. Note that $\boldsymbol{z}^{*}\left(n=0\right)=\boldsymbol{r}$. \label{fig:16}}
\end{figure*}

Eventually, we concluded that it is important and necessary to develop
a decoding method that can infer the target state for any problem instance without
its detailed knowledge. For this purpose, we developed a decoding
method that can infer the physical target state $\boldsymbol{z}_{g}=\boldsymbol{Z}_{g}\otimes\boldsymbol{Z}_{g}$
without errors. This is a novel post-readout
decoding procedure that is applicable to an arbitrary problem instance. We show that this is actually possible
by extending our one-step MVD algorithm to the incorporation of majority
voting of greater than weight-two orthogonal estimators. Let us consider the
following matrix function: 
\begin{equation}
\mathcal{F}\left(\boldsymbol{X}\right)\coloneqq\boldsymbol{X}\left(\boldsymbol{X}-\boldsymbol{I}\right),
\end{equation}
where $\boldsymbol{X}$ is a symmetric square bipolar matrix with
unit diagonal elements and $\boldsymbol{I}$ is the identity matrix
in the matrix space of $\boldsymbol{X}$. Then, the standard one-step
MVD is based on up to weight-two orthogonal estimators
\begin{equation}
\boldsymbol{z}^{*}=\mathrm{sgn}\left[\mathcal{F}\left(\boldsymbol{r}\right)\right].
\end{equation}
Next, we extend this to include up to weight-$2n$ orthogonal estimators
\begin{equation}
z^{*}\left(n\right)=\textrm{sgn}\left[\mathcal{F}^{(n)}\left(\boldsymbol{r}\right)\right],
\label{eq:81}
\end{equation}
where 
\begin{equation}
\mathcal{F}^{(n)}\left(\boldsymbol{X}\right)\coloneqq\mathcal{I}\circ\underset{n}{\underbrace{\mathcal{F}\circ\mathcal{F}\circ\cdots\circ\mathcal{F}}}\left(\boldsymbol{X}\right)
\end{equation}
is an $n$-iteration of $\mathcal{F}$ on $\boldsymbol{X}$. Note that
$\mathcal{F}^{(n)}\left(\boldsymbol{r}\right)$ consists of the
sum of the terms involving up to $2n$-multiplication of matrix elements
$r_{ij}$, where every $r_{ij}$ is treated symmetrically and on an
equal footing, that is, exchanging two arbitrary $r_{ij}$ and $r_{i'j'}$ does
not change $\mathcal{F}^{(n)}\left(\boldsymbol{r}\right)$. In addition, 
the matrix element $\left(\mathcal{F}^{(n)}\left(\boldsymbol{r}\right)\right)_{ij}$
is considered to consist of all possible weight-$2m$ $\left(m=1,\ldots,n\right)$
estimators orthogonal on $z_{ij}$, which would ensure fair majority
voting. It has been pointed out that majority voting involving such 
high-weight syndromes may improve the performance at the expense 
of increasing the calculation cost \cite{Pastawski2016}. We compared the estimator obtained from Eq. (\ref{eq:81}) for an increasing
number of iterations $n$. Figure \ref{fig:16} depicts the matrix
plots of the error pattern of $\boldsymbol{z}^{*}\left(n\right)$
for the two readouts $\boldsymbol{r}$ shown in Figs. \ref{fig:11}
and \ref{fig:15}. It is clear that the errors in $\boldsymbol{z}^{*}\left(n\right)$
are reduced as $n$ increases and reaches the target state
$\boldsymbol{z}_{g}$. We investigated numerous examples to
check the trend of $\boldsymbol{z}^{*}\left(n\right)$ against an 
increasing $n$ and confirmed similar behaviors for all the examples. Therefore,
we concluded that Eq. (\ref{eq:81}) is a practical and effective estimator
for the physical target state $\boldsymbol{z}_{g}$ for a large $n$. By combining this technique
with the energy calculation, we expect that this technique can assist
in sampling the physical target states $\boldsymbol{z}_{g}$
using various sampling techniques. For example, we can consider two scenarios, as follows. We calculate the physical energy $H^{phys}\left(\boldsymbol{z}^{*}\left(n\right)\right)$
for the estimated state $\boldsymbol{z}^{*}\left(n\right)$ for a given
$n$. We can keep track of the lowest possible value of $H^{phys}\left(\boldsymbol{z}^{*}\left(n\right)\right)$
and the associated $\boldsymbol{z}^{*}\left(n\right)$
during the sampling process in bookkeeping, and output the best physical
state $\boldsymbol{z}^{*}\left(n\right)=\underset{n}{\min}H^{phys}\left(\boldsymbol{z}^{*}\left(n\right)\right)$
at the end of the simulation. Alternatively, we can calculate the logical
energies $H^{logi}\left(\boldsymbol{z}_{i}^{*}\left(n\right)\right)$
of the estimator $\boldsymbol{z}_{i}^{*}\left(n\right)$ for a given
$n$ and $i=1,\ldots,N,$ where $\boldsymbol{z}_{i}^{*}\left(n\right)=\left(z_{i1}^{*}\left(n\right),\ldots,z_{iN}^{*}\left(n\right)\right)$
is the logical line $i$ (the $i$th row vector) of $\boldsymbol{z}^{*}\left(n\right)$.
We can track the lowest possible value of $H^{logi}\left(\boldsymbol{z}_{i}^{*}\left(n\right)\right)$
and the associated $\boldsymbol{z}_{i}^{*}\left(n\right)$
during the sampling process and output the best logical state $z_{i}^{*}\left(n\right)=\underset{n}{\min}H^{logi}\left(z_{i}^{*}\left(n\right)\right)$
at the end of the simulation. Obviously, both methods consume some classical
computational resources, which should be a polynomial of the size of
the problem $N$ if we can fix $n$. However, it is plausible that
$n$ depends on $N$. In this case, we need to introduce an appropriate
cutoff for $n$ to cope with the problems associated with the performance
tradeoff. Further analysis is required to clarify the validity
of the present method and optimize it for the PE scheme.

Both MAP decoding and one-step MVD after MAP decoding
may infer the correct target state. However, their working principles
are very different. MAP decoding attempts to sample the target state
$\boldsymbol{z}_{g}$ directly by statistically sampling the state
that minimizes the physical Hamiltonian $H^{phys}\left(\boldsymbol{z}\right)$, 
by which the state should satisfy the parity constraints. In contrast,
one-step MVD after MAP decoding samples a
neighboring state of the target state $\boldsymbol{r}\sim\boldsymbol{z}_{g}$, regardless of
whether it violates the parity constraints in the first-stage MAP
decoding, and converts it into the target state $\boldsymbol{z}_{g}$ deterministically 
using the information on syndrome patterns without supplemental information in
the second-stage one-step MVD. Thus, MAP decoding can be regarded
as a purely stochastic algorithm, whereas one-step MVD after the MAP
decoding can be regarded as a hybrid of stochastic and deterministic algorithms.
Our two-stage decoding method for the PE scheme can
be regarded as similar to the proposal of Sourlas for soft annealing, wherein 
computationally difficult problems can be simplified by embedding
them into those in the larger system because one can circumvent 
the barriers using the deterministic algorithm and accelerate the dynamics
of the algorithm while retaining all essential properties of
the problem \cite{Sourlas2005}. Similarly, finite temperature decoding
of QAC may be regarded as an alternate version of soft annealing based
on the classical repetition codes \cite{Matsuura2016,Matsuura2017}.

We should comment on an inconsistency between the previously reported
result and the current result. We concluded that one-step MVD
is very useful in inferring the correct target state from the erroneous
but correctable states that may arise in the PE scheme. Albash et al. reported a contradictory conclusion in their study
\cite{Albash16}, in whch the validity of one-step MVD on the PE scheme
was investigated based on classical Monte Carlo simulations. They
reported that no clear evidence for the validity
of the one-step MVD was observed. They claimed that majority voting
schemes that attempt to exploit a large number of decoding trees are
not beneficial because the error generated during the simulations is not
an uncorrelated random error. Their claim appears to be in conflict with ours,
as we confirmed that our novel one-step MVD using iterative matrix
multiplications is valid for a wide range of error patterns in the readout
$\boldsymbol{r}$, provided that the maximum number of errors $N_{err}$
in every row (column) of $\boldsymbol{r}$ satisfies $N_{err}<\frac{K-1}{2}$.
Note that a row (column) of $\boldsymbol{r}$, i.e., its logical line,
is a star-like spanning tree with $K-1$ variable elements
(recall that all diagonal elements of $\boldsymbol{r}$ are fixed
to $+1$ by assumption). We point out two possible reasons for the
inconsistency between the two studies. First, the estimator used for inferring
the target state differs between the two studies. In our study, we used
the estimator derived from orthogonal syndromes, which would ensure
fair majority voting. In contrast, the previous study used a majority
vote over the subsets of the readouts $\boldsymbol{r}$, describing
the spanning trees on the logical spins to determine the target state
$\boldsymbol{Z}_{g}$ up to a global spin flip fully. However, this approach lacks
theoretical grounding and may be questionable from the viewpoint of
the fairness of the majority vote. Second, the previous study focused on simulations
with specific simulation parameters and an annealing schedule selected 
based on values associated with actual QA devices.
The authors did not necessarily preclude the validity
of post-readout decoding of the PE scheme for any simulation parameter
or schedule. In our opinion, the first reason is rather important.
The majority vote based on the random spanning tree method does not
consider the fair use of the prior knowledge on the parity constraints and cannot be justified.

\section{Conclusions\label{sec:5}}

We have studied two classical error-correcting schemes 
for solving combinatorial optimization problems based on the dynamical
relaxation process of Ising spins towards their ground states. First,
we reviewed the concept of soft annealing introduced by Sourlas
and examined how an Ising spin Hamiltonian is embedded into an extended Hamiltonian
with classical error-correction codes and the tacit assumptions
made therein. It was clarified that the informative prior for
parity constraints is the source of the penalty Hamiltonian, which was 
introduced ad-hoc into the Hamiltonian in many previous studies.

Subsequently, we addressed two classical error-correction schemes used for QA, 
namely the QAC and PE schemes, and investigated their derivation from the
above concept. In QAC, the classical repetition code is merged
to the logical Hamiltonian and encoded into the Hamiltonian describing
a larger system composed of many copied systems (replicas). These replicas
are mutually independent if the penalty Hamiltonian is omitted; otherwise,
they are mutually correlated. It is obvious that if the copies are
independent, the QAC executes sampling and measurement for independent
replicas in parallel in a single-shot operation, instead of repeating
the same operation on a single copy. The results should be consistent
with the measurements of repeated experiments using a single copy. If
the penalty term is introduced, it should produce a positive correlation
among the replicas, which may be utilized to improve the sampling efficiency. This is very similar to extended ensemble
algorithms, such as replica exchange and parallel tempering,
in classical statistical physics \cite{IBA2001} as well as simulated
QA using the path-integral Monte Carlo method \cite{Marto?ak2002,Heim2015,Waidyasooriya2020,Hu2021},
where the dynamics of an ensemble comprising the mutually
interacting replicas are considered \cite{Iba2001-2,Campillo2009,Cappe2004}. In the PE scheme, soft annealing provides
the encoded physical Hamiltonian directly, but its interpretation is not necessarily
straightforward compared with that of the QAC.

The information in the logical Hamiltonian that is embedded in the physical
Hamiltonian should be retrieved by decoding. By exploiting the advantage
of the informative prior for the parity constraints, we can increase
the reliability of the decoding in two manners. In the case of MAP
decoding, the readouts of all physical spins provide the outcome
without post-readout processing. We need to sample the 
code state with the lowest energy to infer the target state. This can only be achieved
by balancing the weights of the penalty and non-penalty Hamiltonians
appropriately. In this case, the informative prior is used only for the
pre-readout process, that is, the stochastic sampling process. In addition,
we can exploit the advantage of the informative prior for the post-readout 
process. By exploiting this advantage, we can retrieve the
target state from the non-code but correctable erroneous state in
which some of the parity constraints may be violated. This can be
accomplished through one-step MVD, which fairly exploits the parity constraints
as an informative prior. We discussed the one-step MVD for QAC
and the PE scheme. One-step MVD for QAC is relatively simple to understand. We identify logical groups as sets of physical spins connected by ferromagnetic penalty couplings in the QAC Hamiltonian
\cite{Jordan2006,Young2013,Pudenz2014,Pudenz2015,Vinci2015,Bookatz2015,Matsuura2016,Matsuura2017,Pearson2019}.
The spins in the logical Hamiltonian are identified with the logical
groups in the QAC Hamiltonian. One-step MVD is equivalent to assigning the average orientation
of the physical spins in the logical group to the orientation of the
associated logical spin. If these assignments are successful for every
logical spin with respect to the target logical spin states, we obtain
a successful inference. If the penalty Hamiltonian is absent, this is
analogous to finite temperature decoding, where thermal fluctuation
contributes to improving the decoding performance. 

In contrast, the post-readout decoding for the PE scheme is more
subtle. In this study, we found that efficient decoding through one-step MVD 
is possible by exploiting the higher-weight syndromes in the optimal
estimators. We demonstrated the algorithm for realizing this using an iteration of
matrix multiplications and sign decisions for each element in the resultant
matrix. We confirmed the validity of our algorithm for numerous
examples with various error patterns. The RF MCMC simulation showed
that a huge number of non-code but correctable erroneous states contributes
to the inference of the target state by exploiting the advantage of the informative
prior in one-step MVD. Although our simulation results strongly
suggest that one-step MVD might be more advantageous than 
MAP decoding in terms of performance, the classical overhead
related to post-readout processing should be considered. The advantageousness of the approach depends on the sampler used for the first stage 
as well as the available classical computing resources. A detailed 
study is required to decide the usefulness of
our technique in practical scenarios. 
\begin{acknowledgments}
The author acknowledges Dr. Yuki Susa, Dr. Yuki Kobayashi, Dr. Ryouji Miyazaki, Dr. Akihiro Yatabe, Dr. Tomohiro Yamaji, 
and Dr. Masayuki Shirane of NEC Corporation for their helpful discussions
and comments. This paper is partly based on the results obtained from
a project, JPNP16007, commissioned by the New Energy and Industrial
Technology Development Organization (NEDO), Japan.
\end{acknowledgments}

\appendix*
\renewcommand{\thefigure}{A.\arabic{figure}}
\renewcommand{\thetable}{A.\arabic{table}}
\setcounter{figure}{0}
\setcounter{table}{0}
\section{Rejection-free Markov chains\label{sec:Appendix}}

In this appendix, we explain rejection-free Markov chains (RFMCs)
and the connection between the stationary distribution for an RFMC and 
Maxwell Boltzmann (MB) distribution, which is the stationary state
for standard reversible Markov chains (standard MCs). The standard
MC is the reasonable classical model for an open system connected
to a thermal environment. We focused on a discrete-time Markov chain (DTMC).
We conducted a Monte Carlo simulation to confirm the theoretical prediction using
a toy model. First, we briefly reviewed the stationary state
in the reversible DTMC and RFMC. The reversible DTMC is characterized
by its transition kernel 
\begin{equation}
\boldsymbol{P}=\left(\begin{array}{ccccc}
p_{11} & p_{12} & p_{13} & \cdots & p_{1n}\\
p_{21} & p_{22} & p_{23} & \cdots & p_{2n}\\
p_{31} & p_{32} & p_{33} & \cdots & p_{3n}\\
\vdots & \vdots & \vdots & \ddots & \vdots\\
p_{n1} & p_{n2} & p_{n3} & \cdots & p_{nn}
\end{array}\right),\label{eq:A1}
\end{equation}
where $p$$_{ij}$ is the transition probability of the system from
state $i$ to state $j$. Because $\boldsymbol{P}$ is a stochastic
matrix, the sum of the matrix elements in each row is unity. 
\begin{equation}
\boldsymbol{P}\boldsymbol{1}=\stackrel[j=1]{n}{\sum}p_{ij}=\boldsymbol{1},\label{eq:A2}
\end{equation}
where $\boldsymbol{1}=\left(1,\ldots,1\right)^{T}$ is the column
vector with all elements having the value 1. Therefore, the vector $\boldsymbol{1}$
is a right eigenvector with eigenvalue $\lambda=1$. Here, the diagonal
elements $p_{ii}$ $\left(i=1,\ldots,n\right)$ represent the probability
of self-loop transition, i.e., a state transitioning to itself, resulting in no change in the state. The probability $\alpha_{i}=\sum_{j\neq i}p_{ij}=1-p_{ii}$
is the transition probability from $i$ to another state $j\neq i$.
In the Markov chains characterized by the transition kernel $\boldsymbol{P}$
(abbreviated chain $\boldsymbol{P}$), a series of events with only
two possible outcomes, i.e., the event that the system escapes from
the state $i$ and its exclusive event that the system stays in the
state $i$, is described using Bernoulli trials, with the probability of success
in each trial equal to $\alpha_{i}$. Thus, the probability $p_{i}\left(t\right)$
that the system remains in the state $i$ for $t$ rounds is given by
the geometric distribution characterized by a parameter $\alpha_{i}$, that is, 
\begin{equation}
p_{i}(t)=\left(1-\alpha_{i}\right)^{t-1}\alpha_{i}.
\end{equation}
The average value of $t$ is called the waiting time 
\begin{equation}
\left\langle t_{i}\right\rangle =\stackrel[t=1]{\infty}{\sum}tp_{i}\left(t\right)=\frac{1}{\alpha_{i}}.
\end{equation}

Consider the probability vector $\boldsymbol{v}=\left(v_{1},\ldots,v_{n}\right)^{T}$
with the element $v_{i}\in\left[0,1\right]\in\mathbb{R}$, which is the probability that
the system is in the state $i$ and satisfies
\begin{equation}
\stackrel[i=1]{n}{\sum}v_{i}=1.
\end{equation}
If the chain $\boldsymbol{P}$ is ergodic and satisfies the detailed
balance conditions $v_{i}p_{ij}=v_{j}p_{ji}$, $\boldsymbol{v}$
approaches a unique stationary distribution $\boldsymbol{\pi}=\left(\pi_{1},\ldots,\pi_{n}\right)^{T}$
that satisfies the global balance equation 
\begin{equation}
\boldsymbol{\pi}^{T}=\boldsymbol{\pi}^{T}\boldsymbol{P}\label{eq:A6}
\end{equation}
after numerous rounds of state transition. Equation (\ref{eq:A1})
indicates that $\boldsymbol{\pi}$ is the left eigenvector of the
eigen-equation with eigenvalue $\lambda=1$. Furthermore, if the chain
$\boldsymbol{P}$ is aperiodic, $\boldsymbol{\pi}$ is also a limiting
distribution, that is, $\boldsymbol{\pi}=\underset{t\rightarrow\infty}{\lim}\boldsymbol{v}$.
If the system is open to the thermal environment, the detailed balance
conditions lead to 
\begin{equation}
\frac{\pi_{i}}{\pi_{j}}=\frac{p_{ji}}{p_{ij}}=\frac{\exp\left(-\beta E_{i}\right)}{\exp\left(-\beta E_{j}\right)}=\exp\left[-\beta\left(E_{i}-E_{j}\right)\right],
\end{equation}
where $\beta$ is the inverse temperature of the environment and $E_{i}$
is the energy of the system when it is in the state $i$. In this
case, $\boldsymbol{\pi}$ is given by the MB distribution
\begin{equation}
\pi_{i}=\frac{\exp\left[-\beta E_{i}\right]}{Z},\label{eq:A8}
\end{equation}
where $Z$ is a normalization constant called the canonical partition function.

Based on the above formulation for the standard MC, we define
the RFMC and clarify how it is different from the standard MC. We
define the RFMC as a reversible MC with the transition kernel given
by the following matrix $\tilde{\boldsymbol{P}}$, whose diagonal elements
are all zero:
\begin{eqnarray}
\tilde{\boldsymbol{P}} & = & \left(\begin{array}{ccccc}
\tilde{p}_{11} & \tilde{p}_{12} & \tilde{p}_{13} & \cdots & \tilde{p}_{1n}\\
\tilde{p}_{21} & \tilde{p}_{22} & \tilde{p}_{23} & \cdots & \tilde{p}_{2n}\\
\tilde{p}_{31} & \tilde{p}_{32} & \tilde{p}_{33} & \cdots & \tilde{p}_{3n}\\
\vdots & \vdots & \vdots & \ddots & \vdots\\
\tilde{p}_{n1} & \tilde{p}_{n2} & \tilde{p}_{n3} & \cdots & \tilde{p}_{nn}
\end{array}\right)\nonumber \\
 & = & \left(\begin{array}{ccccc}
0 & \frac{p_{12}}{\alpha_{1}} & \frac{p_{13}}{\alpha_{1}} & \cdots & \frac{p_{1n}}{\alpha_{1}}\\
\frac{p_{21}}{\alpha_{2}} & 0 & \frac{p_{23}}{\alpha_{2}} & \cdots & \frac{p_{2n}}{\alpha_{2}}\\
\frac{p_{31}}{\alpha_{3}} & \frac{p_{32}}{\alpha_{3}} & 0 & \cdots & \frac{p_{3n}}{\alpha_{3}}\\
\vdots & \vdots & \vdots & \ddots & \vdots\\
\frac{p_{n1}}{\alpha_{n}} & \frac{p_{n2}}{\alpha_{n}} & \frac{p_{n3}}{\alpha_{n}} & \cdots & 0
\end{array}\right)\nonumber \\
 & = & \left(\begin{array}{ccccc}
0 & \left\langle t_{1}\right\rangle p_{12} & \left\langle t_{1}\right\rangle p_{13} & \cdots & \left\langle t_{1}\right\rangle p_{1n}\\
\left\langle t_{2}\right\rangle p_{21} & 0 & \left\langle t_{2}\right\rangle p_{23} & \cdots & \left\langle t_{2}\right\rangle p_{2n}\\
\left\langle t_{3}\right\rangle p_{31} & \left\langle t_{3}\right\rangle p_{32} & 0 & \cdots & \left\langle t_{3}\right\rangle p_{3n}\\
\vdots & \vdots & \vdots & \ddots & \vdots\\
\left\langle t_{n}\right\rangle p_{n1} & \left\langle t_{n}\right\rangle p_{n2} & \left\langle t_{n}\right\rangle p_{n3} & \cdots & 0
\end{array}\right).\nonumber \\
\label{eq:A9}
\end{eqnarray}
Evidently, $\tilde{\boldsymbol{P}}$ is a stochastic matrix. Because
the chain $\tilde{\boldsymbol{P}}$ is reversible, its detailed
balance condition should be 
\begin{equation}
\tilde{\pi}_{i}\tilde{p}_{ij}=\tilde{\pi}_{j}\tilde{p}_{ji},\label{eq:A10}
\end{equation}
where the vector $\hat{\boldsymbol{\pi}}=\left(\hat{\pi}_{1},\ldots,\hat{\pi}_{n}\right)^{T}$
represents the stationary distribution for the chain $\tilde{\boldsymbol{P}}$.
The detailed balance conditions are fulfilled if $\tilde{\pi}_{i}$ is expressed in terms
of $\pi_{i}$ as
\begin{equation}
\tilde{\pi}_{i}=\frac{\pi_{i}\alpha_{i}}{\Sigma_{j=1}^{n}\pi_{j}\alpha_{j}}=\frac{\pi_{i}/\left\langle t_{i}\right\rangle }{\Sigma_{j=1}^{n}\pi_{j}/\left\langle t_{j}\right\rangle }.\label{eq:A11}
\end{equation}
Conversely, $\pi_{i}$ is written in terms of $\tilde{\pi}_{i}$, as follows: 
\begin{equation}
\pi_{i}=\frac{\pi_{i}/\alpha_{i}}{\Sigma_{j=1}^{n}\pi_{j}/\alpha_{j}}=\frac{\pi_{i}\left\langle t_{i}\right\rangle }{\Sigma_{j=1}^{n}\pi_{j}\left\langle t_{j}\right\rangle }.\label{eq:A12}
\end{equation}
Therefore, the reversible chain $\tilde{\boldsymbol{P}}$ should have
the stationary distribution $\tilde{\boldsymbol{\pi}}$ given by Eq. (\ref{eq:A11}).
The zeroes in the diagonal elements of $\tilde{\boldsymbol{P}}$ represent
the chain $\tilde{\boldsymbol{P}}$ with no self-loop transitions.
The matrix $\tilde{\boldsymbol{P}}$ is the transition kernel of the
embedded DTMC that is associated with the continuous-time Markov chains (CTMCs) \cite{Mazumdar2012,Parekh2020,Rosenthal2021}, which tracks only the states visited directly after each transition and moves
from a state to another state according to the transition probabilities
given by $\tilde{\boldsymbol{P}}$. Equations (\ref{eq:A11}) and (\ref{eq:A12})
can be interpreted as follows: Because $\pi_{i}$ considers the states after the self-loop transition occurring with the probability
$p_{ii}=1-\alpha_{i}$, only a fraction $\alpha_{i}$ of $\pi_{i}$
can contribute to $\tilde{\pi}_{i}$ on average. Therefore, it is reasonable
that $\tilde{\pi}_{i}\propto\pi_{i}\alpha_{i}$ and Eq. (\ref{eq:A11})
provide the probability distribution. Note that even if the kernel $\tilde{\boldsymbol{P}}$
has a stationary distribution $\tilde{\boldsymbol{\pi}}$, it 
is not necessarily the limiting distribution,
as $\tilde{\boldsymbol{\pi}}$ may be a periodic function of the round
$t$ of state transition. It is evident that if $\boldsymbol{\pi}$
is an MB distribution, $\tilde{\boldsymbol{\pi}}$ is not an MB distribution
owing to the $i$-dependence of $\alpha_{i}$. Nevertheless,
it should be noted that we can recover the MB distribution $\boldsymbol{\pi}$
from $\tilde{\boldsymbol{\pi}}$ using Eq. (\ref{eq:A12}) and
$\alpha_{i}=\sum_{j\neq i}p_{ij}$, which is, in principle, calculable
if we can calculate the energy differences between the states.

The RFMC is not merely a theoretical model but can actually be implemented
and demonstrated using Monte Carlo simulation. We focus on the RFMC
Monte Carlo simulation based on the Metropolis algorithm. When the standard
algorithm is applied to an $N$-spin system, the transition probability from
the current state $0$ is given by the following Metropolis rule:
\begin{equation}
p_{0i}=\left\{ \begin{array}{cc}
\frac{1}{N}\min\left\{ 1,\exp\left(-\beta\Delta E_{i}\right)\right\}  & i\in S_{0}\\
1-\alpha_{0} & i=0\\
0 & \mathrm{otherwise}
\end{array}\right.,
\end{equation}
where $S_{0}$ is a set of neighboring states for the current state
$i=0$, that is, the states in which a single spin is flipped, and
$\Delta E_{i}=E_{i}-E_{0}$ is the transition energy from the current
state $0$ to the next state in which the spin $i$ is flipped.
Because $p_{00}=1-\alpha_{0}$, it follows that 
\begin{equation}
\alpha_{0}=\underset{i\in S_{0}}{\sum}p_{0i}=\frac{1}{N}\underset{i\in S_{0}}{\sum}\min\left\{ 1,\exp\left(-\beta\Delta E_{i}\right)\right\} .
\end{equation}
Therefore, $\alpha_{0}$ ($p_{00}$) represents the acceptance (rejection)
probability for flipping the spin. We can also regard it as the escaping
(trapping) probability that the standard MC will move away from
(stay in) the current state in the next step. Then, the ensemble sampled 
by the standard Monte Carlo simulation approaches the stationary as well as limiting distribution
given by Eq. (\ref{eq:A8}), i.e., the MB distribution. From Eq. (\ref{eq:A9}),
it follows that the transition kernel of the RFMC is given by the
transition probability 
\begin{equation}
\tilde{p}_{0i}=\left\{ \begin{array}{cc}
\frac{p_{0i}}{\underset{i\in S_{0}}{\sum}p_{0i}} & i\in S_{0}\\
0 & \mathrm{otherwise}
\end{array}\right.
\end{equation}
The random choice of the next state to move with the transition probability
$p_{0i}$ can be implemented by making use of the conventional acceptance-rejection
(AR) method using a pseudo random number generator. In contrast, the
random choice of the next state with the transition probability $\tilde{p}_{0i}$
can be implemented by making use of various algorithms for generating
a non-uniform random variate, as shown in the following, although their
calculations are more expensive than that of the conventional AR method. 

In fact, the RFMC was demonstrated through the Monte Carlo simulation using the following
toy model. Consider $N=3$ $\left(n=2^{N}=8\right)$ spins described
by the following Ising Hamiltonian: 
\begin{equation}
H\left(\left\{ s_{1},s_{2},s_{3}\right\} \right)=\stackrel[i=1]{N}{\sum}h_{i}s_{i}+\stackrel[i,j=1]{N}{\sum}J_{ij}s_{i}s_{j}
\end{equation}
with
\begin{eqnarray}
&&\left\{ h_{1},h_{2},h_{3},J_{12},J_{13},J_{23},\mathrm{others}\right\}\hspace{20mm}\nonumber \\
&&\hspace{10mm}=\left\{ +2,+1,0,+1,-2,-1,0\right\} .
\end{eqnarray}

\begin{figure}[h]
\begin{centering}
\includegraphics[viewport=265bp 150bp 700bp 440bp,clip,scale=0.6]{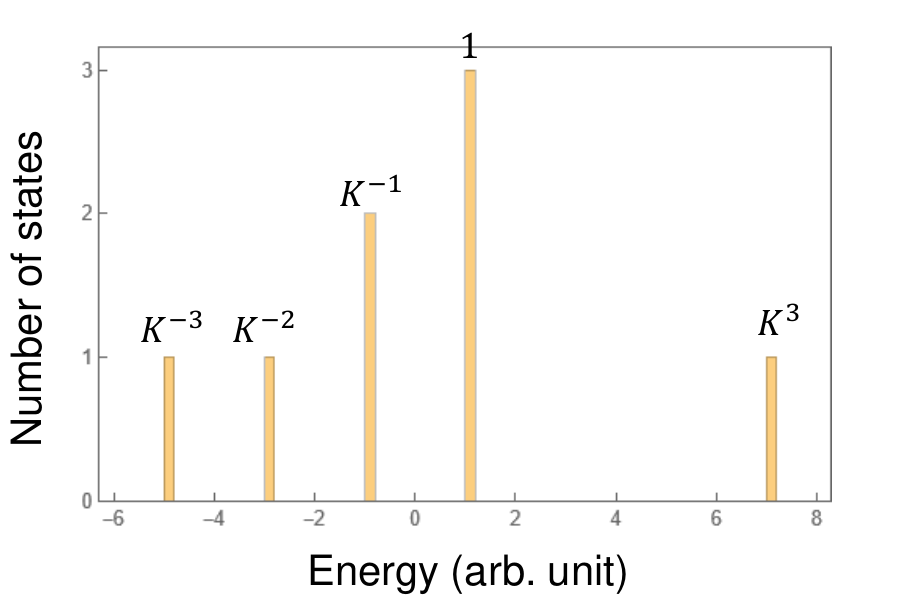}
\caption{Energy spectrum of the associated density of states for our toy model with
$N=3$ spins. The symbols $\left(K^{-3},K^{-2},\ldots,K^{3}\right)$
indicate the relative ratios of the probability distribution for the stationary
MB distribution $\pi$ in Eq. (\ref{eq:A20}). \label{fig:A1}}
\par\end{centering}
\end{figure}

\begin{figure*}
\begin{centering}
\includegraphics[viewport=80bp 120bp 860bp 440bp,clip,scale=0.6]{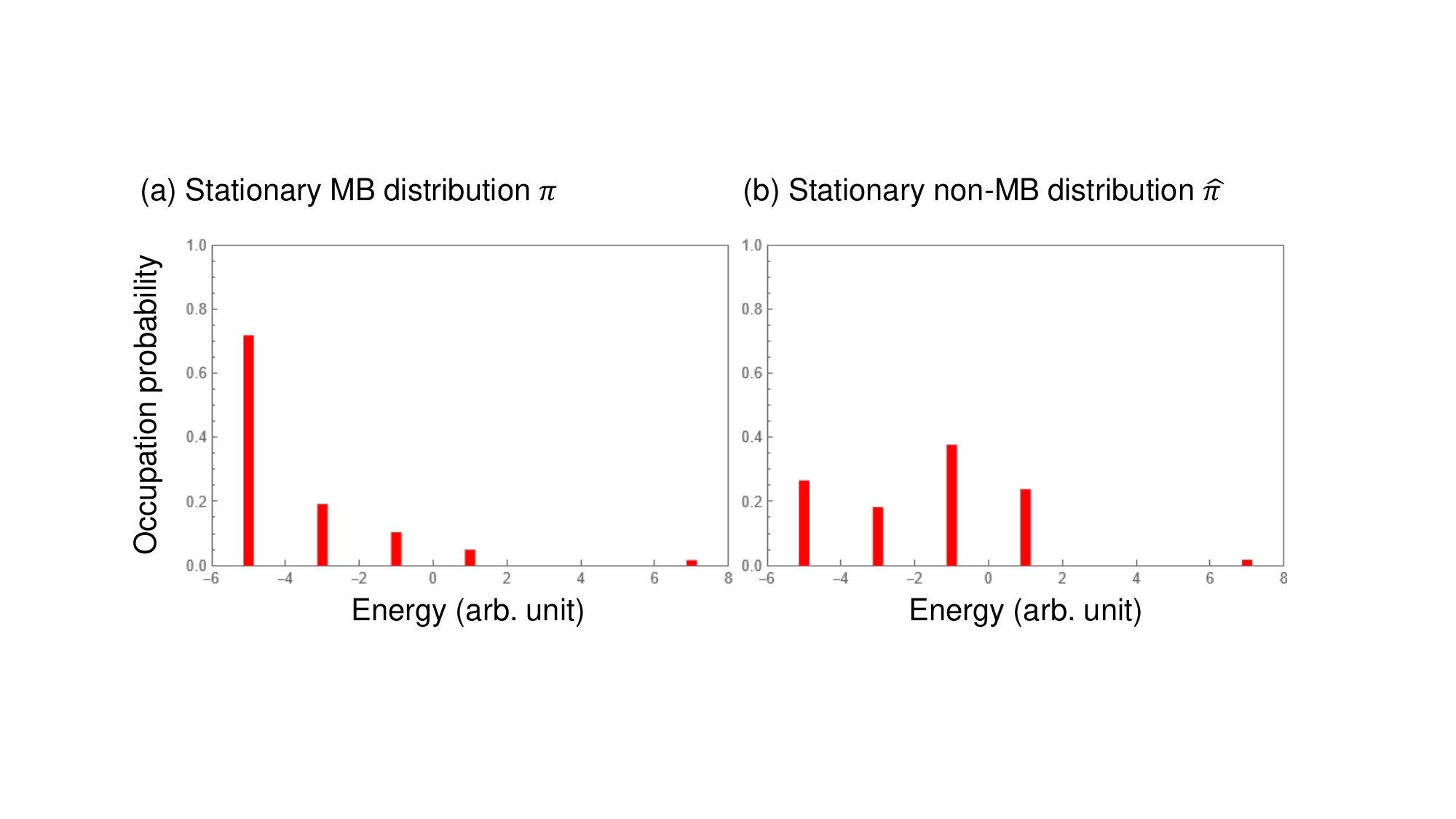}
\caption{Energy spectra of stationary distributions $\boldsymbol{\pi}$ and
$\tilde{\boldsymbol{\pi}}$ when $K=0.25$ is assumed.\label{fig:A2}}
\par\end{centering}
\end{figure*}

Table \ref{tab:1} shows the eight possible spin configurations and associated
energies, and Fig. \ref{fig:A1} depicts the energy spectrum of the associated
density of states $\boldsymbol{\rho}$. The transition kernels of
the standard MC and RFMC are given by
\begin{table}[h]
\begin{centering}
\begin{tabular}{|c|c|c|}
\hline 
$i$ & $\left(s_{1},s_{2},s_{3}\right)$ & Energy\tabularnewline
\hline 
$1$ & $\left(-1,-1,-1\right)$ & $-5$\tabularnewline
\hline 
$2$ & $\left(-1,-1,+1\right)$ & $+1$\tabularnewline
\hline 
$3$ & $\left(-1,+1,-1\right)$ & $-1$\tabularnewline
\hline 
$4$ & $\left(+1,-1,-1\right)$ & $-1$\tabularnewline
\hline 
$5$ & $\left(-1,+1,+1\right)$ & $-3$\tabularnewline
\hline 
$6$ & $\left(+1,-1,+1\right)$ & $+1$\tabularnewline
\hline 
$7$ & $\left(+1,+1,-1\right)$ & $+7$\tabularnewline
\hline 
$8$ & $\left(+1,+1,+1\right)$ & $+1$\tabularnewline
\hline 
\end{tabular}\caption{Possible spin configurations and associated energies for our toy model with $N=3$ spins.\label{tab:1}}
\par\end{centering}
\end{table}
\begin{widetext}
\begin{equation}
\boldsymbol{P}=\frac{1}{3}
{\footnotesize
\left(\begin{array}{cccccccc}
-K^{3}-2K^{2}+3 & \,K^{3} & K^{2} & K^{2} & 0 & 0 & 0 & 0\\
1 & 0 & 0 & 0 & 1 & 1 & 0 & 0\\
1 & 0 & -K^{4}+1 & 0 & 1 & 0 & K^{4} & 0\\
1 & 0 & 0 & -K^{4}-2K+2\, & 0 & K & K^{4} & 0\\
0 & K^{2} & K & 0 & -2K^{2}-K+3 & 0 & 0 & K^{2}\\
0 & 1 & 0 & 1 & 0 & 0 & 0 & 1\\
0 & 0 & 1 & 1 & 0 & 0 & 0 & 1\\
0 & 0 & 0 & 0 & 1 & 1 & K^{3} & -K^{3}+1
\end{array}\right),
}
\end{equation}
\begin{equation}
\tilde{\boldsymbol{P}}=\left(\begin{array}{c}
\begin{array}{cccccccc}
0 & \,\frac{K}{K+2} & \frac{1}{K+2} & \frac{1}{K+2} & 0 & 0 & 0 & 0\\
\frac{1}{3} & 0 & 0 & 0 & \frac{1}{3} & \frac{1}{3} & 0 & 0\\
\frac{1}{K^{4}+2} & 0 & 0 & 0 & \frac{1}{K^{4}+2} & 0 & \frac{K^{4}}{K^{4}+2} & 0\\
\frac{1}{K^{4}+K+1} & 0 & 0 & 0 & 0 & \frac{K}{K^{4}+K+1} & \frac{K^{4}}{K^{4}+K+1} & 0\\
0 & \frac{K}{2K+1} & \frac{1}{2K+1} & 0 & 0 & 0 & 0 & \frac{K}{2K+1}\\
0 & \frac{1}{3} & 0 & \frac{1}{3} & 0 & 0 & 0 & \frac{1}{3}\\
0 & 0 & 1 & 1 & 0 & 0 & 0 & 1\\
0 & 0 & 0 & 0 & \frac{1}{K^{3}+2} & \frac{1}{K^{3}+2} & \frac{K^{3}}{K^{3}+2} & 0
\end{array}\end{array}\right),
\end{equation}
\end{widetext}
respectively, where $K=\exp\left(-2\beta\right)$. The transition
kernels $\boldsymbol{P}$ and $\tilde{\boldsymbol{P}}$ have the left
eigenvectors of the eigen-equation with eigenvalue $\lambda=1$, which
offer the stationary distributions 
\begin{widetext}
\begin{equation}
\boldsymbol{\pi}=\mathcal{N}\left[\left(K^{-3},1,K^{-1},K^{-1},K^{-2},1,K^{3},1\right)\right]^{T},\label{eq:A20}
\end{equation}
\begin{equation}
\tilde{\boldsymbol{\pi}}=\mathcal{N}\left[\left(\frac{1+2K^{-1}}{3},1,\frac{K^{3}+2K^{-1}}{3},\frac{K^{3}+1+K^{-1}}{3},\frac{2+K^{-1}}{3},1,K^{3},\frac{K^{3}+2}{3}\right)\right]^{T},
\end{equation}
\end{widetext}
where $\mathcal{N}${[}$\cdots${]} denotes the consideration of a normalized vector. The stationary distribution $\boldsymbol{\pi}$ is the MB distribution.
Figure \ref{fig:A2} depicts the energy spectra of $\boldsymbol{\pi}$
and $\tilde{\boldsymbol{\pi}}$ when $K=0.25$ is assumed. Note that
the density of states $\boldsymbol{\rho}$ was considered
in these spectra. Evidently, these energy spectra differ. 

\begin{figure*}
\begin{centering}
\includegraphics[viewport=140bp 110bp 860bp 440bp,clip,scale=0.6]{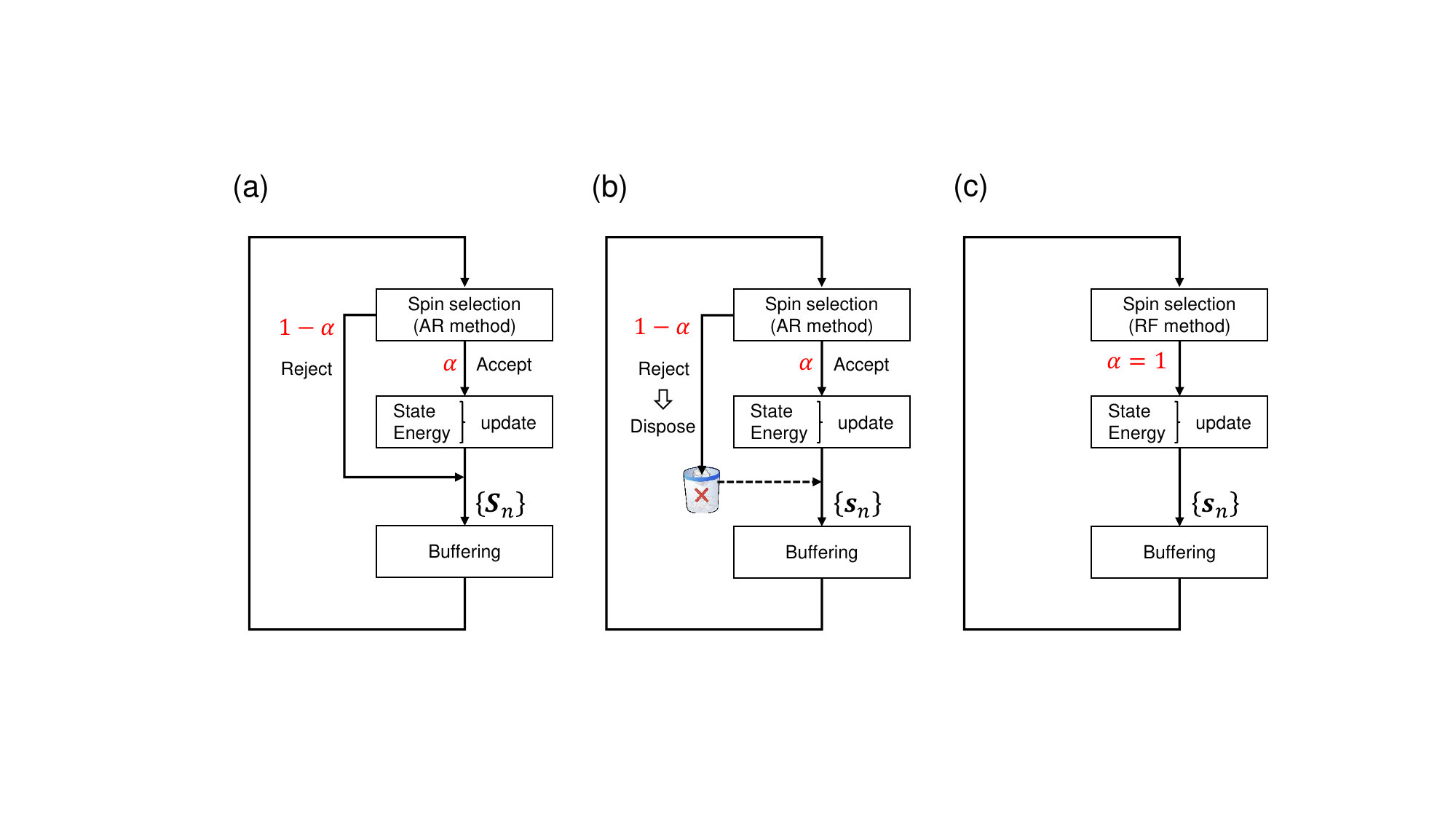}
\par\end{centering}
\caption{Comparison of three algorithms for MCMC simulation: (a) standard MC, (b)
rejection-discarded MC, and (c) RFMC.
\label{fig:A3}}
\end{figure*}

We performed Monte Carlo simulations to confirm whether these stationary states are actually reproduced. We compared the following three algorithms
for the Monte Carlo simulations: (a) Standard MC: Fig. \ref{fig:A3}
(a) shows the block diagram of the standard MC. The Metropolis method was used in this simulation. 
We generated a uniform random number $U(1,N)\in\mathbb{N}$
to select a spin to be flipped in the next state transition
and a uniform random variate $U\left(0,1\right)\in\mathbb{R}$ to
determine whether it was accepted or rejected for flipping its sign. If
the energy $\Delta E_{i}$ required to flip the chosen spin $i$ is
such that $\beta\Delta E_{i}<-\ln U\left(0,1\right)$, it is
accepted; otherwise, it is rejected. We can easily obtain the required elements of the kernel $\boldsymbol{P}$ sweep by sweep
in the standard MC iteration. In the standard MC algorithm, all the
spin states are recorded as the sampled data, regardless of whether the chosen spin is
accepted or rejected for flipping, which is analyzed after the completion
of the simulation. (b) Rejection-discarded MC: Fig. \ref{fig:A3} (b) shows the
modified version of the standard MC. This algorithm differs from (a)
in the recording of the sampled data. Only the spin states after the accepted
transitions are recorded as the sampled data and those after the rejected
transitions are discarded. Because the states after the rejected transitions
are the same as those before it, discarding the state directly after the rejected
transition implies disregarding the states associated with the self-loop
transitions. (c) RFMC: To realize the kernel $\tilde{\boldsymbol{P}}$,
we apply the algorithm for the non-uniform random number generation.
The algorithm generates a set of $N$ exponential random variables
$\left\{ \tau_{1},\ldots,\tau_{N}\right\} $, each of which is characterized
by its own rate parameter $p_{0i}$, e.g., $\tau_{i}\sim\mathrm{Exponential}\left(p_{0i}\right)$.
The random variable $\tau_{i}$ is regarded as the holding time required
to flip the spin $i$ in the CTMC. We accept the smallest of these 
and flip the corresponding spin. According to the properties of the exponential
random variate, the associated probability is given by 
\begin{equation}
\mathrm{Pr}\left(\tau_{i}=\min_{i}\left\{ \tau_{1},\ldots,\tau_{N}\right\} \right)=\frac{p_{0i}}{\stackrel[i=1]{N}{\sum}p_{0i}}.
\end{equation}
This is a well-known non-uniform random number generation (or weighted
random choice) algorithm for selecting one item from $N$ items
according to the weight $\left\{ p_{01},\ldots,p_{0N}\right\} $ \cite{Nambu22}.
Thus, we can obtain the required elements in the transition kernel $\tilde{\boldsymbol{P}}$
recursively in the MC iteration. In this algorithm, all spin
states directly after flipping the selected spin are recorded as the sampled data. 

\begin{figure*}
\begin{centering}
\includegraphics[viewport=80bp 60bp 890bp 490bp,clip,scale=0.62]{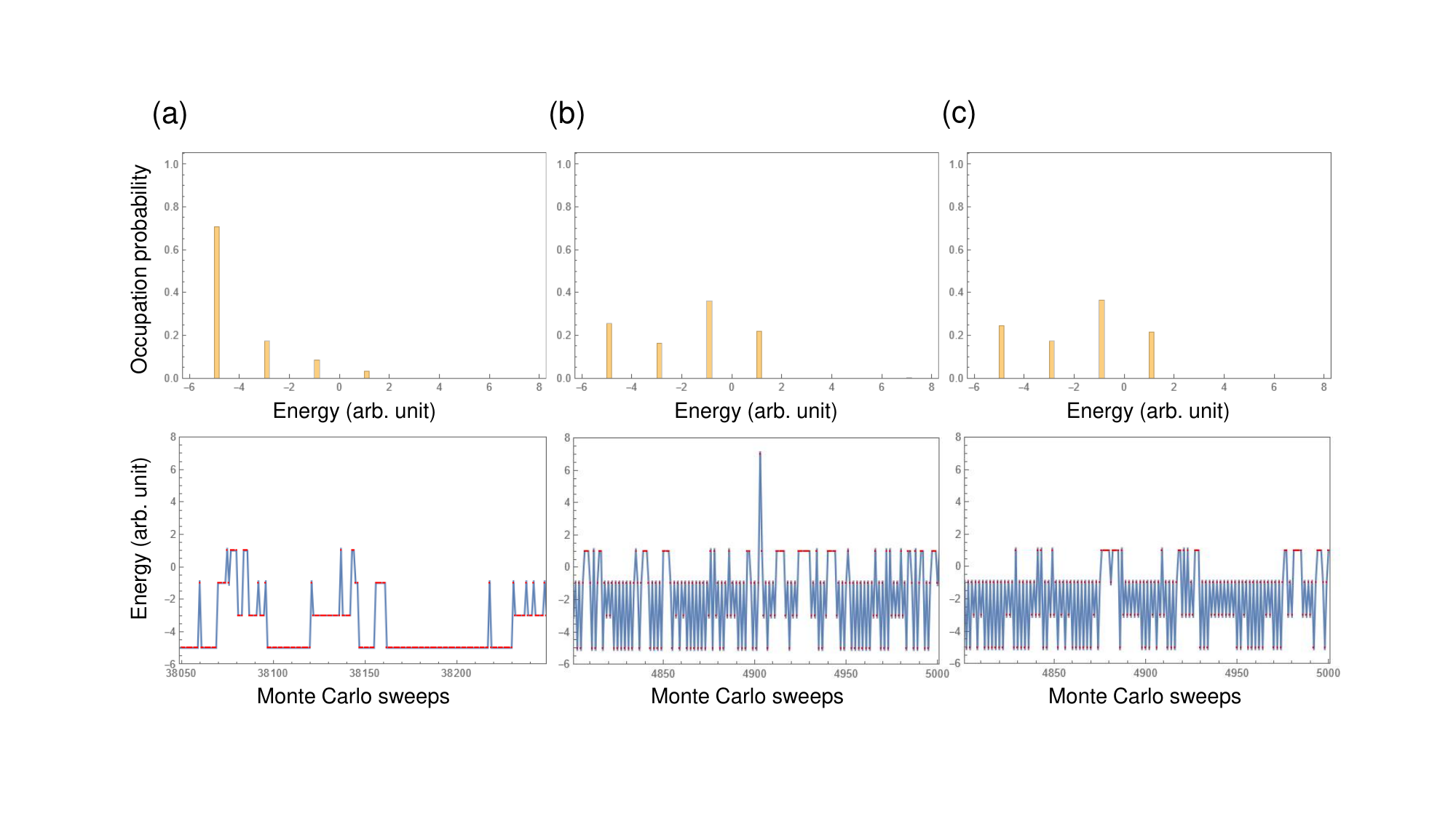}
\par\end{centering}
\caption{Energy spectra of the occupation probability and associated temporal
development for the sampled data associated with algorithms (a)-(c).
\label{fig:A4}}
\end{figure*}

We performed Monte Carlo simulations using the three algorithms to obtain a sufficiently large quantity of sampled data and compared the energy spectra of their stationary distributions with those shown in Fig. \ref{fig:A2}.
Figure \ref{fig:A4} shows the energy spectra
for the stationary distributions and associated temporal developments
for the sampled data. The rounds of MC iterations
were adjusted so that the number of spin flips was approximately 5000 in these simulations. The temporal developments of the energy
of the spin system associated with the last 200 of the sampled data
are shown in the lower plots of Fig. \ref{fig:A4}. The computation
times of algorithms (a) and (b) were comparable but that of 
algorithm (c) was approximately half. 

We can observe that once the spins are trapped at a lower-energy state, numerous trials to flip are required to escape from such a state to the next
state in the temporal profiles in Fig. \ref{fig:A4} (a), which is
accounted for by the self-loop transitions. In contrast, such state trapping is not observed in Fig. \ref{fig:A4} (b) and (c). The
number of trials to flip that is required to escape from the trapped state
is a geometric random variable $\tau\sim\mathrm{\mathrm{Geometric}}\left(\alpha_{0}\right)$
associated with the trapped state $0$ in which the average number is given by $\left\langle \tau\right\rangle =\frac{1}{\alpha_{0}}$.
Note that $\alpha_{0}$ depends on the trapped state 
and changes after every spin flip. In addition, there
are two points to note. The first is that the short trapping
to the states with energy $=+1$ in the lower plots of Fig. \ref{fig:A4}
is not attributed to the self-loop transition, but to the transition
among the energetically degenerated states. The second
is that the chain $\tilde{\boldsymbol{P}}$ has a left eigenvalue
$-1$ so that the stationary distribution $\tilde{\boldsymbol{\pi}}$
is not the limiting distribution \cite{Seabrook2022}. This is because
the MC with no self-loop transition is never aperiodic
and exhibits oscillatory behavior with period $d>1$, as shown in Fig. \ref{fig:A4} (b) and (c). These differences in the temporal developments
for the algorithms are directly reflected in their energy spectra. Figure \ref{fig:A4} also depicts the energy spectra
of the stationary distributions for the three algorithms. The energy spectrum
for algorithm (a) agrees with the MB distribution $\boldsymbol{\pi}$
associated with the transition kernel $\boldsymbol{P}$. In contrast,
the energy spectra for algorithms (b) and (c) agree with the non-MB distribution
$\tilde{\boldsymbol{\pi}}$ associated with the kernel $\tilde{\boldsymbol{P}}$.
Therefore, the temporal developments as well as the energy spectra
of algorithms (b) and (c) agree. This implies that the difference
between two stationary distributions $\boldsymbol{\pi}$ and $\tilde{\boldsymbol{\pi}}$
originates from any immediately repeated state after the transition, i.e.,
self-loop transition. In fact, we can confirm that the MB distribution
$\boldsymbol{\pi}$ reproduces the temporal development and spectra
for the non-MB distribution $\tilde{\boldsymbol{\pi}}$ if we simply
omit these states due to the self-loop transitions. 

We should emphasize that the self-loop transitions also play an important
role in sampling the ground state of the Ising spins by Monte Carlo simulation.
If we simulate a standard reversible MC using the Monte Carlo method,
a sampled series converges to the MB distribution after a sufficiently large
number of transitions. Then, if we maintain the spin systems at a low
temperature, the lower energy portion of the MB distribution $\boldsymbol{\pi}$
increases so that the sampling probability of the ground state increases.
However, this is necessarily accompanied by a decreased acceptance probability
$\alpha_{0}$ for the low lying state and an associated increase in the number
of repeated states immediately after the transition owing to self-loop transition.
The presence of the low-lying excited states may affect the time required
for successfully sampling the ground state, as not only the ground
state but also the low-lying excited states have a large self-loop transition
probability. Once the system is trapped to one such low-lying excited
state, many sweeps would be wasted before a successful escape from
the trapped state. Thus, extra time is required to sample the ground state successfully. This deceleration in sampling the ground state has
been frequently pointed out as a serious problem in simulated
annealing. Although this problem may be relaxed by dynamically controlling
the hyperparameters such as the temperature to a certain extent, the approach has limited benefits. This is a common problem for conventional simulated annealing
that reads out the state at an arbitrary timing as a sample using a trial-and-error
approach. Therefore, mutually incompatible requirements exist regarding
self-loop transitions: increasing the probability of the self-loop
transitions is desirable to increase the sampling efficiency for the
ground state, whereas decreasing it is desirable to quickly approach
the MB distribution as well as the ground state quickly. 

RFMC offers the potential to overcome this situation. Because it has no
self-loop transition, it approaches its stationary distribution more
rapidly than the standard MC method. The drawback of the trial-and-error
sampling approach can be overcome by keeping track of the spin state
as well as its energy sweep by sweep and updating the bookkeeping to hold
the best state to find the ground state. This can be
easily incorporated into the Monte Carlo simulation. We should note that
although the stationary distribution of the RFMC deviates from the
MB distribution, we can recover the MB distribution by implementing the
additional calculation required to recover it. Because the RFMC considers
the same chain as the standard DTMC, except for the omission of any repeated
states owing to self-loop transitions, it loses the information on the
waiting time for a chain of readout states. Suppose that the $k$th state
in an RFMC chain is obtained by omitting $M_{k}$ repetitions of
the states owing to the self-loop transition in the standard DTMC \cite{Rosenthal2021}.
Then, the multiplicity $M_{k}\sim\mathrm{\mathrm{Geometric}}\left(\alpha_{0}\left(k\right)\right)+1$
is lost, where $\alpha_{0}\left(k\right)$ is the acceptance rate for
the $k$th state. This lost information can be easily calculated during
the iterations of the Monte Carlo simulation because the set $\left\{ p_{01},\ldots,p_{0N}\right\}$
of the transition probability 
is necessarily calculated to choose the next spin state in
the RFMC algorithm, which can be accomplished relatively inexpensively by
keeping track of the bookkeeping list $\left\{ \Delta E_{1},\ldots,\Delta E_{n}\right\} $.
Then, we can calculate the acceptance rate $\alpha_{0}=\underset{i\in S_{0}}{\sum}p_{0i}$
and multiplicity $M\sim\left\lfloor \frac{\ln U\left(0,1\right)}{\ln\left(1-\alpha_{0}\right)}\right\rfloor +1$
for the current state sweep by sweep to form the multiplicity list
${M_{k}}$, which is the number of times that the standard chain remains in the
same state after the transition, where $\left\lfloor x\right\rfloor $
denotes the integer part of $x$ and the well-known formulae $\mathrm{Geometric}\left(\lambda\right)=\left\lfloor \mathrm{Exponential}\left(-\ln\left(1-\lambda\right)\right)\right\rfloor $
and $\mathrm{Exponential}\left(\lambda\right)=-\frac{\ln U\left(0,1\right)}{\lambda}$
are used to express $M$\cite{Watanabe2006,Rosenthal2021}.

\nocite{*}
\bibliography{APS}

\providecommand{\noopsort}[1]{}\providecommand{\singleletter}[1]{#1}%
\begin{thebibliography}{75}%
\makeatletter
\providecommand \@ifxundefined [1]{%
 \@ifx{#1\undefined}
}%
\providecommand \@ifnum [1]{%
 \ifnum #1\expandafter \@firstoftwo
 \else \expandafter \@secondoftwo
 \fi
}%
\providecommand \@ifx [1]{%
 \ifx #1\expandafter \@firstoftwo
 \else \expandafter \@secondoftwo
 \fi
}%
\providecommand \natexlab [1]{#1}%
\providecommand \enquote  [1]{``#1''}%
\providecommand \bibnamefont  [1]{#1}%
\providecommand \bibfnamefont [1]{#1}%
\providecommand \citenamefont [1]{#1}%
\providecommand \href@noop [0]{\@secondoftwo}%
\providecommand \href [0]{\begingroup \@sanitize@url \@href}%
\providecommand \@href[1]{\@@startlink{#1}\@@href}%
\providecommand \@@href[1]{\endgroup#1\@@endlink}%
\providecommand \@sanitize@url [0]{\catcode `\\12\catcode `\$12\catcode
  `\&12\catcode `\#12\catcode `\^12\catcode `\_12\catcode `\%12\relax}%
\providecommand \@@startlink[1]{}%
\providecommand \@@endlink[0]{}%
\providecommand \url  [0]{\begingroup\@sanitize@url \@url }%
\providecommand \@url [1]{\endgroup\@href {#1}{\urlprefix }}%
\providecommand \urlprefix  [0]{URL }%
\providecommand \Eprint [0]{\href }%
\providecommand \doibase [0]{http://dx.doi.org/}%
\providecommand \selectlanguage [0]{\@gobble}%
\providecommand \bibinfo  [0]{\@secondoftwo}%
\providecommand \bibfield  [0]{\@secondoftwo}%
\providecommand \translation [1]{[#1]}%
\providecommand \BibitemOpen [0]{}%
\providecommand \bibitemStop [0]{}%
\providecommand \bibitemNoStop [0]{.\EOS\space}%
\providecommand \EOS [0]{\spacefactor3000\relax}%
\providecommand \BibitemShut  [1]{\csname bibitem#1\endcsname}%
\let\auto@bib@innerbib\@empty
\bibitem [{\citenamefont {Kirkpatrick}\ \emph {et~al.}(1983)\citenamefont
  {Kirkpatrick}, \citenamefont {Gelatt},\ and\ \citenamefont
  {Vecchi}}]{Kirkpatrick83}%
  \BibitemOpen
  \bibfield  {author} {\bibinfo {author} {\bibfnamefont {S.}~\bibnamefont
  {Kirkpatrick}}, \bibinfo {author} {\bibfnamefont {C.~D.}\ \bibnamefont
  {Gelatt}}, \ and\ \bibinfo {author} {\bibfnamefont {M.~P.}\ \bibnamefont
  {Vecchi}},\ }\href@noop {} {\bibfield  {journal} {\bibinfo  {journal}
  {Science}\ }\textbf {\bibinfo {volume} {220}},\ \bibinfo {pages} {671}
  (\bibinfo {year} {1983})}\BibitemShut {NoStop}%
\bibitem [{\citenamefont {Kadowaki}\ and\ \citenamefont
  {Nishimori}(1998)}]{Kadowaki98}%
  \BibitemOpen
  \bibfield  {author} {\bibinfo {author} {\bibfnamefont {T.}~\bibnamefont
  {Kadowaki}}\ and\ \bibinfo {author} {\bibfnamefont {H.}~\bibnamefont
  {Nishimori}},\ }\href@noop {} {\bibfield  {journal} {\bibinfo  {journal}
  {Phys. Rev. E}\ }\textbf {\bibinfo {volume} {58}},\ \bibinfo {pages} {5333}
  (\bibinfo {year} {1998})}\BibitemShut {NoStop}%
\bibitem [{\citenamefont {Choi}(2008)}]{Choi2008}%
  \BibitemOpen
  \bibfield  {author} {\bibinfo {author} {\bibfnamefont {V.}~\bibnamefont
  {Choi}},\ }\href {\doibase 10.1007/S11128-008-0082-9/METRICS} {\bibfield
  {journal} {\bibinfo  {journal} {Quantum Inf. Process.}\ }\textbf {\bibinfo
  {volume} {7}},\ \bibinfo {pages} {193} (\bibinfo {year} {2008})}\BibitemShut
  {NoStop}%
\bibitem [{\citenamefont {Choi}(2011)}]{Choi2011}%
  \BibitemOpen
  \bibfield  {author} {\bibinfo {author} {\bibfnamefont {V.}~\bibnamefont
  {Choi}},\ }\href {\doibase 10.1007/s11128-010-0200-3} {\bibfield  {journal}
  {\bibinfo  {journal} {Quantum Inf. Process.}\ }\textbf {\bibinfo {volume}
  {10}},\ \bibinfo {pages} {343} (\bibinfo {year} {2011})}\BibitemShut
  {NoStop}%
\bibitem [{\citenamefont {Lechner}\ \emph {et~al.}(2015)\citenamefont
  {Lechner}, \citenamefont {Hauke},\ and\ \citenamefont {Zoller}}]{Lechner15}%
  \BibitemOpen
  \bibfield  {author} {\bibinfo {author} {\bibfnamefont {W.}~\bibnamefont
  {Lechner}}, \bibinfo {author} {\bibfnamefont {P.}~\bibnamefont {Hauke}}, \
  and\ \bibinfo {author} {\bibfnamefont {P.}~\bibnamefont {Zoller}},\
  }\href@noop {} {\bibfield  {journal} {\bibinfo  {journal} {Sci. Adv.}\
  }\textbf {\bibinfo {volume} {1}},\ \bibinfo {pages} {e1500838} (\bibinfo
  {year} {2015})}\BibitemShut {NoStop}%
\bibitem [{\citenamefont {Vinci}\ \emph {et~al.}(2015)\citenamefont {Vinci},
  \citenamefont {Albash}, \citenamefont {Paz-Silva}, \citenamefont {Hen},\ and\
  \citenamefont {Lidar}}]{Vinci2015}%
  \BibitemOpen
  \bibfield  {author} {\bibinfo {author} {\bibfnamefont {W.}~\bibnamefont
  {Vinci}}, \bibinfo {author} {\bibfnamefont {T.}~\bibnamefont {Albash}},
  \bibinfo {author} {\bibfnamefont {G.}~\bibnamefont {Paz-Silva}}, \bibinfo
  {author} {\bibfnamefont {I.}~\bibnamefont {Hen}}, \ and\ \bibinfo {author}
  {\bibfnamefont {D.~A.}\ \bibnamefont {Lidar}},\ }\href {\doibase
  10.1103/PhysRevA.92.042310} {\bibfield  {journal} {\bibinfo  {journal} {Phys.
  Rev. A}\ }\textbf {\bibinfo {volume} {92}},\ \bibinfo {pages} {042310}
  (\bibinfo {year} {2015})}\BibitemShut {NoStop}%
\bibitem [{\citenamefont {Sourlas}(2005)}]{Sourlas2005}%
  \BibitemOpen
  \bibfield  {author} {\bibinfo {author} {\bibfnamefont {N.}~\bibnamefont
  {Sourlas}},\ }\href {\doibase 10.1103/PhysRevLett.94.070601} {\bibfield
  {journal} {\bibinfo  {journal} {Phys. Rev. Lett.}\ }\textbf {\bibinfo
  {volume} {94}},\ \bibinfo {pages} {070601} (\bibinfo {year}
  {2005})}\BibitemShut {NoStop}%
\bibitem [{\citenamefont {Puri}\ \emph {et~al.}(2017)\citenamefont {Puri},
  \citenamefont {Andersen}, \citenamefont {Grimsmo},\ and\ \citenamefont
  {Blais}}]{Puri2017}%
  \BibitemOpen
  \bibfield  {author} {\bibinfo {author} {\bibfnamefont {S.}~\bibnamefont
  {Puri}}, \bibinfo {author} {\bibfnamefont {C.~K.}\ \bibnamefont {Andersen}},
  \bibinfo {author} {\bibfnamefont {A.~L.}\ \bibnamefont {Grimsmo}}, \ and\
  \bibinfo {author} {\bibfnamefont {A.}~\bibnamefont {Blais}},\ }\href
  {\doibase 10.1038/ncomms15785} {\bibfield  {journal} {\bibinfo  {journal}
  {Nat. Commun.}\ }\textbf {\bibinfo {volume} {8}},\ \bibinfo {pages} {1}
  (\bibinfo {year} {2017})}\BibitemShut {NoStop}%
\bibitem [{\citenamefont {Albash}\ \emph {et~al.}(2016)\citenamefont {Albash},
  \citenamefont {Vinci}, ,\ and\ \citenamefont {Lidar}}]{Albash16}%
  \BibitemOpen
  \bibfield  {author} {\bibinfo {author} {\bibfnamefont {T.}~\bibnamefont
  {Albash}}, \bibinfo {author} {\bibfnamefont {W.}~\bibnamefont {Vinci}}, , \
  and\ \bibinfo {author} {\bibfnamefont {D.~A.}\ \bibnamefont {Lidar}},\
  }\href@noop {} {\bibfield  {journal} {\bibinfo  {journal} {Phys. Rev. A}\
  }\textbf {\bibinfo {volume} {94}},\ \bibinfo {pages} {022327} (\bibinfo
  {year} {2016})}\BibitemShut {NoStop}%
\bibitem [{\citenamefont {Ding}\ \emph {et~al.}(2017)\citenamefont {Ding},
  \citenamefont {Maslennikov}, \citenamefont {Habl\"utzel}, \citenamefont
  {Loh},\ and\ \citenamefont {Matsukevich}}]{Ding2017}%
  \BibitemOpen
  \bibfield  {author} {\bibinfo {author} {\bibfnamefont {S.}~\bibnamefont
  {Ding}}, \bibinfo {author} {\bibfnamefont {G.}~\bibnamefont {Maslennikov}},
  \bibinfo {author} {\bibfnamefont {R.}~\bibnamefont {Habl\"utzel}}, \bibinfo
  {author} {\bibfnamefont {H.}~\bibnamefont {Loh}}, \ and\ \bibinfo {author}
  {\bibfnamefont {D.}~\bibnamefont {Matsukevich}},\ }\href {\doibase
  10.1103/PhysRevLett.119.150404} {\bibfield  {journal} {\bibinfo  {journal}
  {Phys. Rev. Lett.}\ }\textbf {\bibinfo {volume} {119}},\ \bibinfo {pages}
  {150404} (\bibinfo {year} {2017})}\BibitemShut {NoStop}%
\bibitem [{\citenamefont {Grimm}\ \emph {et~al.}(2020)\citenamefont {Grimm},
  \citenamefont {Frattini}, \citenamefont {Puri}, \citenamefont {Mundhada},
  \citenamefont {Touzard}, \citenamefont {Mirrahimi}, \citenamefont {Girvin},
  \citenamefont {Shankar},\ and\ \citenamefont {Devoret}}]{Grimm2020}%
  \BibitemOpen
  \bibfield  {author} {\bibinfo {author} {\bibfnamefont {A.}~\bibnamefont
  {Grimm}}, \bibinfo {author} {\bibfnamefont {N.~E.}\ \bibnamefont {Frattini}},
  \bibinfo {author} {\bibfnamefont {S.}~\bibnamefont {Puri}}, \bibinfo {author}
  {\bibfnamefont {S.~O.}\ \bibnamefont {Mundhada}}, \bibinfo {author}
  {\bibfnamefont {S.}~\bibnamefont {Touzard}}, \bibinfo {author} {\bibfnamefont
  {M.}~\bibnamefont {Mirrahimi}}, \bibinfo {author} {\bibfnamefont {S.~M.}\
  \bibnamefont {Girvin}}, \bibinfo {author} {\bibfnamefont {S.}~\bibnamefont
  {Shankar}}, \ and\ \bibinfo {author} {\bibfnamefont {M.~H.}\ \bibnamefont
  {Devoret}},\ }\href {https://doi.org/10.1038/s41586-020-2587-z} {\bibfield
  {journal} {\bibinfo  {journal} {Nature}\ }\textbf {\bibinfo {volume} {584}},\
  \bibinfo {pages} {205} (\bibinfo {year} {2020})}\BibitemShut {NoStop}%
\bibitem [{\citenamefont {Frattini}\ \emph {et~al.}()\citenamefont {Frattini},
  \citenamefont {Corti{\~n}as}, \citenamefont {Venkatraman}, \citenamefont
  {Xiao}, \citenamefont {Su}, \citenamefont {Lei}, \citenamefont {Chapman},
  \citenamefont {Joshi}, \citenamefont {Girvin}, \citenamefont {Schoelkopf},
  \citenamefont {Puri},\ and\ \citenamefont {Devoret}}]{Frattini2022}%
  \BibitemOpen
  \bibfield  {author} {\bibinfo {author} {\bibfnamefont {N.~E.}\ \bibnamefont
  {Frattini}}, \bibinfo {author} {\bibfnamefont {R.~G.}\ \bibnamefont
  {Corti{\~n}as}}, \bibinfo {author} {\bibfnamefont {J.}~\bibnamefont
  {Venkatraman}}, \bibinfo {author} {\bibfnamefont {X.}~\bibnamefont {Xiao}},
  \bibinfo {author} {\bibfnamefont {Q.}~\bibnamefont {Su}}, \bibinfo {author}
  {\bibfnamefont {C.~U.}\ \bibnamefont {Lei}}, \bibinfo {author} {\bibfnamefont
  {B.~J.}\ \bibnamefont {Chapman}}, \bibinfo {author} {\bibfnamefont {V.~R.}\
  \bibnamefont {Joshi}}, \bibinfo {author} {\bibfnamefont {S.~M.}\ \bibnamefont
  {Girvin}}, \bibinfo {author} {\bibfnamefont {R.~J.}\ \bibnamefont
  {Schoelkopf}}, \bibinfo {author} {\bibfnamefont {S.}~\bibnamefont {Puri}}, \
  and\ \bibinfo {author} {\bibfnamefont {M.~H.}\ \bibnamefont {Devoret}},\
  }\href {https://arxiv.org/abs/2209.03934} {\enquote {\bibinfo {title} {The
  squeezed kerr oscillator: spectral kissing and phase-flip robustness},}\
  }\Eprint {http://arxiv.org/abs/arXiv:2209.03934v1 [quant-ph]}
  {arXiv:2209.03934v1 [quant-ph]} \BibitemShut {NoStop}%
\bibitem [{\citenamefont {Wang}\ \emph {et~al.}(2019)\citenamefont {Wang},
  \citenamefont {Pachal}, \citenamefont {Wollback},\ and\ \citenamefont
  {Arrangoiz-Arriola}}]{Wang2019}%
  \BibitemOpen
  \bibfield  {author} {\bibinfo {author} {\bibfnamefont {Z.}~\bibnamefont
  {Wang}}, \bibinfo {author} {\bibfnamefont {M.}~\bibnamefont {Pachal}},
  \bibinfo {author} {\bibfnamefont {E.~A.}\ \bibnamefont {Wollback}}, \ and\
  \bibinfo {author} {\bibfnamefont {P.}~\bibnamefont {Arrangoiz-Arriola}},\
  }\href {https://journals.aps.org/prx/pdf/10.1103/PhysRevX.9.021049}
  {\bibfield  {journal} {\bibinfo  {journal} {Phys. Rev. X}\ }\textbf {\bibinfo
  {volume} {9}},\ \bibinfo {pages} {021049} (\bibinfo {year}
  {2019})}\BibitemShut {NoStop}%
\bibitem [{\citenamefont {Yamaji}\ \emph {et~al.}(2022)\citenamefont {Yamaji},
  \citenamefont {Kagami}, \citenamefont {Yamaguchi}, \citenamefont {Satoh},
  \citenamefont {Koshino}, \citenamefont {Goto}, \citenamefont {Lin},
  \citenamefont {Nakamura},\ and\ \citenamefont {Yamamoto}}]{Yamaji2022}%
  \BibitemOpen
  \bibfield  {author} {\bibinfo {author} {\bibfnamefont {T.}~\bibnamefont
  {Yamaji}}, \bibinfo {author} {\bibfnamefont {S.}~\bibnamefont {Kagami}},
  \bibinfo {author} {\bibfnamefont {A.}~\bibnamefont {Yamaguchi}}, \bibinfo
  {author} {\bibfnamefont {T.}~\bibnamefont {Satoh}}, \bibinfo {author}
  {\bibfnamefont {K.}~\bibnamefont {Koshino}}, \bibinfo {author} {\bibfnamefont
  {H.}~\bibnamefont {Goto}}, \bibinfo {author} {\bibfnamefont {Z.~R.}\
  \bibnamefont {Lin}}, \bibinfo {author} {\bibfnamefont {Y.}~\bibnamefont
  {Nakamura}}, \ and\ \bibinfo {author} {\bibfnamefont {T.}~\bibnamefont
  {Yamamoto}},\ }\href {\doibase 10.1103/PhysRevA.105.023519} {\bibfield
  {journal} {\bibinfo  {journal} {Phys. Rev. A}\ }\textbf {\bibinfo {volume}
  {105}},\ \bibinfo {pages} {023519} (\bibinfo {year} {2022})}\BibitemShut
  {NoStop}%
\bibitem [{\citenamefont {Yamaji}\ \emph {et~al.}(2023)\citenamefont {Yamaji},
  \citenamefont {Masuda}, \citenamefont {Yamaguchi}, \citenamefont {Satoh},
  \citenamefont {Morioka}, \citenamefont {Igarashi}, \citenamefont {Shirane},\
  and\ \citenamefont {Yamamoto}}]{Yamaji2023}%
  \BibitemOpen
  \bibfield  {author} {\bibinfo {author} {\bibfnamefont {T.}~\bibnamefont
  {Yamaji}}, \bibinfo {author} {\bibfnamefont {S.}~\bibnamefont {Masuda}},
  \bibinfo {author} {\bibfnamefont {A.}~\bibnamefont {Yamaguchi}}, \bibinfo
  {author} {\bibfnamefont {T.}~\bibnamefont {Satoh}}, \bibinfo {author}
  {\bibfnamefont {A.}~\bibnamefont {Morioka}}, \bibinfo {author} {\bibfnamefont
  {Y.}~\bibnamefont {Igarashi}}, \bibinfo {author} {\bibfnamefont
  {M.}~\bibnamefont {Shirane}}, \ and\ \bibinfo {author} {\bibfnamefont
  {T.}~\bibnamefont {Yamamoto}},\ }\href {\doibase
  10.1103/PhysRevApplied.20.014057} {\bibfield  {journal} {\bibinfo  {journal}
  {Phys. Rev. Appl.}\ }\textbf {\bibinfo {volume} {20}},\ \bibinfo {pages}
  {014057} (\bibinfo {year} {2023})}\BibitemShut {NoStop}%
\bibitem [{\citenamefont {Chancellor}\ \emph {et~al.}(2017)\citenamefont
  {Chancellor}, \citenamefont {Zohren},\ and\ \citenamefont
  {Warburton}}]{Chancellor2017}%
  \BibitemOpen
  \bibfield  {author} {\bibinfo {author} {\bibfnamefont {N.}~\bibnamefont
  {Chancellor}}, \bibinfo {author} {\bibfnamefont {S.}~\bibnamefont {Zohren}},
  \ and\ \bibinfo {author} {\bibfnamefont {P.~A.}\ \bibnamefont {Warburton}},\
  }\href {\doibase 10.1038/s41534-017-0022-6} {\bibfield  {journal} {\bibinfo
  {journal} {NPJ Quantum Inf.}\ }\textbf {\bibinfo {volume} {3}},\ \bibinfo
  {pages} {1} (\bibinfo {year} {2017})}\BibitemShut {NoStop}%
\bibitem [{\citenamefont {Jordan}\ \emph {et~al.}(2006)\citenamefont {Jordan},
  \citenamefont {Farhi},\ and\ \citenamefont {Shor}}]{Jordan2006}%
  \BibitemOpen
  \bibfield  {author} {\bibinfo {author} {\bibfnamefont {S.~P.}\ \bibnamefont
  {Jordan}}, \bibinfo {author} {\bibfnamefont {E.}~\bibnamefont {Farhi}}, \
  and\ \bibinfo {author} {\bibfnamefont {P.~W.}\ \bibnamefont {Shor}},\ }\href
  {\doibase 10.1103/PhysRevA.74.052322} {\bibfield  {journal} {\bibinfo
  {journal} {Phys. Rev. A}\ }\textbf {\bibinfo {volume} {74}},\ \bibinfo
  {pages} {052322} (\bibinfo {year} {2006})}\BibitemShut {NoStop}%
\bibitem [{\citenamefont {Young}\ \emph
  {et~al.}(2013{\natexlab{a}})\citenamefont {Young}, \citenamefont
  {Blume-Kohout},\ and\ \citenamefont {Lidar}}]{Young2013}%
  \BibitemOpen
  \bibfield  {author} {\bibinfo {author} {\bibfnamefont {K.~C.}\ \bibnamefont
  {Young}}, \bibinfo {author} {\bibfnamefont {R.}~\bibnamefont {Blume-Kohout}},
  \ and\ \bibinfo {author} {\bibfnamefont {D.~A.}\ \bibnamefont {Lidar}},\
  }\href {\doibase 10.1103/PhysRevA.88.062314} {\bibfield  {journal} {\bibinfo
  {journal} {Phys. Rev. A}\ }\textbf {\bibinfo {volume} {88}},\ \bibinfo
  {pages} {062314} (\bibinfo {year} {2013}{\natexlab{a}})}\BibitemShut
  {NoStop}%
\bibitem [{\citenamefont {Pudenz}\ \emph {et~al.}(2014)\citenamefont {Pudenz},
  \citenamefont {Albash},\ and\ \citenamefont {Lidar}}]{Pudenz2014}%
  \BibitemOpen
  \bibfield  {author} {\bibinfo {author} {\bibfnamefont {K.~L.}\ \bibnamefont
  {Pudenz}}, \bibinfo {author} {\bibfnamefont {T.}~\bibnamefont {Albash}}, \
  and\ \bibinfo {author} {\bibfnamefont {D.~A.}\ \bibnamefont {Lidar}},\ }\href
  {\doibase 10.1038/ncomms4243} {\bibfield  {journal} {\bibinfo  {journal}
  {Nat. Commun.}\ }\textbf {\bibinfo {volume} {5}},\ \bibinfo {pages} {3243}
  (\bibinfo {year} {2014})}\BibitemShut {NoStop}%
\bibitem [{\citenamefont {Pudenz}\ \emph {et~al.}(2015)\citenamefont {Pudenz},
  \citenamefont {Albash},\ and\ \citenamefont {Lidar}}]{Pudenz2015}%
  \BibitemOpen
  \bibfield  {author} {\bibinfo {author} {\bibfnamefont {K.~L.}\ \bibnamefont
  {Pudenz}}, \bibinfo {author} {\bibfnamefont {T.}~\bibnamefont {Albash}}, \
  and\ \bibinfo {author} {\bibfnamefont {D.~A.}\ \bibnamefont {Lidar}},\ }\href
  {\doibase 10.1103/PhysRevA.91.042302} {\bibfield  {journal} {\bibinfo
  {journal} {Phys. Rev. A}\ }\textbf {\bibinfo {volume} {91}},\ \bibinfo
  {pages} {042302} (\bibinfo {year} {2015})}\BibitemShut {NoStop}%
\bibitem [{\citenamefont {Bookatz}\ \emph {et~al.}(2015)\citenamefont
  {Bookatz}, \citenamefont {Farhi},\ and\ \citenamefont {Zhou}}]{Bookatz2015}%
  \BibitemOpen
  \bibfield  {author} {\bibinfo {author} {\bibfnamefont {A.~D.}\ \bibnamefont
  {Bookatz}}, \bibinfo {author} {\bibfnamefont {E.}~\bibnamefont {Farhi}}, \
  and\ \bibinfo {author} {\bibfnamefont {L.}~\bibnamefont {Zhou}},\ }\href
  {\doibase 10.1103/PhysRevA.92.022317} {\bibfield  {journal} {\bibinfo
  {journal} {Phys. Rev. A}\ }\textbf {\bibinfo {volume} {92}},\ \bibinfo
  {pages} {022317} (\bibinfo {year} {2015})}\BibitemShut {NoStop}%
\bibitem [{\citenamefont {Matsuura}\ \emph {et~al.}(2016)\citenamefont
  {Matsuura}, \citenamefont {Nishimori}, \citenamefont {Albash},\ and\
  \citenamefont {Lidar}}]{Matsuura2016}%
  \BibitemOpen
  \bibfield  {author} {\bibinfo {author} {\bibfnamefont {S.}~\bibnamefont
  {Matsuura}}, \bibinfo {author} {\bibfnamefont {H.}~\bibnamefont {Nishimori}},
  \bibinfo {author} {\bibfnamefont {T.}~\bibnamefont {Albash}}, \ and\ \bibinfo
  {author} {\bibfnamefont {D.~A.}\ \bibnamefont {Lidar}},\ }\href {\doibase
  10.1103/PhysRevLett.116.220501} {\bibfield  {journal} {\bibinfo  {journal}
  {Phys. Rev. Lett.}\ }\textbf {\bibinfo {volume} {116}},\ \bibinfo {pages}
  {220501} (\bibinfo {year} {2016})}\BibitemShut {NoStop}%
\bibitem [{\citenamefont {Matsuura}\ \emph {et~al.}(2017)\citenamefont
  {Matsuura}, \citenamefont {Nishimori}, \citenamefont {Vinci}, \citenamefont
  {Albash},\ and\ \citenamefont {Lidar}}]{Matsuura2017}%
  \BibitemOpen
  \bibfield  {author} {\bibinfo {author} {\bibfnamefont {S.}~\bibnamefont
  {Matsuura}}, \bibinfo {author} {\bibfnamefont {H.}~\bibnamefont {Nishimori}},
  \bibinfo {author} {\bibfnamefont {W.}~\bibnamefont {Vinci}}, \bibinfo
  {author} {\bibfnamefont {T.}~\bibnamefont {Albash}}, \ and\ \bibinfo {author}
  {\bibfnamefont {D.~A.}\ \bibnamefont {Lidar}},\ }\href {\doibase
  10.1103/PhysRevA.95.022308} {\bibfield  {journal} {\bibinfo  {journal} {Phys.
  Rev. A}\ }\textbf {\bibinfo {volume} {95}},\ \bibinfo {pages} {22308}
  (\bibinfo {year} {2017})}\BibitemShut {NoStop}%
\bibitem [{\citenamefont {Pearson}\ \emph {et~al.}(2019)\citenamefont
  {Pearson}, \citenamefont {Mishra}, \citenamefont {Hen},\ and\ \citenamefont
  {Lidar}}]{Pearson2019}%
  \BibitemOpen
  \bibfield  {author} {\bibinfo {author} {\bibfnamefont {A.}~\bibnamefont
  {Pearson}}, \bibinfo {author} {\bibfnamefont {A.}~\bibnamefont {Mishra}},
  \bibinfo {author} {\bibfnamefont {I.}~\bibnamefont {Hen}}, \ and\ \bibinfo
  {author} {\bibfnamefont {D.~A.}\ \bibnamefont {Lidar}},\ }\href@noop {}
  {\bibfield  {journal} {\bibinfo  {journal} {NPJ Quantum Inf.}\ }\textbf
  {\bibinfo {volume} {5}} (\bibinfo {year} {2019})}\BibitemShut {NoStop}%
\bibitem [{\citenamefont {Pastawski}\ and\ \citenamefont
  {Preskill}(2016)}]{Pastawski2016}%
  \BibitemOpen
  \bibfield  {author} {\bibinfo {author} {\bibfnamefont {F.}~\bibnamefont
  {Pastawski}}\ and\ \bibinfo {author} {\bibfnamefont {J.}~\bibnamefont
  {Preskill}},\ }\href {\doibase 10.1103/PhysRevA.93.052325} {\bibfield
  {journal} {\bibinfo  {journal} {Phys. Rev. A}\ }\textbf {\bibinfo {volume}
  {93}},\ \bibinfo {pages} {52325} (\bibinfo {year} {2016})}\BibitemShut
  {NoStop}%
\bibitem [{\citenamefont {Massey}(1962)}]{Massey1963}%
  \BibitemOpen
  \bibfield  {author} {\bibinfo {author} {\bibfnamefont {J.~L.}\ \bibnamefont
  {Massey}},\ }\emph {\bibinfo {title} {Threshold decoding}},\ \href@noop {}
  {\bibinfo {type} {{Ph.D.} thesis}},\ \bibinfo  {school} {Massachusetts
  Institute of Technology}, \bibinfo {address} {Research Laboratory of
  Electronics} (\bibinfo {year} {1962})\BibitemShut {NoStop}%
\bibitem [{\citenamefont {Massey}(1968)}]{Massey1968}%
  \BibitemOpen
  \bibfield  {author} {\bibinfo {author} {\bibfnamefont {J.~L.}\ \bibnamefont
  {Massey}},\ }in\ \href@noop {} {\emph {\bibinfo {booktitle} {Advances in
  Communication Systems}}},\ Vol.~\bibinfo {volume} {3},\ \bibinfo {editor}
  {edited by\ \bibinfo {editor} {\bibfnamefont {A.}~\bibnamefont
  {Balakrishnan}}}\ (\bibinfo  {publisher} {Elsevier},\ \bibinfo {year}
  {1968})\ pp.\ \bibinfo {pages} {91--115}\BibitemShut {NoStop}%
\bibitem [{\citenamefont {Laferriere}(1977)}]{Laferriere1977}%
  \BibitemOpen
  \bibfield  {author} {\bibinfo {author} {\bibfnamefont {C.}~\bibnamefont
  {Laferriere}},\ }\href@noop {} {\enquote {\bibinfo {title} {Error correcting
  capability of 1-step majority logic decoding},}\ }\bibinfo {howpublished}
  {\url{http://hdl.handle.net/10393/10473}} (\bibinfo {year}
  {1977})\BibitemShut {NoStop}%
\bibitem [{\citenamefont {Clerk}\ and\ \citenamefont {Cain}(1981)}]{Clerk1981}%
  \BibitemOpen
  \bibfield  {author} {\bibinfo {author} {\bibfnamefont {G.~J.~C.}\
  \bibnamefont {Clerk}}\ and\ \bibinfo {author} {\bibfnamefont {B.~J.}\
  \bibnamefont {Cain}},\ }\href
  {https://vdoc.pub/documents/error-correction-coding-for-digital-communications-1rv49ejkds1o}
  {\emph {\bibinfo {title} {Error-Correction Coding for Digital
  Communications}}}\ (\bibinfo  {publisher} {Plenum Press},\ \bibinfo {year}
  {1981})\BibitemShut {NoStop}%
\bibitem [{\citenamefont {Shu}\ and\ \citenamefont {Costello}(2004)}]{Shu2004}%
  \BibitemOpen
  \bibfield  {author} {\bibinfo {author} {\bibfnamefont {L.}~\bibnamefont
  {Shu}}\ and\ \bibinfo {author} {\bibfnamefont {J.~J.~D.}\ \bibnamefont
  {Costello}},\ }\href@noop {} {\emph {\bibinfo {title} {Error Control Coding:
  fundamentals and applications}}},\ \bibinfo {edition} {2nd}\ ed.\ (\bibinfo
  {publisher} {Pearson Prentice Hall, NJ},\ \bibinfo {year} {2004})\BibitemShut
  {NoStop}%
\bibitem [{\citenamefont {Sourlas}(2001)}]{Sourlas2001}%
  \BibitemOpen
  \bibfield  {author} {\bibinfo {author} {\bibfnamefont {N.}~\bibnamefont
  {Sourlas}},\ }\href {www.elsevier.com/locate/physa} {\bibfield  {journal}
  {\bibinfo  {journal} {Physica A}\ }\textbf {\bibinfo {volume} {302}},\
  \bibinfo {pages} {14} (\bibinfo {year} {2001})}\BibitemShut {NoStop}%
\bibitem [{\citenamefont {Sourlas}(2002)}]{Sourlas2002}%
  \BibitemOpen
  \bibfield  {author} {\bibinfo {author} {\bibfnamefont {N.}~\bibnamefont
  {Sourlas}},\ }in\ \href
  {http://www.ens-lyon.fr/LIP/ISC/WSCS2002/slides/sourlas/WSCS-sourlas.pdf}
  {\emph {\bibinfo {booktitle} {Winter School on Complex Systems}}},\ \bibinfo
  {editor} {edited by\ \bibinfo {editor} {\bibfnamefont {J.}~\bibnamefont
  {Mazoyer}}, \bibinfo {editor} {\bibfnamefont {M.}~\bibnamefont {Morvan}}, \
  and\ \bibinfo {editor} {\bibfnamefont {N.}~\bibnamefont {Schabanel}}}\
  (\bibinfo {organization} {IXXI},\ \bibinfo {address} {Lyon},\ \bibinfo {year}
  {2002})\BibitemShut {NoStop}%
\bibitem [{\citenamefont {Gottesman}(1996)}]{Gottesman1996}%
  \BibitemOpen
  \bibfield  {author} {\bibinfo {author} {\bibfnamefont {D.}~\bibnamefont
  {Gottesman}},\ }\href {\doibase 10.1103/PhysRevA.54.1862} {\bibfield
  {journal} {\bibinfo  {journal} {Phys. Rev. A}\ }\textbf {\bibinfo {volume}
  {54}},\ \bibinfo {pages} {1862} (\bibinfo {year} {1996})}\BibitemShut
  {NoStop}%
\bibitem [{\citenamefont {Vicente}\ \emph {et~al.}(2003)\citenamefont
  {Vicente}, \citenamefont {Saad},\ and\ \citenamefont
  {Kabashima}}]{Vicente2003}%
  \BibitemOpen
  \bibfield  {author} {\bibinfo {author} {\bibfnamefont {R.}~\bibnamefont
  {Vicente}}, \bibinfo {author} {\bibfnamefont {D.}~\bibnamefont {Saad}}, \
  and\ \bibinfo {author} {\bibfnamefont {Y.}~\bibnamefont {Kabashima}},\ }in\
  \href {\doibase 10.1016/S1076-5670(02)80018-0} {\emph {\bibinfo {booktitle}
  {Advances in Imaging and Electron Physics}}},\ Vol.\ \bibinfo {volume} {125}\
  (\bibinfo  {publisher} {Elsevier},\ \bibinfo {year} {2003})\ pp.\ \bibinfo
  {pages} {231--353}\BibitemShut {NoStop}%
\bibitem [{\citenamefont {Nambu}(2022)}]{Nambu22}%
  \BibitemOpen
  \bibfield  {author} {\bibinfo {author} {\bibfnamefont {Y.}~\bibnamefont
  {Nambu}},\ }\href@noop {} {\bibfield  {journal} {\bibinfo  {journal} {IEEE
  Access}\ }\textbf {\bibinfo {volume} {10}},\ \bibinfo {pages} {84279}
  (\bibinfo {year} {2022})}\BibitemShut {NoStop}%
\bibitem [{\citenamefont {Rocchetto}\ \emph {et~al.}(2016)\citenamefont
  {Rocchetto}, \citenamefont {Benjamin},\ and\ \citenamefont
  {Li}}]{Rocchetto2016}%
  \BibitemOpen
  \bibfield  {author} {\bibinfo {author} {\bibfnamefont {A.}~\bibnamefont
  {Rocchetto}}, \bibinfo {author} {\bibfnamefont {S.~C.}\ \bibnamefont
  {Benjamin}}, \ and\ \bibinfo {author} {\bibfnamefont {Y.}~\bibnamefont
  {Li}},\ }\href {https://www.science.org/doi/10.1126/sciadv.1601246}
  {\bibfield  {journal} {\bibinfo  {journal} {Sci. Adv.}\ }\textbf {\bibinfo
  {volume} {2}} (\bibinfo {year} {2016})}\BibitemShut {NoStop}%
\bibitem [{\citenamefont {Ender}\ \emph {et~al.}(2022)\citenamefont {Ender},
  \citenamefont {Messinger}, \citenamefont {Fellner}, \citenamefont {Dlaska},\
  and\ \citenamefont {Lechner}}]{Ender2022}%
  \BibitemOpen
  \bibfield  {author} {\bibinfo {author} {\bibfnamefont {K.}~\bibnamefont
  {Ender}}, \bibinfo {author} {\bibfnamefont {A.}~\bibnamefont {Messinger}},
  \bibinfo {author} {\bibfnamefont {M.}~\bibnamefont {Fellner}}, \bibinfo
  {author} {\bibfnamefont {C.}~\bibnamefont {Dlaska}}, \ and\ \bibinfo {author}
  {\bibfnamefont {W.}~\bibnamefont {Lechner}},\ }\href {\doibase
  10.1103/PRXQuantum.3.030304} {\bibfield  {journal} {\bibinfo  {journal} {PRX
  Quantum}\ }\textbf {\bibinfo {volume} {3}},\ \bibinfo {pages} {030304}
  (\bibinfo {year} {2022})}\BibitemShut {NoStop}%
\bibitem [{\citenamefont {Fellner}\ \emph {et~al.}(2022)\citenamefont
  {Fellner}, \citenamefont {Messinger}, \citenamefont {Ender},\ and\
  \citenamefont {Lechner}}]{Fellner2022}%
  \BibitemOpen
  \bibfield  {author} {\bibinfo {author} {\bibfnamefont {M.}~\bibnamefont
  {Fellner}}, \bibinfo {author} {\bibfnamefont {A.}~\bibnamefont {Messinger}},
  \bibinfo {author} {\bibfnamefont {K.}~\bibnamefont {Ender}}, \ and\ \bibinfo
  {author} {\bibfnamefont {W.}~\bibnamefont {Lechner}},\ }\href {\doibase
  10.1103/PhysRevLett.129.180503} {\bibfield  {journal} {\bibinfo  {journal}
  {Phys. Rev. Lett.}\ }\textbf {\bibinfo {volume} {129}},\ \bibinfo {pages}
  {180503} (\bibinfo {year} {2022})}\BibitemShut {NoStop}%
\bibitem [{\citenamefont {Fuchs}\ \emph {et~al.}(2022)\citenamefont {Fuchs},
  \citenamefont {Lye}, \citenamefont {Nilsen}, \citenamefont {Stasik},\ and\
  \citenamefont {Sartor}}]{Fuchs2022}%
  \BibitemOpen
  \bibfield  {author} {\bibinfo {author} {\bibfnamefont {F.~G.}\ \bibnamefont
  {Fuchs}}, \bibinfo {author} {\bibfnamefont {K.~O.}\ \bibnamefont {Lye}},
  \bibinfo {author} {\bibfnamefont {H.~M.}\ \bibnamefont {Nilsen}}, \bibinfo
  {author} {\bibfnamefont {A.~J.}\ \bibnamefont {Stasik}}, \ and\ \bibinfo
  {author} {\bibfnamefont {G.}~\bibnamefont {Sartor}},\ }\href {\doibase
  10.3390/A15060202} {\bibfield  {journal} {\bibinfo  {journal} {Algorithms}\
  }\textbf {\bibinfo {volume} {15}},\ \bibinfo {pages} {202} (\bibinfo {year}
  {2022})}\BibitemShut {NoStop}%
\bibitem [{\citenamefont {Weidinger}\ \emph {et~al.}(2023)\citenamefont
  {Weidinger}, \citenamefont {Mbeng},\ and\ \citenamefont
  {Lechner}}]{Weidinger2023}%
  \BibitemOpen
  \bibfield  {author} {\bibinfo {author} {\bibfnamefont {A.}~\bibnamefont
  {Weidinger}}, \bibinfo {author} {\bibfnamefont {G.~B.}\ \bibnamefont
  {Mbeng}}, \ and\ \bibinfo {author} {\bibfnamefont {W.}~\bibnamefont
  {Lechner}},\ }\href {http://arxiv.org/abs/2301.05042} {\  (\bibinfo {year}
  {2023})},\ \Eprint {http://arxiv.org/abs/arXiv:2301.05042v1 [quant-ph]}
  {arXiv:2301.05042v1 [quant-ph]} \BibitemShut {NoStop}%
\bibitem [{\citenamefont {Messinger}\ \emph {et~al.}(2023)\citenamefont
  {Messinger}, \citenamefont {Fellner},\ and\ \citenamefont
  {Lechner}}]{Messinger2023}%
  \BibitemOpen
  \bibfield  {author} {\bibinfo {author} {\bibfnamefont {A.}~\bibnamefont
  {Messinger}}, \bibinfo {author} {\bibfnamefont {M.}~\bibnamefont {Fellner}},
  \ and\ \bibinfo {author} {\bibfnamefont {W.}~\bibnamefont {Lechner}},\ }\href
  {http://arxiv.org/abs/2303.08602} {\enquote {\bibinfo {title} {Constant depth
  code deformations in the parity architecture},}\ } (\bibinfo {year} {2023}),\
  \Eprint {http://arxiv.org/abs/arXiv:2303.08602v1 [quant-ph]}
  {arXiv:2303.08602v1 [quant-ph]} \BibitemShut {NoStop}%
\bibitem [{\citenamefont {Konz}\ \emph {et~al.}(2019)\citenamefont {Konz},
  \citenamefont {Mazzola}, \citenamefont {Ochoa}, \citenamefont {Katzgraber},\
  and\ \citenamefont {Troyer}}]{Konz2019}%
  \BibitemOpen
  \bibfield  {author} {\bibinfo {author} {\bibfnamefont {M.~S.}\ \bibnamefont
  {Konz}}, \bibinfo {author} {\bibfnamefont {G.}~\bibnamefont {Mazzola}},
  \bibinfo {author} {\bibfnamefont {A.~J.}\ \bibnamefont {Ochoa}}, \bibinfo
  {author} {\bibfnamefont {H.~G.}\ \bibnamefont {Katzgraber}}, \ and\ \bibinfo
  {author} {\bibfnamefont {M.}~\bibnamefont {Troyer}},\ }\href@noop {}
  {\bibfield  {journal} {\bibinfo  {journal} {Phys. Rev. A}\ }\textbf {\bibinfo
  {volume} {100}} (\bibinfo {year} {2019})}\BibitemShut {NoStop}%
\bibitem [{\citenamefont {Marto\v{n}\'{a}k}\ \emph {et~al.}(2002)\citenamefont
  {Marto\v{n}\'{a}k}, \citenamefont {Santoro},\ and\ \citenamefont
  {Tosatti}}]{Marto?ak2002}%
  \BibitemOpen
  \bibfield  {author} {\bibinfo {author} {\bibfnamefont {R.}~\bibnamefont
  {Marto\v{n}\'{a}k}}, \bibinfo {author} {\bibfnamefont {G.~E.}\ \bibnamefont
  {Santoro}}, \ and\ \bibinfo {author} {\bibfnamefont {E.}~\bibnamefont
  {Tosatti}},\ }\href {\doibase 10.1103/PhysRevB.66.094203} {\bibfield
  {journal} {\bibinfo  {journal} {Phys. Rev. B}\ }\textbf {\bibinfo {volume}
  {66}},\ \bibinfo {pages} {094203} (\bibinfo {year} {2002})}\BibitemShut
  {NoStop}%
\bibitem [{\citenamefont {Heim}\ \emph {et~al.}(2015)\citenamefont {Heim},
  \citenamefont {Ronnow}, \citenamefont {Isakov},\ and\ \citenamefont
  {Troyer}}]{Heim2015}%
  \BibitemOpen
  \bibfield  {author} {\bibinfo {author} {\bibfnamefont {B.}~\bibnamefont
  {Heim}}, \bibinfo {author} {\bibfnamefont {T.~F.}\ \bibnamefont {Ronnow}},
  \bibinfo {author} {\bibfnamefont {S.~V.}\ \bibnamefont {Isakov}}, \ and\
  \bibinfo {author} {\bibfnamefont {M.}~\bibnamefont {Troyer}},\ }\href
  {\doibase 10.1126/science.aaa4170} {\bibfield  {journal} {\bibinfo  {journal}
  {Science}\ }\textbf {\bibinfo {volume} {348}},\ \bibinfo {pages} {215}
  (\bibinfo {year} {2015})}\BibitemShut {NoStop}%
\bibitem [{\citenamefont {Waidyasooriya}\ and\ \citenamefont
  {Hariyama}(2020)}]{Waidyasooriya2020}%
  \BibitemOpen
  \bibfield  {author} {\bibinfo {author} {\bibfnamefont {H.~M.}\ \bibnamefont
  {Waidyasooriya}}\ and\ \bibinfo {author} {\bibfnamefont {M.}~\bibnamefont
  {Hariyama}},\ }\href {\doibase 10.1109/ACCESS.2020.2985699} {\bibfield
  {journal} {\bibinfo  {journal} {IEEE Access}\ }\textbf {\bibinfo {volume}
  {8}},\ \bibinfo {pages} {67929} (\bibinfo {year} {2020})}\BibitemShut
  {NoStop}%
\bibitem [{\citenamefont {Hu}\ and\ \citenamefont {Wang}(2021)}]{Hu2021}%
  \BibitemOpen
  \bibfield  {author} {\bibinfo {author} {\bibfnamefont {J.}~\bibnamefont
  {Hu}}\ and\ \bibinfo {author} {\bibfnamefont {Y.}~\bibnamefont {Wang}},\
  }\href {\doibase 10.1080/10618600.2020.1814787} {\bibfield  {journal}
  {\bibinfo  {journal} {J. Comput. Graph. Stat.}\ }\textbf {\bibinfo {volume}
  {30}},\ \bibinfo {pages} {284} (\bibinfo {year} {2021})}\BibitemShut
  {NoStop}%
\bibitem [{\citenamefont {Vinci}\ \emph {et~al.}(2016)\citenamefont {Vinci},
  \citenamefont {Albash},\ and\ \citenamefont {Lidar}}]{Vinci2016}%
  \BibitemOpen
  \bibfield  {author} {\bibinfo {author} {\bibfnamefont {W.}~\bibnamefont
  {Vinci}}, \bibinfo {author} {\bibfnamefont {T.}~\bibnamefont {Albash}}, \
  and\ \bibinfo {author} {\bibfnamefont {D.~A.}\ \bibnamefont {Lidar}},\ }\href
  {\doibase 10.1038/npjqi.2016.17} {\bibfield  {journal} {\bibinfo  {journal}
  {NPJ Quantum Inf.}\ }\textbf {\bibinfo {volume} {2}},\ \bibinfo {pages} {1}
  (\bibinfo {year} {2016})}\BibitemShut {NoStop}%
\bibitem [{\citenamefont {Dodds}\ \emph {et~al.}(2019)\citenamefont {Dodds},
  \citenamefont {Kendon}, \citenamefont {Adams},\ and\ \citenamefont
  {Chancellor}}]{Dodds2019}%
  \BibitemOpen
  \bibfield  {author} {\bibinfo {author} {\bibfnamefont {A.~B.}\ \bibnamefont
  {Dodds}}, \bibinfo {author} {\bibfnamefont {V.}~\bibnamefont {Kendon}},
  \bibinfo {author} {\bibfnamefont {C.~S.}\ \bibnamefont {Adams}}, \ and\
  \bibinfo {author} {\bibfnamefont {N.}~\bibnamefont {Chancellor}},\ }\href
  {\doibase 10.1103/PHYSREVA.100.032320/FIGURES/4/MEDIUM} {\bibfield  {journal}
  {\bibinfo  {journal} {Phys. Rev. A}\ }\textbf {\bibinfo {volume} {100}},\
  \bibinfo {pages} {032320} (\bibinfo {year} {2019})}\BibitemShut {NoStop}%
\bibitem [{\citenamefont {Nishimori}(1993)}]{Nishimori1993}%
  \BibitemOpen
  \bibfield  {author} {\bibinfo {author} {\bibfnamefont {H.}~\bibnamefont
  {Nishimori}},\ }\href {\doibase 10.1143/JPSJ.62.2973} {\bibfield  {journal}
  {\bibinfo  {journal} {J. of Phys. Soc. Japan}\ }\textbf {\bibinfo {volume}
  {62}},\ \bibinfo {pages} {2973} (\bibinfo {year} {1993})}\BibitemShut
  {NoStop}%
\bibitem [{\citenamefont {Rujan}(1993)}]{Rujan1993}%
  \BibitemOpen
  \bibfield  {author} {\bibinfo {author} {\bibfnamefont {P.}~\bibnamefont
  {Rujan}},\ }\href {\doibase 10.1103/PhysRevLett.70.2968} {\bibfield
  {journal} {\bibinfo  {journal} {Phys. Rev. Lett.}\ }\textbf {\bibinfo
  {volume} {70}},\ \bibinfo {pages} {2968} (\bibinfo {year}
  {1993})}\BibitemShut {NoStop}%
\bibitem [{\citenamefont {Nishimori}\ and\ \citenamefont
  {Wong}(1999)}]{Nishimori1999}%
  \BibitemOpen
  \bibfield  {author} {\bibinfo {author} {\bibfnamefont {H.}~\bibnamefont
  {Nishimori}}\ and\ \bibinfo {author} {\bibfnamefont {K.~Y.~M.}\ \bibnamefont
  {Wong}},\ }\href {\doibase 10.1103/PhysRevE.60.132} {\bibfield  {journal}
  {\bibinfo  {journal} {Phys. Rev. E}\ }\textbf {\bibinfo {volume} {60}},\
  \bibinfo {pages} {132} (\bibinfo {year} {1999})}\BibitemShut {NoStop}%
\bibitem [{\citenamefont {Chancellor}\ \emph {et~al.}(2016)\citenamefont
  {Chancellor}, \citenamefont {Szoke}, \citenamefont {Vinci}, \citenamefont
  {Aeppli},\ and\ \citenamefont {Warburton}}]{Chancellor2016}%
  \BibitemOpen
  \bibfield  {author} {\bibinfo {author} {\bibfnamefont {N.}~\bibnamefont
  {Chancellor}}, \bibinfo {author} {\bibfnamefont {S.}~\bibnamefont {Szoke}},
  \bibinfo {author} {\bibfnamefont {W.}~\bibnamefont {Vinci}}, \bibinfo
  {author} {\bibfnamefont {G.}~\bibnamefont {Aeppli}}, \ and\ \bibinfo {author}
  {\bibfnamefont {P.~A.}\ \bibnamefont {Warburton}},\ }\href {\doibase
  10.1038/srep22318} {\bibfield  {journal} {\bibinfo  {journal} {Sci. Rep.}\
  }\textbf {\bibinfo {volume} {6}},\ \bibinfo {pages} {22318} (\bibinfo {year}
  {2016})}\BibitemShut {NoStop}%
\bibitem [{\citenamefont {Tanaka}\ \emph {et~al.}(1980)\citenamefont {Tanaka},
  \citenamefont {Furusawa},\ and\ \citenamefont {Kaneku}}]{Tanaka1980}%
  \BibitemOpen
  \bibfield  {author} {\bibinfo {author} {\bibfnamefont {H.}~\bibnamefont
  {Tanaka}}, \bibinfo {author} {\bibfnamefont {K.}~\bibnamefont {Furusawa}}, \
  and\ \bibinfo {author} {\bibfnamefont {S.}~\bibnamefont {Kaneku}},\
  }\href@noop {} {\bibfield  {journal} {\bibinfo  {journal} {IEEE Trans. Inf.
  Theory}\ }\textbf {\bibinfo {volume} {IT-26}},\ \bibinfo {pages} {244}
  (\bibinfo {year} {1980})}\BibitemShut {NoStop}%
\bibitem [{\citenamefont {Rudolph}\ and\ \citenamefont
  {Robbins}(1972)}]{Rudolph1972}%
  \BibitemOpen
  \bibfield  {author} {\bibinfo {author} {\bibfnamefont {L.}~\bibnamefont
  {Rudolph}}\ and\ \bibinfo {author} {\bibfnamefont {W.}~\bibnamefont
  {Robbins}},\ }\href {\doibase 10.1109/TIT.1972.1054807} {\bibfield  {journal}
  {\bibinfo  {journal} {IEEE Trans. Inf. Theory}\ }\textbf {\bibinfo {volume}
  {IT-18}},\ \bibinfo {pages} {446} (\bibinfo {year} {1972})}\BibitemShut
  {NoStop}%
\bibitem [{\citenamefont {Lidar}(2019)}]{Lidar2019}%
  \BibitemOpen
  \bibfield  {author} {\bibinfo {author} {\bibfnamefont {D.~A.}\ \bibnamefont
  {Lidar}},\ }\href {\doibase 10.1103/PhysRevA.100.022326} {\bibfield
  {journal} {\bibinfo  {journal} {Phys. Rev. A}\ }\textbf {\bibinfo {volume}
  {100}},\ \bibinfo {pages} {022326} (\bibinfo {year} {2019})}\BibitemShut
  {NoStop}%
\bibitem [{\citenamefont {Young}\ \emph
  {et~al.}(2013{\natexlab{b}})\citenamefont {Young}, \citenamefont {Sarovar},\
  and\ \citenamefont {Blume-Kohout}}]{Young2013-2}%
  \BibitemOpen
  \bibfield  {author} {\bibinfo {author} {\bibfnamefont {K.~C.}\ \bibnamefont
  {Young}}, \bibinfo {author} {\bibfnamefont {M.}~\bibnamefont {Sarovar}}, \
  and\ \bibinfo {author} {\bibfnamefont {R.}~\bibnamefont {Blume-Kohout}},\
  }\href {\doibase 10.1103/PhysRevX.3.041013} {\bibfield  {journal} {\bibinfo
  {journal} {Phys. Rev. X}\ }\textbf {\bibinfo {volume} {3}},\ \bibinfo {pages}
  {041013} (\bibinfo {year} {2013}{\natexlab{b}})}\BibitemShut {NoStop}%
\bibitem [{\citenamefont {Sarovar}\ and\ \citenamefont
  {Young}(2013)}]{Sarovar2013}%
  \BibitemOpen
  \bibfield  {author} {\bibinfo {author} {\bibfnamefont {M.}~\bibnamefont
  {Sarovar}}\ and\ \bibinfo {author} {\bibfnamefont {K.~C.}\ \bibnamefont
  {Young}},\ }\href {http://arxiv.org/abs/1307.5892
  http://dx.doi.org/10.1088/1367-2630/15/12/125032} {\bibfield  {journal}
  {\bibinfo  {journal} {New J. Phys.}\ }\textbf {\bibinfo {volume} {15}},\
  \bibinfo {pages} {125032} (\bibinfo {year} {2013})}\BibitemShut {NoStop}%
\bibitem [{\citenamefont {Amin}(2015)}]{Amin2015}%
  \BibitemOpen
  \bibfield  {author} {\bibinfo {author} {\bibfnamefont {M.~H.}\ \bibnamefont
  {Amin}},\ }\href {\doibase 10.1103/PhysRevA.92.052323} {\bibfield  {journal}
  {\bibinfo  {journal} {Phys. Rev. A}\ }\textbf {\bibinfo {volume} {92}},\
  \bibinfo {pages} {052323} (\bibinfo {year} {2015})}\BibitemShut {NoStop}%
\bibitem [{\citenamefont {Benedetti}\ \emph {et~al.}(2016)\citenamefont
  {Benedetti}, \citenamefont {Realpe-Gomez}, \citenamefont {Biswas},\ and\
  \citenamefont {Perdomo-Ortiz}}]{Benedetti2016}%
  \BibitemOpen
  \bibfield  {author} {\bibinfo {author} {\bibfnamefont {M.}~\bibnamefont
  {Benedetti}}, \bibinfo {author} {\bibfnamefont {J.}~\bibnamefont
  {Realpe-Gomez}}, \bibinfo {author} {\bibfnamefont {R.}~\bibnamefont
  {Biswas}}, \ and\ \bibinfo {author} {\bibfnamefont {A.}~\bibnamefont
  {Perdomo-Ortiz}},\ }\href {\doibase 10.1103/PhysRevA.94.022308} {\bibfield
  {journal} {\bibinfo  {journal} {Phys. Rev. A}\ }\textbf {\bibinfo {volume}
  {94}},\ \bibinfo {pages} {22308} (\bibinfo {year} {2016})}\BibitemShut
  {NoStop}%
\bibitem [{\citenamefont {Marshall}\ \emph {et~al.}(2019)\citenamefont
  {Marshall}, \citenamefont {Venturelli}, \citenamefont {Hen},\ and\
  \citenamefont {Rieffel}}]{Marshall2019}%
  \BibitemOpen
  \bibfield  {author} {\bibinfo {author} {\bibfnamefont {J.}~\bibnamefont
  {Marshall}}, \bibinfo {author} {\bibfnamefont {D.}~\bibnamefont
  {Venturelli}}, \bibinfo {author} {\bibfnamefont {I.}~\bibnamefont {Hen}}, \
  and\ \bibinfo {author} {\bibfnamefont {E.~G.}\ \bibnamefont {Rieffel}},\
  }\href {\doibase 10.1103/PhysRevApplied.11.044083} {\bibfield  {journal}
  {\bibinfo  {journal} {Phys. Rev. Appl.}\ }\textbf {\bibinfo {volume} {11}},\
  \bibinfo {pages} {044083} (\bibinfo {year} {2019})}\BibitemShut {NoStop}%
\bibitem [{\citenamefont {Watanabe}\ \emph {et~al.}(2006)\citenamefont
  {Watanabe}, \citenamefont {Yukawa}, \citenamefont {Novotny},\ and\
  \citenamefont {Ito}}]{Watanabe2006}%
  \BibitemOpen
  \bibfield  {author} {\bibinfo {author} {\bibfnamefont {H.}~\bibnamefont
  {Watanabe}}, \bibinfo {author} {\bibfnamefont {S.}~\bibnamefont {Yukawa}},
  \bibinfo {author} {\bibfnamefont {M.~A.}\ \bibnamefont {Novotny}}, \ and\
  \bibinfo {author} {\bibfnamefont {N.}~\bibnamefont {Ito}},\ }\href {\doibase
  10.1103/PHYSREVE.74.026707/FIGURES/13/MEDIUM} {\bibfield  {journal} {\bibinfo
   {journal} {Phys. Rev. E}\ }\textbf {\bibinfo {volume} {74}},\ \bibinfo
  {pages} {026707} (\bibinfo {year} {2006})}\BibitemShut {NoStop}%
\bibitem [{\citenamefont {Mazumdar}(2012)}]{Mazumdar2012}%
  \BibitemOpen
  \bibfield  {author} {\bibinfo {author} {\bibfnamefont {R.~R.}\ \bibnamefont
  {Mazumdar}},\ }\href@noop {} {\enquote {\bibinfo {title} {Continuous-time
  markov chains (ctmc)},}\ }\bibinfo {howpublished}
  {\url{https://ece.uwaterloo.ca/~mazum/ECE605_2012/Notes/MarkovChains_CTMC.pdf}}
  (\bibinfo {year} {2012})\BibitemShut {NoStop}%
\bibitem [{\citenamefont {Parekh}(2020)}]{Parekh2020}%
  \BibitemOpen
  \bibfield  {author} {\bibinfo {author} {\bibfnamefont {S.}~\bibnamefont
  {Parekh}},\ }\href@noop {} {\enquote {\bibinfo {title} {Continuous time
  markov chains},}\ }\bibinfo {howpublished}
  {\url{https://inst.eecs.berkeley.edu/~ee126/fa20/notes/ctmcs.pdf}} (\bibinfo
  {year} {2020})\BibitemShut {NoStop}%
\bibitem [{\citenamefont {Rosenthal}\ \emph {et~al.}(2021)\citenamefont
  {Rosenthal}, \citenamefont {Dote}, \citenamefont {Dabiri}, \citenamefont
  {Tamura}, \citenamefont {Chen},\ and\ \citenamefont
  {Sheikholeslami}}]{Rosenthal2021}%
  \BibitemOpen
  \bibfield  {author} {\bibinfo {author} {\bibfnamefont {J.~S.}\ \bibnamefont
  {Rosenthal}}, \bibinfo {author} {\bibfnamefont {A.}~\bibnamefont {Dote}},
  \bibinfo {author} {\bibfnamefont {K.}~\bibnamefont {Dabiri}}, \bibinfo
  {author} {\bibfnamefont {H.}~\bibnamefont {Tamura}}, \bibinfo {author}
  {\bibfnamefont {S.}~\bibnamefont {Chen}}, \ and\ \bibinfo {author}
  {\bibfnamefont {A.}~\bibnamefont {Sheikholeslami}},\ }\href {\doibase
  10.1007/S00180-021-01095-2} {\bibfield  {journal} {\bibinfo  {journal}
  {Comput. Stat.}\ }\textbf {\bibinfo {volume} {36}},\ \bibinfo {pages} {2789}
  (\bibinfo {year} {2021})}\BibitemShut {NoStop}%
\bibitem [{\citenamefont {Iba}(2001{\natexlab{a}})}]{IBA2001}%
  \BibitemOpen
  \bibfield  {author} {\bibinfo {author} {\bibfnamefont {Y.}~\bibnamefont
  {Iba}},\ }\href {\doibase 10.1142/S0129183101001912} {\bibfield  {journal}
  {\bibinfo  {journal} {Int. J. Mod. Phys.}\ }\textbf {\bibinfo {volume} {C
  12}},\ \bibinfo {pages} {623} (\bibinfo {year}
  {2001}{\natexlab{a}})}\BibitemShut {NoStop}%
\bibitem [{\citenamefont {Iba}(2001{\natexlab{b}})}]{Iba2001-2}%
  \BibitemOpen
  \bibfield  {author} {\bibinfo {author} {\bibfnamefont {Y.}~\bibnamefont
  {Iba}},\ }\href {\doibase 10.1527/tjsai.16.279} {\bibfield  {journal}
  {\bibinfo  {journal} {Trans. Jpn. Soc. Artif. Intell.}\ }\textbf {\bibinfo
  {volume} {16}},\ \bibinfo {pages} {279} (\bibinfo {year}
  {2001}{\natexlab{b}})}\BibitemShut {NoStop}%
\bibitem [{\citenamefont {Campillo}\ \emph {et~al.}(2009)\citenamefont
  {Campillo}, \citenamefont {Rakotozafy},\ and\ \citenamefont
  {Rossi}}]{Campillo2009}%
  \BibitemOpen
  \bibfield  {author} {\bibinfo {author} {\bibfnamefont {F.}~\bibnamefont
  {Campillo}}, \bibinfo {author} {\bibfnamefont {R.}~\bibnamefont
  {Rakotozafy}}, \ and\ \bibinfo {author} {\bibfnamefont {V.}~\bibnamefont
  {Rossi}},\ }\href {\doibase 10.1016/J.MATCOM.2009.04.010} {\bibfield
  {journal} {\bibinfo  {journal} {Math. Comput. Simul.}\ }\textbf {\bibinfo
  {volume} {79}},\ \bibinfo {pages} {3424} (\bibinfo {year}
  {2009})}\BibitemShut {NoStop}%
\bibitem [{\citenamefont {Cappe}\ \emph {et~al.}(2004)\citenamefont {Cappe},
  \citenamefont {Guillin}, \citenamefont {Marin},\ and\ \citenamefont
  {Robert}}]{Cappe2004}%
  \BibitemOpen
  \bibfield  {author} {\bibinfo {author} {\bibfnamefont {O.}~\bibnamefont
  {Cappe}}, \bibinfo {author} {\bibfnamefont {A.}~\bibnamefont {Guillin}},
  \bibinfo {author} {\bibfnamefont {J.~M.}\ \bibnamefont {Marin}}, \ and\
  \bibinfo {author} {\bibfnamefont {C.~P.}\ \bibnamefont {Robert}},\ }\href
  {\doibase 10.1198/106186004X12803} {\bibfield  {journal} {\bibinfo  {journal}
  {J. Comput. Graph. Stat.}\ }\textbf {\bibinfo {volume} {13}},\ \bibinfo
  {pages} {907} (\bibinfo {year} {2004})}\BibitemShut {NoStop}%
\bibitem [{\citenamefont {Seabrook}\ and\ \citenamefont
  {Wiskott}(2022)}]{Seabrook2022}%
  \BibitemOpen
  \bibfield  {author} {\bibinfo {author} {\bibfnamefont {E.}~\bibnamefont
  {Seabrook}}\ and\ \bibinfo {author} {\bibfnamefont {L.}~\bibnamefont
  {Wiskott}},\ }\href {http://arxiv.org/abs/2207.02296} {\enquote {\bibinfo
  {title} {A tutorial on the spectral theory of markov chains},}\ } (\bibinfo
  {year} {2022}),\ \Eprint {http://arxiv.org/abs/arXiv:2207.02296v2 [cs.LG]}
  {arXiv:2207.02296v2 [cs.LG]} \BibitemShut {NoStop}%
\bibitem [{\citenamefont {Hopfield}\ and\ \citenamefont
  {Tank}(1986)}]{Hopfield86}%
  \BibitemOpen
  \bibfield  {author} {\bibinfo {author} {\bibfnamefont {J.~J.}\ \bibnamefont
  {Hopfield}}\ and\ \bibinfo {author} {\bibfnamefont {D.~W.}\ \bibnamefont
  {Tank}},\ }\href@noop {} {\bibfield  {journal} {\bibinfo  {journal}
  {Science}\ }\textbf {\bibinfo {volume} {233}},\ \bibinfo {pages} {625}
  (\bibinfo {year} {1986})}\BibitemShut {NoStop}%
\bibitem [{\citenamefont {Sourlas}(1989)}]{Sourlas1989}%
  \BibitemOpen
  \bibfield  {author} {\bibinfo {author} {\bibfnamefont {N.}~\bibnamefont
  {Sourlas}},\ }\href {\doibase 10.1038/339693a0} {\bibfield  {journal}
  {\bibinfo  {journal} {Nature}\ }\textbf {\bibinfo {volume} {339}},\ \bibinfo
  {pages} {693} (\bibinfo {year} {1989})}\BibitemShut {NoStop}%
\bibitem [{\citenamefont {Goto}\ \emph {et~al.}(2019)\citenamefont {Goto},
  \citenamefont {Tatsumura},\ and\ \citenamefont {Dixon}}]{Goto19}%
  \BibitemOpen
  \bibfield  {author} {\bibinfo {author} {\bibfnamefont {H.}~\bibnamefont
  {Goto}}, \bibinfo {author} {\bibfnamefont {K.}~\bibnamefont {Tatsumura}}, \
  and\ \bibinfo {author} {\bibfnamefont {A.~R.}\ \bibnamefont {Dixon}},\
  }\href@noop {} {\bibfield  {journal} {\bibinfo  {journal} {Sci. Adv.}\
  }\textbf {\bibinfo {volume} {5}},\ \bibinfo {pages} {eaav2372} (\bibinfo
  {year} {2019})}\BibitemShut {NoStop}%
\bibitem [{\citenamefont {Inagaki}\ \emph {et~al.}(2016)\citenamefont
  {Inagaki}, \citenamefont {Haribara}, \citenamefont {Igarashi}, \citenamefont
  {Sonobe}, \citenamefont {Tamate}, \citenamefont {Honjo}, \citenamefont
  {Marandi}, \citenamefont {McMahon}, \citenamefont {Umeki}, \citenamefont
  {Enbutsu}, \citenamefont {Tadanaga}, \citenamefont {Takenouchi},
  \citenamefont {Aihara}, \citenamefont {Kawarabayashi}, \citenamefont {Inoue},
  \citenamefont {Utsunomiya}, ,\ and\ \citenamefont {Takesue}}]{Inagaki16}%
  \BibitemOpen
  \bibfield  {author} {\bibinfo {author} {\bibfnamefont {T.}~\bibnamefont
  {Inagaki}}, \bibinfo {author} {\bibfnamefont {Y.}~\bibnamefont {Haribara}},
  \bibinfo {author} {\bibfnamefont {K.}~\bibnamefont {Igarashi}}, \bibinfo
  {author} {\bibfnamefont {T.}~\bibnamefont {Sonobe}}, \bibinfo {author}
  {\bibfnamefont {S.}~\bibnamefont {Tamate}}, \bibinfo {author} {\bibfnamefont
  {T.}~\bibnamefont {Honjo}}, \bibinfo {author} {\bibfnamefont
  {A.}~\bibnamefont {Marandi}}, \bibinfo {author} {\bibfnamefont {P.~L.}\
  \bibnamefont {McMahon}}, \bibinfo {author} {\bibfnamefont {T.}~\bibnamefont
  {Umeki}}, \bibinfo {author} {\bibfnamefont {K.}~\bibnamefont {Enbutsu}},
  \bibinfo {author} {\bibfnamefont {O.}~\bibnamefont {Tadanaga}}, \bibinfo
  {author} {\bibfnamefont {H.}~\bibnamefont {Takenouchi}}, \bibinfo {author}
  {\bibfnamefont {K.}~\bibnamefont {Aihara}}, \bibinfo {author} {\bibfnamefont
  {K.}~\bibnamefont {Kawarabayashi}}, \bibinfo {author} {\bibfnamefont
  {K.}~\bibnamefont {Inoue}}, \bibinfo {author} {\bibfnamefont
  {S.}~\bibnamefont {Utsunomiya}}, , \ and\ \bibinfo {author} {\bibfnamefont
  {H.}~\bibnamefont {Takesue}},\ }\href@noop {} {\bibfield  {journal} {\bibinfo
   {journal} {Science}\ }\textbf {\bibinfo {volume} {354}},\ \bibinfo {pages}
  {603} (\bibinfo {year} {2016})}\BibitemShut {NoStop}%
\bibitem [{\citenamefont {Marvian}\ and\ \citenamefont
  {Lidar}(2014)}]{Marvian2014}%
  \BibitemOpen
  \bibfield  {author} {\bibinfo {author} {\bibfnamefont {I.}~\bibnamefont
  {Marvian}}\ and\ \bibinfo {author} {\bibfnamefont {D.~A.}\ \bibnamefont
  {Lidar}},\ }\href {\doibase 10.1103/PhysRevLett.113.260504} {\bibfield
  {journal} {\bibinfo  {journal} {Phys. Rev. Lett.}\ }\textbf {\bibinfo
  {volume} {113}},\ \bibinfo {pages} {260504} (\bibinfo {year}
  {2014})}\BibitemShut {NoStop}%
\bibitem [{\citenamefont {Marvian}\ and\ \citenamefont
  {Lidar}(2017)}]{Marvian2017}%
  \BibitemOpen
  \bibfield  {author} {\bibinfo {author} {\bibfnamefont {M.}~\bibnamefont
  {Marvian}}\ and\ \bibinfo {author} {\bibfnamefont {D.~A.}\ \bibnamefont
  {Lidar}},\ }\href {\doibase 10.1103/PhysRevLett.118.030504} {\bibfield
  {journal} {\bibinfo  {journal} {Phys. Rev. Lett.}\ }\textbf {\bibinfo
  {volume} {118}},\ \bibinfo {pages} {030504} (\bibinfo {year}
  {2017})}\BibitemShut {NoStop}%
\end{thebibliography}%

\end{document}